\documentclass[journal,draftcls,onecolumn]{IEEEtran}

\usepackage{times}
\usepackage{epsfig}
\usepackage{colortbl}
\usepackage{float}
\usepackage{subfig}
\usepackage{verbatim}

\usepackage[latin1]{inputenc}
\usepackage[T1]{fontenc}
\usepackage{lmodern}
\usepackage{amsmath}
\usepackage{amsthm}
\usepackage{amssymb}
\usepackage{bbm}

\usepackage{graphicx}
\usepackage{tikz}
\usepackage{pgfplots}
\usepackage[right]{eurosym}
\usepackage{multirow}
\usepackage{color}
\usepackage{psfrag}
\usepackage{todonotes}
\usepackage[normalem]{ulem} 
\usepackage{cite} 

\usepackage{cleveref} 
\usepackage{pdfpages} 

\usetikzlibrary{shapes,decorations,patterns}
\usetikzlibrary{arrows,backgrounds,fit}
\tikzstyle{block} = [draw, align=center, rectangle, minimum height=2.5em, minimum width=2.5em]
\tikzstyle{dmc_block} = [draw, align=center, rectangle, rounded corners, minimum height=6em, minimum width=5em]
\tikzstyle{receiver_block} = [draw, align=center, rectangle, rounded corners, minimum height=2.5em, minimum width=5em]
\tikzstyle{transmitter_block} = [draw, align=center, rectangle, rounded corners, minimum height=4em, minimum width=5em]
\tikzstyle{mul} = [draw, circle, inner sep=0pt]
\tikzstyle{dot} = [draw, circle, minimum size=0.2pt,scale=0.3, fill=black,black]

\pgfdeclarepatternformonly{soft horizontal lines}{\pgfpointorigin}{\pgfqpoint{100pt}{1pt}}{\pgfqpoint{100pt}{3pt}}%
{
  \pgfsetstrokeopacity{0.3}
  \pgfsetlinewidth{0.1pt}
  \pgfpathmoveto{\pgfqpoint{0pt}{0.5pt}}
  \pgfpathlineto{\pgfqpoint{100pt}{0.5pt}}
  \pgfusepath{stroke}
}

\pgfdeclarepatternformonly{soft crosshatch}{\pgfqpoint{-1pt}{-1pt}}{\pgfqpoint{4pt}{4pt}}{\pgfqpoint{3pt}{3pt}}%
{
  \pgfsetstrokeopacity{0.3}
  \pgfsetlinewidth{0.4pt}
  \pgfpathmoveto{\pgfqpoint{3.1pt}{0pt}}
  \pgfpathlineto{\pgfqpoint{0pt}{3.1pt}}
  \pgfpathmoveto{\pgfqpoint{0pt}{0pt}}
  \pgfpathlineto{\pgfqpoint{3.1pt}{3.1pt}}
  \pgfusepath{stroke}
}

\newtheorem{theorem}{Theorem}
\newtheorem{proposition}[theorem]{Proposition}
\newtheorem{corollary}[theorem]{Corollary}
\newtheorem{lemma}[theorem]{Lemma}
\newtheorem{remark}[theorem]{Remark}

\newfont{\bbb}{msbm10 scaled 500}

\newfont{\bb}{msbm10 scaled 1100}

\newcommand{\RR}{\mbox{\bb R}}

\newcommand{\Prob}{\textrm{Pr}}

\newcommand{\Mc}{{\cal M}}

\newcommand{\Xc}{{\cal X}}
\newcommand{\Yc}{{\cal Y}}

\makeatletter
\newcommand*{\rom}[1]{\expandafter\@slowromancap\romannumeral #1@}
\makeatother

\definecolor{OXO-emph}{RGB}{153,0,0}

\graphicspath{{./fig/}}
\DeclareGraphicsExtensions{.pdf} 

\begin{document}


\title{Individual Secrecy for Broadcast Channels \\with Receiver Side Information 
\thanks{This paper was presented in part at International Zurich Seminar on Communications, Zurich, Switzerland, Feb. 2014 and IEEE International Symposium on Information Theory, Honolulu, HI, Jun. 2014.}
\thanks{
Y. Chen and A. Sezgin are with the Institute of Digital Communication Systems, Ruhr University Bochum, Germany (e-mail: yanling.chen-q5g@rub.de, aydin.sezgin@rub.de).
O. O. Koyluoglu is with the Department of Electrical and Computer
Engineering, The University of Arizona, Tucson, AZ 85721, USA (e-mail:
ozan@email.arizona.edu).}
}

\author{\IEEEauthorblockN{Yanling Chen, O.~Ozan~Koyluoglu, and Aydin Sezgin}}

\maketitle


\begin{abstract}

This paper studies the problem of secure communication over the broadcast channel with receiver side information under the lens of individual secrecy constraints. That is, the transmitter wants to send two independent messages to two receivers which have, respectively, the desired message of the other receiver as side information, while keeping the eavesdropper ignorant of {\it each} message (i.e., the information leakage from {\em each} message to the eavesdropper is made vanishing). 
Building upon one-time pad, secrecy coding, and broadcasting schemes, achievable rate regions are investigated, and the capacity region for special cases of either a {\it weak} or {\it strong} eavesdropper (compared to both legitimate receivers) 
are characterized. Interestingly, the capacity region for the former corresponds to a line and the latter corresponds to a square with missing corners; a phenomenon occurring due to the coupling between user's rates. Moreover, the individual secrecy capacity region is also fully characterized for the case where the eavesdropper's channel is {\em deterministic}.  
In addition to discrete memoryless setup, Gaussian scenarios are studied. For the Gaussian model, in addition to the strong and weak eavesdropper cases, the capacity region is characterized for the low and high SNR regimes when the eavesdropper's channel is stronger than one receiver but weaker than the other. Remarkably, positive secure transmission rates are always guaranteed under the individual secrecy constraint, unlike the case of the joint secrecy constraint (i.e., the information leakage from {\it both} messages to the eavesdropper is made vanishing). Thus, this notion of secrecy serves as an appropriate candidate for trading off secrecy level and transmission rate; making secrecy more {\it affordable} but still {\it acceptable} to the end user.  

\end{abstract}


\section{Introduction}

\subsection{Background}

The broadcast channel is a fundamental communication model that involves transmission of independent messages to different users. However, the broadcast nature makes the communication very susceptible to eavesdropping. Therefore, it is desirable to offer a reliable communication with a certain level of security guarantee, especially for ensuring sensitive information to be protected from unauthorized parties. 

The problem of secure communication from an information theoretic point of view was first studied by Shannon \cite{src:Shannon1949}. In this work, a cipher system was introduced under the assumption that the transmitter and the intended receiver share a secret random key which is out of the eavesdropper's knowledge. For the purpose of a secure communication, the message is first encrypted into a ciphertext before being transmitted, and it is assumed that the eavesdropper has full access to the ciphertext as the intended receiver. A cipher system with {\it perfect secrecy} demands that knowing the ciphertext, however, gives no clue about the message. Such a perfect cipher system is shown to be possible via the so-called {\it one-time pad} scheme \cite{src:Shannon1949} (previously porposed by Vernam \cite{src:Vernam1926}), provided that the secret key is sufficient to randomize the message. 

Wyner, in his seminal paper \cite{src:Wyner1975}, introduced the wiretap channel, where he addressed the problem of secret message transmission from a transmitter to a legitimate receiver (without sharing keys beforehand) over a degraded broadcast channel in the presence of an eavesdropper. It is shown that the secure communication is still possible when the eavesdropper observes a degraded version of the legitimate receiver's observation. The fundamental limit of  secure communication, i.e., {\it secrecy capacity}, is defined to be the maximum rate under a weak secrecy constraint, where the rate of information leaked to the eavesdropper is made vanishing. Later on, Csisz\'{a}r and K\"{o}rner \cite{Csisz'ar:Broadcast78} extended Wyner's work by considering a setup of transmitting secret and common message over a general broadcast channel, and provided a single-letter characterization of the secrecy capacity. Notably, the secrecy capacity results hold also under a strong secrecy constraint, where the total amount of information leaked to the eavesdropper is made vanishing, as demonstrated in \cite{src:Maurer2000}.

For those wiretap channels where the legitimate receiver does not have any advantage over the eavesdropper, interestingly, Maurer \cite{src:Maurer1993} demonstrated that it is still possible to achieve a positive secret rate if a public feedback channel is made available. In parallel, Csisz\'{a}r and Ahlswede \cite{src:Ahlswede1993} recognized that correlated source observations could be explored for generating secret key that could be used further for secret message transmission via one-time pad. These offer alternative solutions to achieve information theoretic secrecy, which are especially interesting in cases that the legitimate users have no advantage against the eavesdropper on the communication channels. 

Inspired by these pioneering works, there has been a body of growing literature studying the problem of secret message transmission and/or secret key generation by exploring the resources available in different settings. Extensive types of resources have been taken into account in order to establish secret communications without much sacrifice, or turn the disadvantages into advantages so as to make the impossible possible or even improve the overall performance. Such resources include channel state information \cite{src:Mitrpant2006, src:Chen2004, src:Liuwei2007, src:Chia2012, src:mustafa2012}, side information \cite{src:Boche2013}, feedback \cite{src:Ahlswede2006, src:Gunduz2008, src:Lai2008, src:Ardestanizadeh2009, src:Rezki2014}, correlated sources \cite{src:Khisti2012, src:Prabhakaran2012, src:Chen2014Cai} or shared keys \cite{Yamamoto:Rate97, src:Merhav2008, Kang:Wiretap10}, and so on. In the meantime, the channel, still serves as one of the most significant resource for secure communication. Several communication channels of particular practical interest have received intense research attention. Instances include but not limited to the broadcast channels \cite{src:Ruoheng2013, Bagherikaram:Secrecy09, src:Ekrem2012MIMO, src:Wyrembelski2013}, multiple access channels \cite{src:Liang2008MAC}, two-way channel \cite{src:Pierrot2011, src:ElGamal2013}, the interference channels\cite{src:Ruoheng2008, Koyluoglu:Cooperative11}, and compound channels\cite{src:Ekrem2012Compound}. 

\subsection{Contributions}
In this paper, we consider the problem of secure communication over the broadcast channel with receiver side information (BC-RSI). The model is different from the wiretap channel with side information due to the broadcast nature of the communication channel. That is, in this model, the transmitter wants to send two independent messages to two receivers which have, respectively, the desired message of the other receiver (already available in their possession, e.g., due to previous communications) as side information. (See Fig. \ref{fig: wiretap channel with receiver side info}.) This is a simple setup of a general scenario, which consists of more than two legitimate receivers, each having a piece of partial information about the transmitted message. In the following, we summarize the main contributions of the paper:

\begin{itemize}
\item The linear deterministic model is studied and corresponding individual secrecy capacity region is characterized. 
Due to its relevance to the corresponding Gaussian case, study of this specific model provides insight into the individual secrecy capacity region of Gaussian case especially in the high SNR regime.

\item To investigate the fundamental limits of communication under individual secrecy constraints, constructions building upon one-time pad, wiretap coding, superposition coding, and Marton's coding are proposed. 
\begin{itemize}
	\item First construction, referred to as secret key approach, utilizes side information at receivers as secret keys of one-time pad signals, which further is encoded as \emph{cloud centers} in broadcast coding schemes. This approach is shown to be capacity achieving for a {\em strong} eavesdropper (compared to both legitimate receivers).
	\item Secret key approach is extended with secrecy coding, where the one-time pad signal is  utilized as a \emph{part of the randomization} to confuse the eavesdropper (i.e., to limit her ability to obtain information regarding each message). This approach is shown to be capacity achieving for a {\em weak} eavesdropper (compared to both legitimate receivers).
	\item The proposed superposition coding can be considered as an extension of secret key approach and combined secret key and secrecy coding approach. It takes advantage of the rate splitting of one-time pad signals such that they serve for two distinct purposes: 1) as a cloud center; and 2) as a part of randomization within the satellite codewords to confuse the eavesdropper. Also, it is shown that the suggested rate splitting is sufficient within superpostion coding since further rate splitting does not improve the established region. Remarkably, superposition coding is shown to be optimal for special cases of either a {\em strong} or {\em weak} eavesdropper (compared to both legitimate receivers), and in case that the eavesdropper has a {\em deterministic} channel. 
	\item The proposed Marton's coding approach is built on the superposition coding but with one additional coding layer that employs Marton's coding. The idea is to further explore the advantage of rate splitting at the encoding phase (with introduction of joint distributed satellite codewords which carry independent message pieces intended for each legitimate receiver); and at the decoding phase only the individual satellite codewords will be decoded. As a result, a general achievable individual secrecy rate region is established, which not only includes but further improves the region obtained by superposition coding approach. The improvement is demonstrated for the {\em mixed} case where the eavesdropper's channel is weaker than one of the legitimate receivers channels but stronger than the other.
	\item As a by-product, two achievable {\em joint secrecy} rate regions are also obtained by the proposed superposition coding approach and Marton's coding approach, respectively; in which the former is included and potentially improved by the latter, i.e., Marton's coding approach.
\end{itemize}
\item Gaussian model is studied. And, in addition to strong and weak eavesdropper scenarios, the capacity region for low and high SNR regimes are characterized for the {\em mixed} case when the eavesdropper is stronger than one legitimate receiver but weaker than the other.
\end{itemize}

\subsection{Related Work}

Our model can be thought of as a broadcast phase of a relay network after a multiple access phase where the nodes transmit their messages to the relay in the first phase. Remarkably, this two-way relay setting simply illustrates how the information are shared in today's networked world. To maximize the broadcasting throughput, the technique employed at the relay node is very relevant to network coding. As demonstrated in \cite{Kramer:Capacity07}, the relay node (i.e., the transmitter in our model) can broadcast the XORed messages. Then, the legitimate receivers, utilizing the side information they have, can decode their intended message. The broadcasting capacity region (Fig. \ref{fig: wiretap channel with receiver side info} without an eavesdropper) is characterized in \cite{Kramer:Capacity07}. 

In addition to the broadcasting to share information in the most efficient way, the secrecy aspect of the communication has been a growing concern. Considering the existence of an external eavesdropper in the model of the broadcast channel with receiver side information (BC-RSI), the authors in~\cite{Wyrembelski:Secrecy11} proposed achievable rate regions and outer bounds subject to a \emph{joint} secrecy constraint, whereby the information leakage from \emph{both} messages to the eavesdropper is made vanishing. Differently from \cite{Wyrembelski:Secrecy11}, we focus on the problem under \emph{individual} secrecy constraints that aims to minimize the information leakage from \emph{each} message to the eavesdropper. Other relevant works include \cite{Wyrembelski:Physical12, Wyrembelski:Privacy12, Mansour:Secrecy14, Mansour:Capacity15}. The work \cite{Wyrembelski:Physical12} considered transmitting common and private messages to each user for the BC with side information model in addition to transmitting a confidential message to one of the users while treating the other as an eavesdropper. The same setting without common messages was considered in~\cite{Wyrembelski:Privacy12} and the secrecy capacity was characterized. Recently, in a parallel work \cite{Mansour:Secrecy14, Mansour:Capacity15}, Mansour et al. considered discrete memoryless broadcast channels with degraded message sets and message cognition. The model in \cite{Mansour:Capacity15}, when the common messages are removed and individual secrecy constraint is imposed, reduces to the model considered in this paper. In particular, the scenarios of weak and stronger eavesdroppers (as characterized in Theorem~\ref{thm: secret key CS} and Theorem~\ref{thm: Ind SC by secrecy coding} here) overlaps with the corresponding propositions in \cite{Mansour:Capacity15}, in which the authors consider more capable/less noisy scenarios as well. Our initial results on this topic are presented in \cite{Koyluoglu:Broadcast14, src:Chen2014-IndS}, and, in addition to stronger/weaker eavesdropper cases, we focus on other DMC models, deterministic channels, and Gaussian scenarios for BC-RSI with individual secrecy constraints in this paper.

Although the individual secrecy constraint is by definition weaker than the joint one, this notion nevertheless provides a security level that keeps each legitimate receiver away from non-negligible information leakage on its intended message, therefore acceptable to the end user. In addition, a joint secrecy constraint can be difficult or even impossible to fulfill in certain cases. For instance, when the eavesdropper has the same or a better channel observation than at least one of the legitimate receivers, imposing joint secrecy constraints result in a vanishing communication rate to the respective receiver. In this paper, we devote a particular attention to these \emph{mixed} scenarios, where the eavesdropper can be stronger than one receiver but weaker than the other. In such cases, individual secrecy serves as a practical security solution that is attainable. In fact, such a weaker security constraint is shown to be preferable in large-scale networks. For instance, this notion has the same spirit as the concept of {\em weak security} as defined in \cite{src:Bhattad2005} to guarantee that the eavesdropper is unable to get any {\em meaningful} information about the source in a multicast network scenario. In addition, a similar security criterion is considered to be sufficient for distributed storage systems. For instance, one can find its application in the design of secure cloud storage systems as proposed in \cite{Dau:Block13, src:JianLiu2013}.


\section{System model}\label{sec:SystemModel}

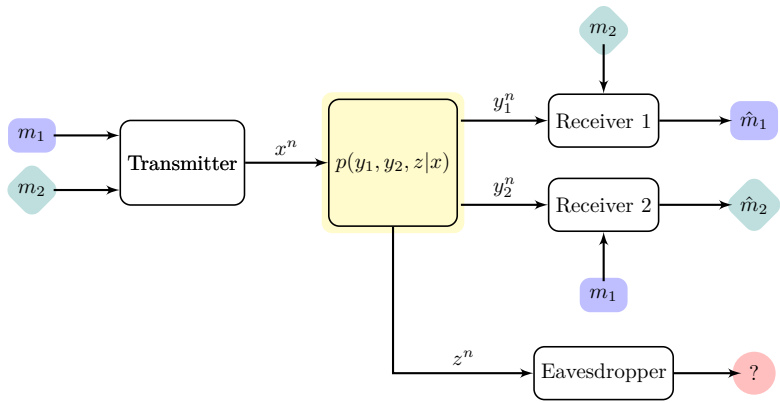
\begin{figure}[!h]
 \centering
 \begin{tikzpicture}[semithick, >=latex',scale=0.8, every node/.style={transform shape}]
 
  \begin{scope}[local bounding box=tx]
  
 	\node (enc) at (1.5,0) [transmitter_block] {Transmitter};
  \end{scope}
  \node (enc) at (1.5,0) {Transmitter};
  
  \begin{scope}[local bounding box=Receiver 1]
  \node (dec1) at (8.5,0.7) [receiver_block] {Receiver 1};
  \end{scope} 
  
  \begin{scope}[local bounding box=Receiver 2]
  \node (dec2) at (8.5,-0.7) [receiver_block] {Receiver 2};
  \end{scope} 
   
  \begin{scope}[local bounding box=Eavesdropper]  
  \node (eve) at (8.5,-3.5) [receiver_block] {Eavesdropper};
  \end{scope} 
  
 \begin{scope}[local bounding box=channel]
 \node (dmc) at (5,0) [dmc_block] {$p(y_1,y_2,z|x)$};
 \end{scope}
 
 \node (dmc1) at ([yshift=0.7cm, xshift=+1cm]dmc) {};
 \node (dmc2) at ([yshift=-0.7cm, xshift=+1cm]dmc){};
 
 \begin{scope}[local bounding box=sideinfo1]
 	\node (sidedec1) at (8.5,2.2) {$m_2$};
 \end{scope}
 
 \begin{scope}[local bounding box=sideinfo2]
 	\node (sidedec2) at (8.5,-2.2) {$m_1$};
 \end{scope}
 
  \draw[->,thick] (enc) to node[above] {$x^n$} (dmc);
  \draw[->,thick] (dmc1) to node[above] {$y_1^n$} (dec1);
  \draw[->,thick] (dmc2) to node[above] {$y_2^n$} (dec2);
  
  \draw[->,thick] (sidedec1) to (dec1);
  \draw[->,thick] (sidedec2) to (dec2);
  
  \draw[->,thick] (dmc) |- node[above, pos=0.75] {$z^n$} (eve);
  
  \begin{scope}[local bounding box=message1]

  \node (mes1) at ([yshift=3ex, xshift=-2.5cm]enc) {$m_1$};
  \end{scope}
  
  \begin{scope}[local bounding box=message2]

  \node (mes2) at ([yshift=-3ex, xshift=-2.5cm]enc) {$m_2$};
  \end{scope}
  
  \path[->,thick] (mes1) edge ([yshift=3ex]enc.west);
  \path[->,thick] (mes2) edge ([yshift=-3ex]enc.west);
  
  \begin{scope}[local bounding box=decmessage1]
  \node (mess1out) at ([xshift=2.5cm]dec1) {$\hat{m}_1$};
  \end{scope}
  
  \begin{scope}[local bounding box=decmessage2]
  \node (mess2out) at ([xshift=2.5cm]dec2) {$\hat{m}_2$};
  \end{scope}
  
  \path[->,thick] (dec1.east) edge (mess1out);
  \path[->,thick] (dec2.east) edge (mess2out);

  \begin{scope}[local bounding box=evemessage]
  \node (eveOut) at ([xshift=2.5cm]eve) {?};
  
  \end{scope}
  \draw[->,thick] (eve) to (eveOut);

  \begin{pgfonlayer}{background}
  
 	\node [fill=yellow!25,fit=(channel),rounded corners] {};
  
 	\node [fill=blue!25,fit=(message1),rounded corners, scale=0.85] {};
  
 	\node [shape=diamond, fill=teal!25,fit=(message2),rounded corners, scale=0.6] {};
  
 	\node [fill=blue!25,fit=(decmessage1),rounded corners, scale=0.85] {};
  
 	\node [shape=diamond, fill=teal!25,fit=(decmessage2),rounded corners, scale=0.6] {};
  
 	\node [shape=diamond, fill=teal!25,fit=(sideinfo1),rounded corners, scale=0.6] {};
  
 	\node [fill=blue!25,fit=(sideinfo2),rounded corners, scale=0.85] {};
  
 	\node [shape=circle, fill=red!25,fit=(evemessage),rounded corners, scale=0.75] {};
  \end{pgfonlayer}
 \end{tikzpicture}
  \caption{BC-RSI with an external eavesdropper.} \label{fig: wiretap channel with receiver side info}
 \end{figure}

Consider a discrete memoryless broadcast channel given by $p(y_1,y_2,z|x)$ with two legitimate receivers and one passive eavesdropper, as shown in Fig. \ref{fig: wiretap channel with receiver side info}. The transmitter aims to send messages $m_1,m_2$ to the legitimate receiver $1,2,$ respectively. Suppose $x^n$ is the channel input to convey $m_1, m_2$ in $n$ channel uses, whilst $y_1^n$ (at receiver 1), $y_2^n$ (at receiver 2) and $z^n$ (at eavesdropper), are the channel outputs. Besides, $m_2$ (available at receiver 1) and $m_1$ (available at receiver 2) serve as side information that help to decode the desired message. (Unless otherwise specified, we use capital letters for random variables, the corresponding calligraphic letters for their alphabets and small cases for their realizations.)

Encoder employed by the transmitter is a mapping $f:\Mc_1\times\Mc_2 \to \Xc^n$, where $m_1\in\Mc_1$, $m_2\in\Mc_2$, and $x^n\in\Xc^n$. (Here, the channel input alphabet is $\Xc$). Decoder employed at receiver $i$ is a mapping $g_i:\Yc_i^n\times \Mc_j \to \Mc_i$, where $j\neq i$, and $y_i^n\in\Yc_i^n$  for $i=1,2$. (Here, the channel output alphabet at receiver $i$ is $\Yc_i$.) Denote the average probability of decoding error at receiver $i$ as $P_{e,i}=\Prob\{m_i\neq g_i(y_i^n,m_j)\}$ with $j\neq i$. The rate pair $(R_1,R_2)$ is said to be \emph{achievable}, if for any $\epsilon>0,$ there exists an encoder-decoder tuple $(f,g_1,g_2)$ such that
\begin{align}
	\frac{1}{n}H (M_i) 	\geq & R_i-\epsilon \label{eq:R_i}\\
	P_{e,i} 					\leq & \epsilon \label{eq:P_e,i}\\
	\frac{1}{n}I(M_i;Z^n) 	\leq & \epsilon, \label{eq:IndSec}
\end{align}
for $i=1,2$ (and, for sufficiently large $n$). Note that \eqref{eq:R_i} corresponds to the targeted transmission \emph{rate}; \eqref{eq:P_e,i} corresponds to the \emph{reliability} constraint at the legitimate receivers; while \eqref{eq:IndSec} corresponds to the \emph{individual} secrecy constraints against the eavesdropper. If the coding scheme fulfils a stronger condition that
\begin{equation}\label{eq:JointSec}
\frac{1}{n}I(M_1,M_2;Z^n) \leq \epsilon,
\end{equation}
then it is said to satisfy the \emph{joint} secrecy constraint.  Clearly, the joint secrecy constraint implies the individual secrecy constraints.


\section{An illustrative example}\label{sec: example}

\begin{figure}[!h]
 \centering
 \begin{tikzpicture}[semithick, >=latex',scale=0.8, every node/.style={transform shape}]
 
  \begin{scope}[local bounding box=tx]
  	\node (enc) at (1.5,0) [transmitter_block] {Transmitter};
  \end{scope}
   
   \begin{scope}[local bounding box=message1]
   	 \node (mes1) at ([xshift=-2.5cm]enc) {$u^k$};
   \end{scope}
   
  \begin{scope}[local bounding box=Receiver 1]
  	\node (dec1) at (6,3.5) [receiver_block] {Receiver 1};
  \end{scope} 
  
  \begin{scope}[local bounding box=Receiver 2]
  	\node (dec2) at (6,1.5) [receiver_block] {Receiver 2};
  \end{scope} 
   
  \begin{scope}[local bounding box=Receiver k]
    \node (dec3) at (6,-2) [receiver_block] {Receiver $k$};
  \end{scope} 
  
   \begin{scope}[local bounding box=sideinfo1]
   	\node (sidedec1) at (6,4.5) {$u_1$};
   \end{scope}
   
   \begin{scope}[local bounding box=sideinfo2]
   	\node (sidedec2) at (6,2.5) {$u_2$};
   \end{scope}
   
   \begin{scope}[local bounding box=sideinfo3]
     \node (sidedec3) at (6,-1) {$u_k$};
   \end{scope}
   
   \draw[->,thick] (sidedec1) to (dec1);
   \draw[->,thick] (sidedec2) to (dec2); 
   \draw[->,thick] (sidedec3) to (dec3);
   
    \begin{scope}[local bounding box=decmessage1]
    	\node (mess1out) at ([xshift=2.5cm]dec1) {$u^k\backslash\{u_1\}$};
    \end{scope}
    
    \begin{scope}[local bounding box=decmessage2]
    	\node (mess2out) at ([xshift=2.5cm]dec2) {$u^k\backslash\{u_2\}$};
    \end{scope}
    
    \begin{scope}[local bounding box=decmessage3]
     	\node (mess3out) at ([xshift=2.5cm]dec3) {$u^k\backslash\{u_k\}$};
    \end{scope} 
   
  \begin{scope}[local bounding box=Eavesdropper]  
  	\node (eve) at (6,-3.5) [receiver_block] {Eavesdropper};
  \end{scope} 

  \begin{scope}[local bounding box=evemessage]
  	\node (eveOut) at ([xshift=2.5cm]eve) {?};
  \end{scope}
    
  	\node (dots) at (6,0.2) {$\vdots$};
  	\node (dots) at (8,0.2) {$\vdots$};
 	\node (dot1) at (4,0) [dot] {};

 	\draw[->,thick] (enc) to node[above] {$x^n$} (dot1);
   	\draw[->,thick] (dot1) |- (dec1);
   	\draw[->,thick] (dot1) |- (dec2);
   	\draw[->,thick] (dot1) |- (dec3);
   	\draw[->,thick] (dot1) |- (eve);
 	\path[->,thick] (mes1) edge (enc.west);
	\path[->,thick] (dec1.east) edge (mess1out);
  	\path[->,thick] (dec2.east) edge (mess2out);
  	\path[->,thick] (dec3.east) edge (mess3out);
	\draw[->,thick] (eve) to (eveOut);

  \begin{pgfonlayer}{background}
    
 	\node [fill=blue!25,fit=(message1),rounded corners, scale=0.85] {};
    
 	\node [fill=blue!25,fit=(decmessage1),rounded corners, scale=0.85] {};
 	\node [fill=blue!25,fit=(decmessage2),rounded corners, scale=0.85] {};
 	\node [fill=blue!25,fit=(decmessage3),rounded corners, scale=0.85] {};
  
 	\node [shape=diamond, fill=teal!25,fit=(sideinfo1),rounded corners, scale=0.6] {};
 	\node [shape=diamond, fill=teal!25,fit=(sideinfo2),rounded corners, scale=0.6] {};
 	\node [shape=diamond, fill=teal!25,fit=(sideinfo3),rounded corners, scale=0.6] {};
  
 	\node [shape=circle, fill=red!25,fit=(evemessage),rounded corners, scale=0.75] {};
  \end{pgfonlayer}
 \end{tikzpicture}
  \caption{Secure communication over $1$-to-$k$ broadcast channel with receiver side information.} \label{fig: an example 1 to k broadcast channel with an external eavesdropper}
 \end{figure}
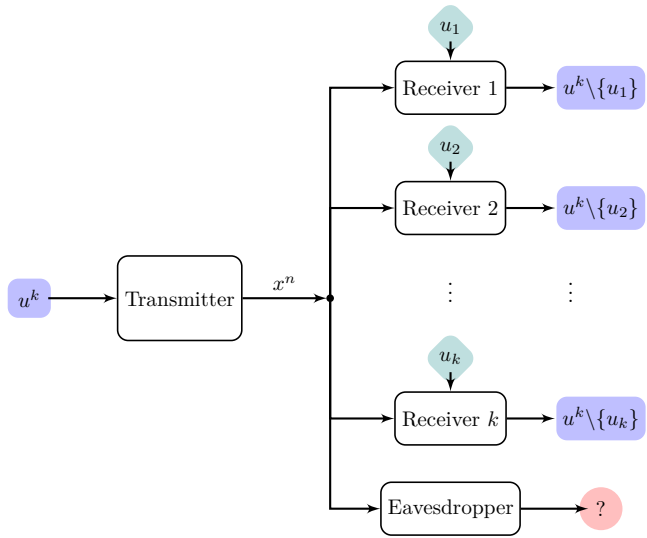

In this section, we motivate the {\em individual secrecy} constraint by using the scenario of $1$-to-$k$ broadcasting as shown in Fig. \ref{fig: an example 1 to k broadcast channel with an external eavesdropper}. The model consists of one transmitter, $k$ legitimate receivers, and one passive eavesdropper. The transmitter aims to broadcast $k$ information bits  $U^k=(U_1, U_2, \cdots, U_k)$ to $k$ legitimate receivers with $U_i\sim \mathrm{Bern}(1/2)$; whilst each receiver $i$ holds already one piece of information $U_i$ as side information. Suppose that $U^k$ is encoded into $X^n=(X_1, X_2, \cdots, X_n)$ and consider that this channel input is transmitted over noiseless channels. Then, for the purpose of broadcasting, each legitimate receiver $i$ (which holds $U_i$ and receives $X^n$) shall be able to recover the $k-1$ information bits $U^k\backslash\{U_i\}$, i.e.,
		\begin{equation}\label{eg: reliability}
			H(U^k|X^n, U_i)=0.
		\end{equation}
Thus, we have 
		\begin{align}
			H(U^k|X^n)&=H(U^k, U_i|X^n)=H(U_i|X^n)+H(U^k|X^n, U_i)\nonumber\\
						&\stackrel{\eqref{eg: reliability}}{=}H(U_i|X^n). \label{eg: joint vs individual}
		\end{align}

Let us now consider the secrecy aspect of broadcasting by imposing the \emph{joint} and \emph{individual} secrecy constraints, respectively. We note that the eavesdropper also receives a perfect copy of $X^n.$

1) For the \emph{joint} secrecy constraint, we have that
			\begin{equation}\label{eg: joint}
				H(U^k|X^n)=H(U^k).
			\end{equation}
		Recall (\ref{eg: joint vs individual}). We obtain 
			\begin{equation*}
				H(U^k|X^n)=H(U_i|X^n)\leq H(U_i)<H(U^k),
			\end{equation*}
		where the last strict inequality follows since $U_i \sim \mathrm{Bern}(1/2).$ Thus, equality in (\ref{eg: joint}) is not possible. That is, for this example, no broadcasting scheme could fulfill the \emph{joint} secrecy constraint.

2) For the \emph{individual} secrecy constraint, we have that
			\begin{equation}\label{eg: individual}
				H(U_i|X^n)=H(U_i), \quad \mbox{for}\ 1\leq i\leq k.
			\end{equation}
Suppose there is a coding scheme that fulfills both purposes of broadcasting, i.e., (\ref{eg: reliability}), and the individual secrecy, i.e., (\ref{eg: individual}). Then, we have 
			\begin{align}
			H(U^k, X^n)&= H(U_i, X^n)+H(U_1^{i-1}, U_{i+1}^n|U_i, X^n)\nonumber\\
						&\stackrel{(a)}{=} H(X^n)+H(U_i|X^n)\nonumber\\
						&\stackrel{(b)}{=} H(X^n)+H(U_i),\label{eg: indproof}
		\end{align}
where $(a)$ is due to (\ref{eg: reliability}); and $(b)$ is due to (\ref{eg: individual}).	
Using $H(U^k, X^n)\geq H(U^k)$ in \eqref{eg: indproof}, we obtain that
		\begin{equation*}
			H(X^n)\geq H(U^k)-H(U_i)=k-1.
		\end{equation*}
So to say, the optimal encoding scheme (with respect to the overall transmission rate $k/n$) from $U^k$ to $X^n$ is such that $H(X^n)=k-1.$ Thus, to obtain the optimal rate, one shall take $n=k-1$. This is feasible. In fact, there are many coding schemes that could achieve this. One of the options is to take
	\begin{equation*}
		x_i=u_1\oplus u_{i+1}, \quad \mbox{for}\quad 1\leq i\leq k-1.
	\end{equation*}
The decoding at each receiver $i$ is straightforward. Since $u_i$ is available at receiver $i$ as side information, it could first help to recover $u_1$ by $u_1\triangleq x_{i-1}\oplus u_i$ if $i>1;$ and then sequentially recover $u_j$ by $u_j\triangleq x_{j-1}\oplus u_1$ for $j\neq 1, i.$    And, the transmission rate $R_i$ to each receiver $i$, for $1\leq i\leq k,$ is equal to 1, since $k-1$ bits are received in $n=k-1$ channel uses. Noting that the capacity for a binary noiseless channel is one, we conclude that the above scheme actually achieves the individual secrecy capacity for all receivers. 

The following insights immediately follow from this example: 
\begin{itemize}
	\item Joint secrecy might be impossible to achieve.
	\item Individual secrecy could be the highest secrecy level to offer (as shown in \eqref{eg: joint vs individual} on the equivocation at the eavesdropper). 
	\item Individual secrecy could be achieved without any rate degradation (as compared to the capacity region without security constraints)!
\end{itemize} 

In fact, joint secrecy could be impossible for a more general set-up as demonstrated in the following proposition.
\begin{proposition}\label{thm:joint secrecy with strong Eve}
For the communication model as shown in Fig. \ref{fig: wiretap channel with receiver side info} under the joint secrecy constraint, any rate pair $(R_1,R_2)\in\RR^+$ is infeasible if the channel to at least one of the receivers is more noisy than the channel to the eavesdropper. 
\end{proposition}
\begin{IEEEproof}
	Assume that receiver 2 receives $Y_2^n$ as a more noisy version of $Z^n,$ the channel output at the eavesdropper. From the following analysis, we show that $R_2>0$ is not possible.
	\begin{align*}
		nR_2	&= H(M_2)=I(M_2;M_1,Y_2^n)+H(M_2|M_1,Y_2^n)\\    	
		 		&\stackrel{(a)}{\leq} I(M_2;Y_2^n|M_1)+n\epsilon'
		 		\leq I(M_1,M_2;Y_2^n)+n\epsilon'\\
				&\stackrel{(b)}{\leq} I(M_1,M_2; Z^n)+n\epsilon'\\
				&\stackrel{(c)}{\leq} n(\epsilon+\epsilon'),	
	\end{align*}
where 
	$(a)$ is due to Fano's inequality (implying that $H(M_2|M_1,Y_2^n)\leq n\epsilon'$ for some $\epsilon'\to0$ as $n\to\infty$) and the fact that $I(M_2;M_1,Y_2^n)=I(M_2;M_1)+I(M_2;Y_2^n|M_1)=I(M_2;Y_2^n|M_1)$ as $M_1$ and $M_2$ are independent;
	$(b)$ is due to $I(M_1,M_2;Y_2^n)\leq I(M_1,M_2;Z^n)$ which follows from the fact that receiver 2 has a more noisy observation $Y_2^n$ than the eavesdropper's observation $Z^n;$
 	and 
	$(c)$ is due to the joint secrecy constraint \eqref{eq:JointSec}. 
	
	This implies that $R_2\leq \epsilon+\epsilon'$, which is arbitrarily small for an arbitrarily small $P_{e,2}$ (i.e., $\epsilon'$) and an arbitrarily small information leakage rate $\epsilon$ to the eavesdropper.
\end{IEEEproof}

Nevertheless, an achievable rate region was established in \cite{Wyrembelski:Secrecy11} for the BC-RSI under joint secrecy constraint. In the following sections, we will focus on deriving the individual secrecy capacity or achievable rate regions for different BC-RSI models. In particular, we will start with a specific linear deterministic case, where we establish the individual secrecy capacity region. Then, we will address the general discrete memoryless model, where we obtain achievable rate regions with characterization of the capacity region in special cases. Finally, we look into the Gaussian case and obtain inner and upper bounds for the individual secrecy capacity region.


\section{Linear Deterministic BC-RSI}\label{sec:deterministic}

Motivated by the success of the linear deterministic approach \cite{src:avestimehr2011wireless, src:mustafa2012} in approximating the (secrecy) capacity region within constant bits regardless of the received signal-to-noise ratio and its relevance particularly in the high SNR regime, we first take a look at the linear deterministic broadcast channel \cite{src:avestimehr2011wireless} with receiver side information. In this specific model, the received signals at the legitimate receivers and the eavesdropper are given by 
\begin{align}\label{deter: system function}
	\begin{split}
			{Y}_1&=D^{q-n_1} {X},\\
			{Y}_2&=D^{q-n_2} {X},\\
			Z&=D^{q-n_e} {X},
	\end{split}
\end{align}
where ${X}$ is the binary input vector of length $q=\max\{n_1,n_2,n_e\}$; $D$ is the $q\times q$ down-shift matrix; $n_1, n_2$ and $n_e$ are the integer channel gains of the channels from the transmitter to receiver 1, receiver 2, and the eavesdropper, respectively. Note that 
\begin{enumerate}
	\item as $q=n_1\geq n_2\geq n_e,$ the channel is degraded in the manner that $X\to Y_1\to Y_2\to Z$ forms a Markov chain;
	\item as $q=n_1\geq n_e \geq n_2,$ the channel is degraded in the manner that $X\to Y_1\to Z \to Y_2$ forms a Markov chain;
	\item as $q=n_e\geq n_1 \geq n_2,$ the channel is degraded in the manner that $X\to Z\to Y_1 \to Y_2$ forms a Markov chain.
\end{enumerate}
In all cases, we have the following theorem:
\begin{theorem}\label{thm: IndS linear deterministic}
The individual secrecy capacity region of the linear deterministic broadcast channel with receiver side information is the set of the rate pairs $(R_1, R_2)$ defined by
\begin{align*}
	 R_1&\leq \min\{n_1, [n_1-n_e]^+ +R_2\};\\
	 R_2&\leq \min\{n_2, [n_2-n_e]^+ +R_1\},
\end{align*}
where $[a]^{+}=\max\{0, a\}.$
\end{theorem}

\begin{IEEEproof}
The converse follows directly from \cite[Theorem 1]{Kramer:Capacity07} and Proposition \ref{lemma: upper bound} in Appendix~\ref{sec:UpperBoundAppendix}, where the former is the capacity region of the BC-RSI without any secrecy constraints; and the latter is an upper bound of the secrecy capacity region of BC-RSI.

The achievability follows by considering different scenarios, each is classified according to the relation between the channel gains $n_1, n_2, n_e$, and the relation between the rates $R_1, R_2$. For a given scenario, we consider the construction of the codeword $X$ as a function of $m_1, m_2$. Note that, at the receiver side, according to the system input-output relation as defined in \eqref{deter: system function}, receiver 1 receives the first $n_1$ bits of $X$, receiver 2 gets the first $n_2$ bits of $X$, and the eavesdropper gets the first $n_e$ bits of $X$. This holds for all scenarios. In order to achieve a {\em reliable} and {\em secure} communication under the individual secrecy constraint, $X$ should be designed in such a way that both legitimate receivers could decode the desired message with the help of the side information (i.e., the other message); while the eavesdropper can only observe bits either in the form of $m_1\oplus m_2$, or mixture of part of the messages, and/or random bits. This gives no information on $m_1$ and $m_2$ individually. In Appendix~\ref{sec:AppDeterministic}, we provide a specific coding scheme for each scenario, achieving the corresponding individual secrecy capacity region. Putting all pieces together establishes the achievability of the stated region.
\end{IEEEproof}

  \begin{figure}[!ht]
  \centering
    \subfloat[$n_1\geq n_2\geq n_e$\label{Determinstic: SC1}]{
      \begin{tikzpicture}[scale=0.38]
      \begin{axis}[
      xlabel={\Large{$\mathbf{R_{1}}$}},
      ylabel={\Large{$\mathbf{R_{2}}$}},
      xtick=\empty,ytick=\empty,
      xmin=0,xmax=5,
      ymin=0,ymax=4,
      x=2cm,y=1.5cm,
      grid=major,
      extra y ticks={1,3,4},extra x ticks={0,1,4,5},
      extra y tick labels={\LARGE{$n_e$}, \LARGE{$n_2-n_e$}, \LARGE{$n_2$}},extra x tick labels={\Large{0}, \LARGE{$n_e$}, \LARGE{$n_1-n_e$}, \LARGE{$n_1$}}]
      \addplot[blue, fill=yellow!20, very thick] plot coordinates
      { (0.05, 0.05) (0.05, 3) (1, 3.95) (4.95, 3.95) (4.95, 1) (4, 0.05) (0.05, 0.05)};
      \end{axis}
      \draw[->,color=black, semithick] (0,0) -- (11,0);
      \draw[->,color=black, semithick] (0,0) -- (0,7);
      \end{tikzpicture}
    }
    \hfill
    \subfloat[$n_1\geq n_e\geq n_2$\label{Determinstic: SC2}]{
      \begin{tikzpicture}[scale=0.45]
      \begin{axis}[
      xlabel={\Large{$\mathbf{R_{1}}$}},
      ylabel={\Large{$\mathbf{R_{2}}$}},
      xtick=\empty,ytick=\empty,
      xmin=0,xmax=4.5,
      ymin=0,ymax=3,
      x=2cm,y=1.5cm,
      grid=major,
      extra y ticks={3},extra x ticks={0,1.5,4.5},
      extra y tick labels={\LARGE{$n_2$}},extra x tick labels={\Large{0}, \LARGE{$n_1-n_e$}, \LARGE{$n_1-n_e+n_2$}}]
      \addplot[blue, fill=yellow!20, very thick] plot coordinates
      { (0.05, 0.05) (3, 3) (4.5, 3) (1.5, 0.05) (0.05, 0.05)};
      \end{axis}
      \draw[->,color=black, semithick] (0,0) -- (10,0);
      \draw[->,color=black, semithick] (0,0) -- (0,5.5);
      \end{tikzpicture}
    }
    \hfill
    \subfloat[$n_e\geq n_1\geq n_2$\label{Determinstic: SC3}]{
          \begin{tikzpicture}[scale=0.45]
          \begin{axis}[
          xlabel={\Large{$\mathbf{R_{1}}$}},
          ylabel={\Large{$\mathbf{R_{2}}$}},
          xtick=\empty,ytick=\empty,
          xmin=0,xmax=4,
          ymin=0,ymax=3,
          x=2cm,y=1.5cm,
          grid=major,
          extra y ticks={3},extra x ticks={0,3,4},
          extra y tick labels={\LARGE{$n_2$}},
          extra x tick labels={\Large{0},\LARGE{$n_2$},\LARGE{$n_1$}},
          legend style={at={(0.05, 0.9)},
          anchor=north west
          },
          ]
          \addplot[blue, very thick] plot coordinates
          { (0.05, 0.05) (3, 3)};
          \legend{
          $R_1=R_2$
          }
          \end{axis}
          \node (h) at (7, 4.8) {{\scriptsize $\min\{n_1, n_2\}$}};
          \draw[red, semithick] (6,4.5) circle (0.05cm);
          \draw[->,color=black, semithick] (0,0) -- (9,0);
          \draw[->,color=black, semithick] (0,0) -- (0,5.5);
          \end{tikzpicture}
        }
    \caption{Individual secrecy capacity region of the linear deterministic BC-RSI}
    \label{fig:deterministic SC}
  \end{figure}
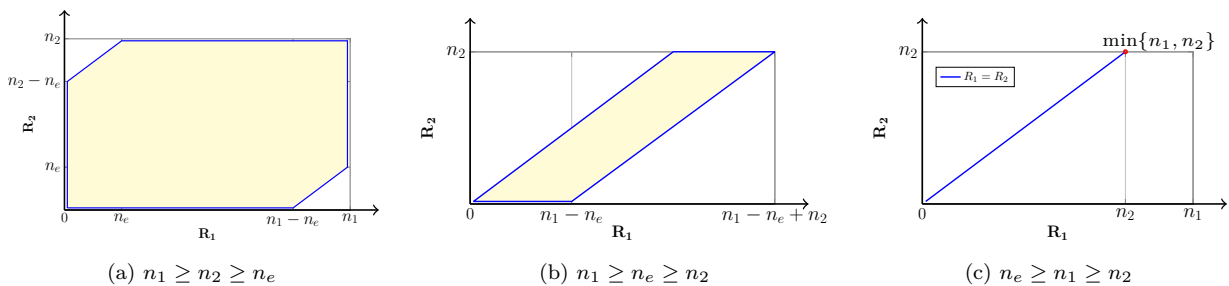

The individual secrecy capacity region of the linear deterministic BC-RSI is depicted in Fig. \ref{fig:deterministic SC}. We remark that the capacity region is 
\begin{itemize}
	\item a {\em rectangle} with two missing corners in case of $n_1\geq n_2\geq n_e;$ 
	\item a {\em parallelogram} in case of $n_1\geq n_e\geq n_2;$ and
	\item a {\em line} in case of $n_e\geq n_1\geq n_2.$
\end{itemize}
Compared to the capacity region of the BC-RSI (following from \cite[Theorem 1]{Kramer:Capacity07}, this region is given by $R_1\leq n_1$ and $R_2\leq n_2$ as shown in Fig.~\ref{fig:deterministic SC}), the missing parts reflect the loss in the transmission rates due to the individual secrecy constraints. And, as the eavesdropper gets stronger, the loss increases. Nevertheless, in the worst case, positive secrecy rate pairs are still possible under the individual secrecy constraint (as shown in Fig. \ref{Determinstic: SC3}), unlike the case under the joint secrecy constraint (as demonstrated in Propostion \ref{thm:joint secrecy with strong Eve}).


\section{Discrete Memoryless BC-RSI}

In this section, we consider the discrete memoryless BC-RSI with an external eavesdropper (Fig. \ref{fig: wiretap channel with receiver side info}). When none of the secrecy constraints are taken into account, this model reduces to discrete memoryless BC-RSI, for which the capacity region is given by the union of rate pairs $(R_1, R_2)$ satisfying $R_i\leq I(X;Y_i)$ for $i=1,2$, where the union is taken among all possible input probability distributions $p(x)$ \cite[Theorem 1]{Kramer:Capacity07}. Here, we focus on coding schemes that can achieve not only reliability but also (individual) secrecy for this model. In particular, we investigate to what extend this capacity region has to be modified in order to accommodate (individual) secrecy.

In order to investigate the fundamental limits of communication under individual secrecy constraints, we utilize coding approaches including one-time pad, wiretap coding, superposition coding, and Marton's coding, which have been proposed for  communication scenarios such as Shannon's cipher system, wiretap channel, and broadcast channel \cite{ElGamal:2012}. The key ingredient of our proposed schemes is the utilization of side information at receivers as secret keys of one-time pad signals, which further is encoded as \emph{cloud centers} in broadcast coding schemes. That is, one-time pad signals are constructed such that they can be decoded at both receivers, which can then extract their desired information utilizing the side information, whereas the eavesdropper will be left with full ambiguity regarding the information content for each message individually. We refer this signaling technique as the \emph{secret key} approach.

As detailed in this section, we observe, for the case of a strong eavesdropper, that the secret key approach (i.e., coding via one-time pad by mixing the messages) is the best one can do; while, in case of a weak eavesdropper, the combined secret key and secrecy coding approach is required in order to achieve higher rates. (Here, we use the phrase \emph{secrecy coding} in order to refer to the extension of wiretap coding technique to our broadcast model, where both users randomize their signals in order to confuse the eavesdropper.) After a characterization of achievable rates and special case capacity results with these strategies, we detail a universal approach by employing superposition coding and Marton's coding to establish (general) achievable individual secrecy rate regions.


\subsection{Secret key approach and the capacity region for BC-RSI with a stronger eavesdropper}\label{subsec: key approach}

Consider the symmetric secret rate region where $R_1=R_2=R,$ i.e., $M_1$ and $M_2$ are of the same entropy. Under these conditions, communicating the message $M_1\oplus M_2$ readily provides individual secrecy, i.e., the following rate region is achievable.

\begin{proposition}\label{thm:Ach0}
Any $(R_1,R_2)\in\RR^+$ satisfying 
\begin{equation}
R_1=R_2 \leq \min\{I(X;Y_1),I(X;Y_2)\},
\end{equation} 
for any $p(x)$ is achievable.
\end{proposition}

\begin{IEEEproof}
Randomly generate $2^{nR}$ codewords $x^n$ according to $p(x).$ Given $(m_1, m_2),$ send $x^n(m_k)$ with $m_k=m_1\oplus m_2$ to the channel. See Fig. \ref{fig: SK coding} for the construction of $X^n$. Both receivers can decode reliably by utilizing their side information to extract intended messages if $R_1=R_2\leq\min\{I(X;Y_1),I(X;Y_2)\}$. For the secrecy of $M_i$, $i=1,2$ we have
\begin{eqnarray}
I(M_i;Z^n)\stackrel{(a)}{\leq} I(M_i;Z^n,M_k)\stackrel{(b)}{=}I(M_i;M_k)\stackrel{(c)}{=}0,
\end{eqnarray}
where (a) is due to the non-negativity of the conditional mutual information, i.e., $I(M_i;M_k|Z^n)\geq 0$; (b) is due to Markov chain $M_i\to M_k\to Z^n$, i.e., $I(M_i;Z_n|M_k)=0$; and (c) follows as $M_i$ is secured with a one-time pad $M_{j}$ ($j\neq i$) in $M_k$.
			\begin{figure}[h]
				\centering
					\begin{tabular}{rcl}
							$m_1:$ &	& 
									$\begin{tikzpicture}
										\node[minimum height=1.6em, minimum width=5.5em, anchor=base, fill=blue!25] {$m_{1}$}; 
									\end{tikzpicture}$\\
							$m_2:$ &	& 
									$\begin{tikzpicture}
										\node[minimum height=1.6em,minimum width=5.5em, anchor=base, fill=teal!25] {$m_{2}$}; 
									\end{tikzpicture}$\\
							$x^n:$ &	&
									$\underbrace{
									\begin{tikzpicture}
										\node[minimum height=1.6em, minimum width=5.5em, anchor=base, fill=red!25] {$m_{1}\oplus m_{2}$};
									\end{tikzpicture}
									}_{nR}$
							\end{tabular}
							\caption{Secret key approach: Encoding.}
							\label{fig: SK coding}											
						\end{figure}
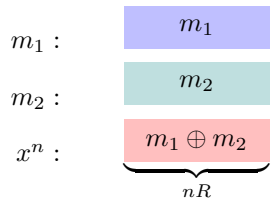
\end{IEEEproof}

Note that the above achievable region is limited by the capacity of the worse channel of the legitimate receivers. Nevertheless, it serves as the individual secrecy capacity region when the eavesdropper has an advantage on the channel over both legitimate receivers.

\begin{theorem}\label{thm: secret key CS}
If the channels to the legitimate receivers are degraded with respect to the channel to the eavesdropper, then the individual secrecy capacity region is given by the union of non-negative rate pairs $(R_1,R_2)$ satisfying 
\begin{equation}
	R_1=R_2\leq \min \{I(X;Y_1),I(X;Y_2)\},
\end{equation} 
where the union is taken over $p(x)$. 
\end{theorem}
\begin{IEEEproof} The achievablity follows from the proof of Proposition \ref{thm:Ach0}. Here, we detail the converse.
\begin{align*}
	nR_1	&= H(M_1)=I(M_1;M_2,Y_1^n)+H(M_1|M_2,Y_1^n)\\    	
		 		&\stackrel{(a)}{\leq} I(M_1;Y_1^n|M_2)+n\epsilon'
		 		\leq I(M_1,M_2;Y_1^n)+n\epsilon';\\
				&\stackrel{(b)}{\leq} I(X^n;Y_1^n)+n\epsilon'
				\stackrel{(c)}{=}\sum_{i=1}^n I(X_i;Y_{1,i})+n\epsilon'\\
				&\stackrel{(d)}{=}n I(X_Q;Y_{1,Q}|Q)+n\epsilon'
				\stackrel{(e)}{\leq} nI(X;Y_1)+n\epsilon';	
\end{align*}
and, continuing from (a), we have
\begin{align*}				
	nR_1	&\leq I(M_1;Y_1^n|M_2)+n\epsilon'  \leq I(M_1,M_2;Y_1^n)+n\epsilon'\\
			&\stackrel{(f)}{\leq} I(M_1,M_2; Z^n)+n\epsilon' 
			\stackrel{(g)}{\leq} I(M_2;Z^n|M_1)+n(\epsilon'+\epsilon)\\
			&\leq H(M_2)+n(\epsilon'+\epsilon)
			\stackrel{(h)}{=} nR_2+n(\epsilon'+\epsilon)\\
			&\stackrel{(i)}{\leq} nI(X;Y_2)+n(\epsilon'+\epsilon)
\end{align*}
where 
	$(a)$ is due to Fano's inequality and the fact that $I(M_1;M_2)=0$; 
	$(b)$ is due to Markov chain $(M_1,M_2)\to X^n\to Y_1^n;$ 
	$(c)$ follows as the channel is memoryless; 
	$(d)$ is by introducing a time-sharing random variable $Q$ which is uniform over $1,2,\dots, n;$ 
	$(e)$ is by taking $X=X_Q, Y_1=Y_{1, Q};$ 
	$(f)$ is due to the channel degradedness, i.e., Markov chain $(M_1, M_2) \to Z^n \to Y_1^n;$ 
	$(g)$ is by the individual secrecy constraint \eqref{eq:IndSec}; 
	$(h)$ is due to $H(M_2)=nR_2;$ and 
	$(i)$ is derived by applying a proof similar to $nR_1\leq nI(X;Y_1)+n\epsilon'$ and by taking $Y_2=Y_{2,Q}.$ At this point, from $(h)$, we have $R_1\leq R_2;$ and $R_1\leq \min\{I(X;Y_1), I(X;Y_2)\}$. By symmetry, we have $R_2\leq R_1$ and $R_2\leq \min\{I(X;Y_1), I(X;Y_2)\}$. Thus, we establish that $R_1=R_2\leq \min\{I(X;Y_1), I(X;Y_2)\}$.
\end{IEEEproof}

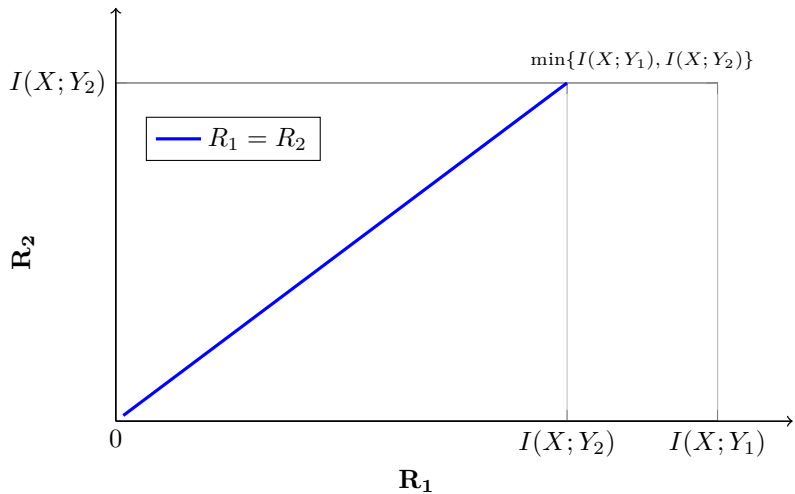
\begin{figure}
\centering
\begin{tikzpicture}[scale=1]
\begin{axis}[
xlabel={$\mathbf{R_{1}}$},
ylabel={$\mathbf{R_{2}}$},
xtick=\empty,ytick=\empty,
xmin=0,xmax=4,
ymin=0,ymax=3,
x=2cm,y=1.5cm,
grid=major,
extra y ticks={3},extra x ticks={0,3,4},
extra y tick labels={$I(X;Y_2)$},
extra x tick labels={0,$I(X;Y_2)$,$I(X;Y_1)$},
legend style={at={(0.05, 0.9)}, anchor=north west}
]
\addplot[blue, very thick] plot coordinates
{ (0.05, 0.05) (3, 3)};
\legend{%
$R_1=R_2$
}
\end{axis}
\node (h) at (7, 4.8) {{\scriptsize $\min\{I(X;Y_1), I(X;Y_2)\}$}};
\draw[->,color=black, semithick] (0,0) -- (9,0);
\draw[->,color=black, semithick] (0,0) -- (0,5.5);
\end{tikzpicture}
\caption{Individual secrecy capacity region in case of a strong eavesdropper.}
\label{fig:IndSec_strong eavesdropper}
\end{figure}

The individual secrecy capacity described in Theorem \ref{thm: secret key CS} is depicted in Fig. \ref{fig:IndSec_strong eavesdropper}. That is, in case of a strong eavesdropper, the best transmission strategy is to send the one-time pad of the messages to both receivers, where both of them could recover its desired message with the help of side information; while the eavesdropper gets only the mixed copy, which gives no clue for each message individually.

\subsection{Combined secret key and secrecy coding approach and the capacity region for BC-RSI with a weaker eavesdropper}\label{sec: secrecy coding}

Although the secret key approach is optimal in case of a strong eavesdropper, this scheme can be strictly suboptimal for other scenarios. In fact, a counter-example follows from the linear deterministic model, for the case where the eavesdropper is weak. In general, consider channel inputs $p(x)$ such that $I(X;Z)\leq \min\{I(X;Y_1), I(X;Y_2)\}$. We show in this section that, asymmetric rate pairs beyond the secret key approach can be achieved if we combine secret key with a secrecy coding approach. That is, besides using the receiver side information as secret key, one can further take the advantage over the channel against the eavesdropper by employing secrecy coding approach \cite{src:Wyner1975, Csisz'ar:Broadcast78}. First, we have the following proposition.
\begin{proposition}\label{thm:Ach1}
Any $(R_1,R_2)\in\RR^+$ satisfying 
	\begin{align*} 
		I(X;Z)&\leq R_1 \leq I(X;Y_1)\\
		I(X;Z)&\leq R_2 \leq I(X;Y_2)
	\end{align*} 
for $p(x)$ such that $I(X;Z)\leq \min\{I(X;Y_1),I(X;Y_2)\}$ is achievable.
\end{proposition}

\begin{IEEEproof}
{\em Rate splitting:} Assume that $R_2\leq R_1.$ As illustrated in Fig. \ref{fig: csksc rate splitting}, we split $M_1$ into two parts, i.e., $M_1=(M_{1k}, M_{1s})$ with $M_{1k}$ of entropy $nR_2,$ the same as $M_2;$ whilst $M_{1s}$ of entropy $nR_{1s}.$ Note that $R_1=R_{1k}+R_{1s}.$
			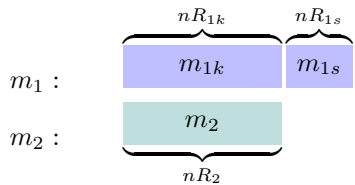
\begin{figure}[h]
			\centering	
				\begin{tabular}{rcl}		
							$m_1:$ &	& 
								$\overbrace{	
									\begin{tikzpicture}
										\node[minimum height=1.6em, minimum width=6em, anchor=base, fill=blue!25] {$m_{1k}$}; 
									\end{tikzpicture}
									}^{nR_{1k}}
									\overbrace{
									\begin{tikzpicture}
										\node[minimum height=1.6em, minimum width=2.5em, anchor=base, fill=blue!25] {$m_{1s}$}; 
									\end{tikzpicture}
									}^{nR_{1s}}$\\
							$m_2:$ & 	& 
									$\underbrace{
									\begin{tikzpicture}
										\node[minimum height=1.6em,minimum width=6em, anchor=base, fill=teal!25] {$m_{2}$}; 
									\end{tikzpicture}
									}_{nR_2}$
				\end{tabular}
				\caption{Combined secret key and secrecy coding approach: Rate splitting.}
				\label{fig: csksc rate splitting}	
			\end{figure}

{\em Codebook generation:} Randomly generate $2^{nR_1}$ codewords $x^n$ according to $p(x).$ Throw them into $2^{nR_{1s}}$ bins \cite{ElGamal:2012} and index them by $x^n(i_k, i_{1s})$ with $(i_k,i_{1s})\in [1:2^{nR_2}]\times [1:2^{nR_{1s}}].$   

{\em Encoding:} To send messages $(m_1, m_2),$ choose $x^n(m_k, m_{1s})$ with $m_k=m_{1k}\oplus m_2$ and transmit it to the channel. The choice of the codeword $x^n$ for given $(m_1, m_2)$ is illustrated in Fig. \ref{fig: SC coding}. 
				\begin{figure}[h]
				\centering
						\begin{tabular}{rcl}
							$x^n:$ &	&
									$\underbrace{									
									\overbrace{
									\begin{tikzpicture}
										\node[minimum height=1.6em, minimum width=6em, anchor=base, fill=red!25] {$m_{1k}\oplus m_{2}$};
									\end{tikzpicture}
									}^{nR_2}
									\underbrace{
									\begin{tikzpicture}
										\node[minimum height=1.6em, minimum width=2.5em, anchor=base, fill=blue!25] {$m_{1s}$};
									\end{tikzpicture}
									}_{nR_{1s}}
									}_{nR_1}$
						\end{tabular}
						\caption{Combined secret key and secrecy coding approach: Encoding}
						\label{fig: SC coding}											
						\end{figure}
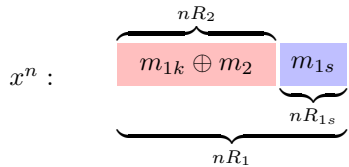
						
{\em Decoding:} Receiver 2 can decode $m_k$ reliably using typical set decoding if 
	\begin{equation}\label{eqn: SC coding cond1}
		R_2<I(X; Y_2)
	\end{equation} 
with the knowledge of $m_1$, and thus extract $m_2$. Receiver 1 can decode both $m_k$ and $m_{1s}$ if 
	\begin{equation}\label{eqn: SC coding cond2}
		R_1<I(X;Y_1)
	\end{equation}
and extract $m_{1k}$ from the former with the knowledge of $m_2$.

{\em Individual secrecy:} At the eavesdropper, we see that $M_{1k}$ is secured by capsuling with $M_2$ as a one-time pad (thus $M_2$ is also secured as in Section \ref{subsec: key approach}), while $M_{1s}$ is secured by using secrecy coding for classical wiretap channels under the condition that 
	\begin{equation}\label{eqn: SC coding cond3}
		R_2\geq I(X;Z). 
	\end{equation}

More specifically, the secrecy of $M_2$ follows from
\begin{equation*}
	I(M_2;Z^n)\leq I(M_2;Z^n,M_k, M_{1s})=I(M_2;M_k, M_{1s})=0.
\end{equation*}
And, the secrecy of $M_{1}$ is shown as follows. Since $R_2\geq I(X;Z),$ for a fixed $i_{1s},$ one can further bin the codewords $x^n$ and index them as $x^n(i_{kr}, i_{ks}, i_{1s})$ with $i_k=(i_{kr}, i_{ks})\in [1: 2^{n(I(X;Z)-\epsilon)}]\times[1: 2^{n(R_2-I(X;Z)+\epsilon)}],$ as illustrated in Fig. \ref{fig: SC secrecy proof}. 
					\begin{figure}[h]
						\centering
						\begin{tabular}{rcl}
							$x^n:$ &	&
								$
									\underbrace{
									\overbrace{
									\begin{tikzpicture}
										\node[minimum height=1.6em, minimum width=4em, anchor=base, fill=red!25] {$i_{kr}$};
									\end{tikzpicture}
									}^{\approx nI(X;Z)}
									\begin{tikzpicture}
										\node[minimum height=1.6em, minimum width=2em, anchor=base, fill=red!25] {$i_{ks}$};
									\end{tikzpicture}
									}_{nR_2}
									\underbrace{
									\begin{tikzpicture}
										\node[minimum height=1.6em, minimum width=2.5em, anchor=base, fill=blue!25] {$i_{1s}$};
									\end{tikzpicture}
									}_{nR_{1s}}
								$
						\end{tabular}
						\caption{secret key and secrecy coding approach: Secrecy analysis.}		
						\label{fig: SC secrecy proof}											
						\end{figure}
Correspondingly, split $M_k=(M_{kr},M_{ks}).$ We have
	\begin{align*}
		H(M_{1s}, M_{ks}|Z^n)
		=&	H(M_{1s}, M_{ks}, X^n|Z^n)-H(X^n|M_{1s}, M_{ks}, Z^n)\\
		\stackrel{(a)}{\geq}& H(M_{1s}, M_{ks}, X^n, Z^n) - H(Z^n)-n\epsilon_1 \\
		=& H(X^n) +H(Z^n|X^n)- H(Z^n)-n\epsilon_1 \\
		\stackrel{(b)}{\geq}& nR_1 + nH(Z|X) - nH(Z)-n\epsilon_1 \\
		\stackrel{(c)}{\geq}& H(M_{1s}, M_{ks}) -n\delta(\epsilon),
	\end{align*}
where 
	$(a)$ follows as $H(X^n|M_{1s}, M_{ks}, Z^n)\leq n\epsilon_1$ due to Fano's inequality and that the eavesdropper can decode $X^n$ reliably by using typical set decoder, given $(M_{ks}, M_{1s}, Z^n);$ 
	$(b)$ is due to the fact that $H(X^n)=nR_1$; $H(Z^n|X^n)=nH(Z|X)$ since the channel is memoryless; and $H(Z^n)=\sum_{i=1}^n H(Z_i|Z_1^{i-1})\leq \sum_{i=1}^n H(Z_i)=nH(Z);$ 
	$(c)$ follows that $H(M_{1s}, M_{ks})=nR_{1s}+n(R_2-I(X;Z)+\epsilon)=nR_{1}-nI(X;Z)+n\epsilon$ and $\delta(\epsilon)=\epsilon_1+\epsilon.$

Above inequality implies $I(M_{1s};Z^n)\leq n\delta(\epsilon)$. Besides, due to Markov chain $M_{1k}\to (M_k, M_{1s}) \to Z^n,$ we can bound $I(M_{1k};Z^n|M_{1s})\leq I(M_{1k};Z^n,M_{1s},M_k) = I(M_{1k};M_k, M_{1s})=0.$ Therefore, we obtain
\begin{equation*}
I(M_1;Z^n)=I(M_{1s};Z^n)+I(M_{1k};Z^n|M_{1s})\leq n\delta(\epsilon).
\end{equation*}
This concludes the proof of the individual secrecy. 

{\em Achievable rate region:} Combining the sufficient conditions for reliable transmission to both receivers, i.e., \eqref{eqn: SC coding cond1} and \eqref{eqn: SC coding cond2}, and the condition for individual secrecy, i.e., \eqref{eqn: SC coding cond3}, we obtain  
	\begin{align*}
		I(X;Z)\leq & R_2 \leq I(X;Y_2)\\
		R_2\leq & R_1\leq I(X;Y_1),
	\end{align*}
as the achievable rate region for the case $R_2\leq R_1.$ Furthermore, one can apply a similar proof to establish the rate region for the case $R_2>R_1.$ Putting them together completes the proof of the proposition. 
\end{IEEEproof}

\begin{figure}
\centering
\begin{tikzpicture}[scale=1]
\begin{axis}[
xlabel={$\mathbf{R_{1}}$},
ylabel={$\mathbf{R_{2}}$},
xtick=\empty,ytick=\empty,
xmin=0,xmax=5,
ymin=0,ymax=4,
x=2cm,y=1.5cm,
grid=major,
extra y ticks={1,3,4},extra x ticks={0,1,4,5},
extra y tick labels={$I(X;Z)$, $I(X;Y_2)-I(X;Z)$, $I(X;Y_2)$},extra x tick labels={0, $I(X;Z)$, $I(X;Y_1)-I(X;Z)\quad\quad\quad\quad$, $I(X;Y_1)$}]
\addplot[blue, fill=yellow!20, very thick, dotted] plot coordinates
{(0.05, 0.05)(1, 1)(1, 3.95) (4.95, 3.95) (4.95, 1) (1,1)};
\addplot[blue, pattern=dots, pattern color=red!50, very thick, dotted] plot coordinates
{(0.05, 0.05)(1, 1)(1, 3.95) (0.05, 3) (0.05, 0.05)};
\addplot[blue, pattern=soft crosshatch, pattern color=gray, very thick, dotted] plot coordinates
{(0.05, 0.05)(1, 1)(4.95, 1) (4, 0.05) (0.05, 0.05)};
\addplot[blue, very thick] plot coordinates
{ (0.05, 0.05) (0.05, 3) (1, 3.95) (4.95, 3.95) (4.95, 1) (4, 0.05) (0.05, 0.05)};
\node (h) at (axis cs:3, 2.5) {Region \rom{1}};
\node (i) at (axis cs:2.5,0.5){Region \rom{3}};
\node (j) at (axis cs:0.5,2){Region \rom{2}};
\end{axis}
\draw[->,color=black, semithick] (0,0) -- (11,0);
\draw[->,color=black, semithick] (0,0) -- (0,7);
\end{tikzpicture}
\caption{Individual secrecy capacity region in case of a weak eavesdropper.}
\label{fig:IndSec_more capable}
\end{figure}
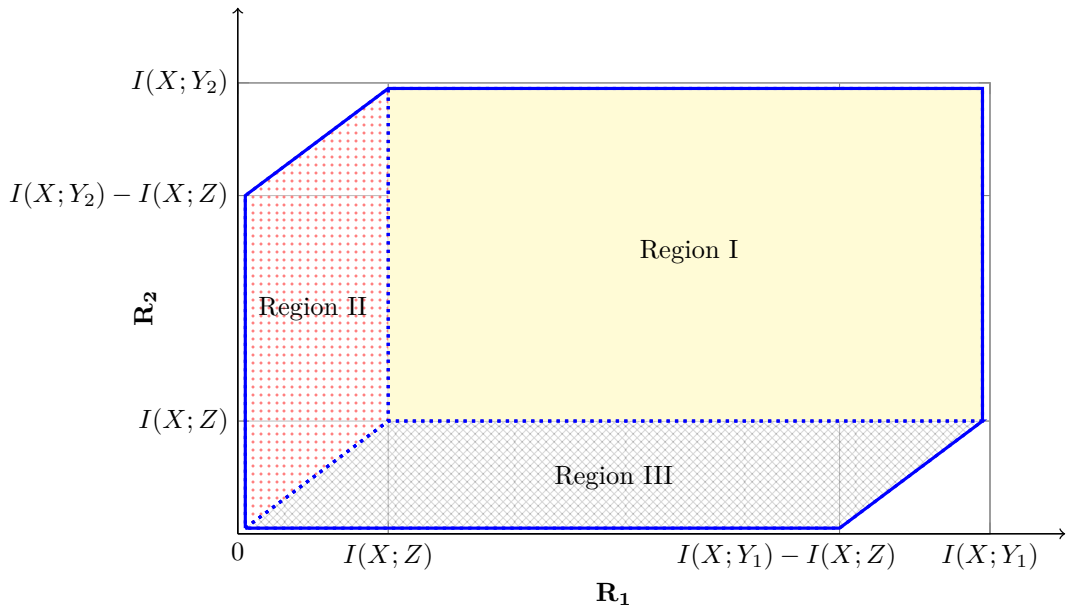

\begin{theorem}\label{thm: Ind SC by secrecy coding}
If the channel to the eavesdropper is degraded with respect to the channels to both legitimate receivers, then the individual secrecy capacity region is given by the union of the non-negative rate pairs $(R_1,R_2)$ satisfying 
\begin{align*}
	R_1& \leq \min\{I(X;Y_1)-I(X;Z)+R_2,\ I(X;Y_1)\};\\
	R_2& \leq \min\{I(X;Y_2)-I(X;Z)+R_1,\ I(X;Y_2)\},
\end{align*} 
where the union is taken over $p(x).$
\end{theorem}

\begin{IEEEproof}
Under the degradedness condition, we have that $I(X;Z)\leq \min\{I(X;Y_1), I(X;Y_2)\}$ holds for any $p(x)$. Utilizing the scheme in Proposition~\ref{thm:Ach1}, Region \rom{1} in Fig.~\ref{fig:IndSec_more capable} is achievable. To show Region \rom{2} is achievable, one can employ secrecy coding \cite[Theorem 3]{Csisz'ar:Broadcast78} to achieve rate pairs $(R_1, R_2)$ such that $R_1=0$ and $R_2\leq I(X;Y_2)-I(X;Z).$ Then, applying time sharing with the left boundary rate pairs of Region \rom{1}, one obtains the remaining rate pairs of Region \rom{2}. A similar proof applies to establish the achievability of Region \rom{3}.    

The converse follows from the fact that the achievable region is equal to the intersection of upper bounds given in \cite[Theorem 1]{Kramer:Capacity07}, which is the capacity region of the BC-RSI without an external eavesdropper, and the upper bound given in Proposition \ref{lemma: upper bound}, which is a partial upper bound by applying the results for wiretap channel with shared key for one receiver (while ignoring the requirement of reliable and secure communication for the other).
\end{IEEEproof}

As shown in Fig. \ref{fig:IndSec_more capable}, the individual secrecy capacity region for a weak eavesdropper is a rectangle with missing corners. Due to the symmetric roles of receiver 1 and receiver 2, the rate region is bounded in a symmetric manner as well. But, unlike the case of a strong eavesdropper, for which the individual secrecy capacity region is given in Fig. \ref{fig:IndSec_strong eavesdropper}, asymmetric rate pairs are possible. Note that both receivers could benefit from each other due to the possession of the message of the other as side information. On one hand, higher rate for one receiver indicates more side information for the other. As a result, there is no loss in the high rate pair region (i.e., $R_1, R_2\geq I(X;Z)$), compared to \cite[Theorem 1]{Kramer:Capacity07} which gives the capacity region of the BC-RSI without any secrecy constraints. That is, individual secrecy to each legitimate receiver is offered for free in high rate region. On the other hand, lower rate for one receiver implies less side information for the other. In this case, the side information might be insufficient to facilitate the secure communication of the other message at a high transmission rate and additional randomness might be necessary. This results in a loss in the rate region, i.e., the missing corners. Another interesting observation is that, for communication under individual secrecy constraint, one may not claim that if $(R_1,R_2)$ is achievable, then $(R_1-c_1,R_2-c_2)$ is achievable for any $c_1\leq R_1,c_2\leq R_2$. This follows as the individual secrecy rates are \emph{coupled} in the BC-RSI setting.

\subsection{Superposition coding}\label{subsec: superposition}

It is well-known that superposition coding is optimal for a degraded broadcast channel where $X\to Y_1\to Y_2$ forms a Markov chain, wherein one can transmit a cloud center to the weak receiver and both the cloud center and satellite codewords to the strong receiver~\cite{ElGamal:2012}. For the BC-RSI model, we consider utilizing the one-time pad signal as the cloud center and additional information on both messages being carried in the satellite codeword. This approach generalizes the aforementioned secret key approach and the combined secret key and secrecy coding approach, and thus achieves the optimality for stronger and weaker eavesdropper scenarios. In the following, we first provide the corresponding achievability region and then discuss the details of the proposed scheme together with the special cases.

\begin{theorem}\label{pro: superposition}
The individual secrecy rate region for the BC-RSI with an external eavesdropper is achievable for the set of the non-negative rate pairs $(R_1, R_2)$ such that
\begin{align}\label{eqn: superp SR}
	\begin{split}
			R_1	&\leq \min\{I(V;Y_1), \ \ I(V;Y_1|U)-I(V;Z|U)+R_2\};\\
			R_2	&\leq \min\{I(V;Y_2), \ \ I(V;Y_2|U)-I(V;Z|U)+R_1\}.
	\end{split}
\end{align} 
over all $p(u)p(v|u)p(x|v)$ subject to $I(V;Y_i|U)\geq I(V;Z|U)$ for $i=1,2.$
\end{theorem}

\begin{IEEEproof}
The proof is given in Appendix~\ref{sec:AppSuperposition}.
\end{IEEEproof}

The coding approach we develop here utilizes \emph{cloud centers}, i.e., the $U^n$ codewords, to carry a one-time pad signal constructed from parts of the messages. In particular, the message intended for receiver $i$ is splitted into $M_i=(M_{ik},M_{isk},M_{is})$, and the one-time pad signal carried by $u^n(m_k)$ is constructed as $M_k=M_{1k}\oplus M_{2k}$. This cloud center is designed to be decodable at both receivers, which then extract their desired messages utilizing the corresponding side information available. In addition, the code design utilizes \emph{satellite codewords}, i.e., the $V^n$ codewords, that are not only superimposed on the cloud centers but also carry additional information represented as $(M_{sk},M_{1s},M_{2s},M_{r})$. Here, $M_{sk}$ is an additional one-time pad signal injected into $V^n$, and given by $M_{sk}=M_{1sk}\oplus M_{2sk}$, and $M_{r}$ is additional randomness. We remark that both $M_{sk}$ and $M_{r}$ serve as randomness to confuse the eavesdropper in this scheme, in order to achieve secrecy of $(M_{1s},M_{2s})$. 

An interesting aspect of our superposition coding approach lies in the role of one-time pad signals. On one hand, one-time pad signal is utilized as the message of the cloud centers (i.e., $M_k$). On the other hand, it is also utilized as a part of randomization within the satellite codewords (i.e., $M_{sk}$). In other words, the coding scheme takes advantage of the rate splitting of one-time pad signals, in order to serve for these two distinct purposes. 

One may wonder, whether further rate splitting helps to improve the current region or not. For instance, split $M_i$ into $M_i=(M_{ik},M_{isk},M_{is},M_{im}),$ with an additional layer in the coding scheme, say $T^n$ which carries information on $M_{im}$ that is secured by employing secrecy coding. Interestingly, the answer is no if still using superposition coding. For a detailed proof of this, one can refer to Appendix \ref{sec:App_Superposition_Rate Splitting}. However, if combining with Marton's coding, further rate splitting may improve the achievable rate region as we demonstrate in Section \ref{sec: Marton's coding}.

Furthermore, we have the following observations:
\begin{itemize}
	\item Setting $Y_2=\emptyset,$ the region coincides with the secrecy capacity region of the wiretap channel \cite{Csisz'ar:Broadcast78};
	\item Letting $U=\emptyset$ and $R_k=R_{sk}=0$ in the proof as given in Appendix~\ref{sec:AppSuperposition} and applying Fourier-Motzkin procedure, an achievable region under the joint secrecy constraint (follows from the secrecy proof in Appendix~\ref{sec:AppSuperposition}) can be obtained. And this region (i.e., \eqref{eqn: JoS Superposition}) coincides with the one established in \cite{Wyrembelski:Secrecy11}. 
	\item Superposition coding remains optimal in the following cases.
		\begin{enumerate}
			\item A {\em strong} eavesdropper, where the eavesdropper's channel is {\em less noisy} than both of the legitimate receivers. In this case, the individual secrecy capacity as shown in Fig. \ref{fig:IndSec_strong eavesdropper}, can be achieved by taking $U=V=X,$ whereby the superposition coding reduces to the secret key approach. (See Theorem~\ref{thm: secret key CS}.)
			
			\item A {\em weak} eavesdropper, where both of the legitimate receivers channels are {\em less noisy} than the one to the eavesdropper. In this case, the individual secrecy capacity is as shown in Fig. \ref{fig:IndSec_more capable}. Here, the left boundary rate pairs of Region \rom{2} in Fig. \ref{fig:IndSec_more capable} can be achieved by taking $R_2=0,$ 
			whereby the superposition coding reduces to the secrecy coding approach as proposed in \cite{src:Wyner1975, Csisz'ar:Broadcast78}. Similarly, the bottom boundary rate pairs of Region \rom{3} in Fig. \ref{fig:IndSec_more capable} can be achieved by taking $R_1=0.$ 
			Region \rom{1} in Fig. \ref{fig:IndSec_more capable} can be achieved by taking $U=\emptyset$ and $V=X$, whereby the superposition coding reduces to the combined secret key and secrecy coding as given in Section \ref{sec: secrecy coding}. These achievable points together with their time sharing provide the individual secrecy capacity region as shown in Fig. \ref{fig:IndSec_more capable}.
			
			\item If the eavesdropper's channel is {\em deterministic} in the manner that $Z$ is a function of $X,$ superposition coding is optimal for achieving individual secrecy capacity as we demonstrate in the Theorem \ref{thm: deterministic Z IndS}.
		\end{enumerate} 	
\end{itemize} 

\begin{theorem}\label{thm: deterministic Z IndS}
For the BC-RSI channel with an external eavesdropper, if the eavesdropper's channel is {\em deterministic} in the manner that $Z$ is a function of $X$, then the individual secrecy capacity region is given by the convex hull of the non-negative rate pairs $(R_1, R_2)$ satisfying
			\begin{align}\label{eqn: deter Z IndS region}
				\begin{split}
					R_1 &\leq \min\{I(X; Y_1), \ \ I(X; Y_1|Z)+R_2\},\\
					R_2 &\leq \min\{I(X; Y_2), \ \ I(X; Y_2|Z)+R_1\}.
				\end{split}
			\end{align} 
\end{theorem}
\begin{IEEEproof}
The achievability is obtained by taking $U=Z$ and $V=X$ in \eqref{eqn: superp SR}. The proof of the converse is given in Appendix \ref{app: Proof of Converse for deterministic BC-RSI with an eavesdropper}.
\end{IEEEproof}

In particular, if $Y_1, Y_2$ and $Z$ are all functions of $X,$ the above corollary simplifies to the following:
\begin{corollary}\label{cor: deterministic IndS}
If the BC-RSI channel with an external eavesdropper is {\em deterministic} in the manner that $Z$, $Y_1$, and $Y_2$ are deterministic functions of $X$, then the individual secrecy capacity region is given by the convex hull of the non-negative rate pairs $(R_1, R_2)$ satisfying
			\begin{align}\label{eqn: deter-IndS region}
				\begin{split}
					R_1 &\leq \min\{H(Y_1), \ \ H(Y_1|Z)+R_2\},\\
					R_2 &\leq \min\{H(Y_2), \ \ H(Y_2|Z)+R_1\}.
				\end{split}
			\end{align} 
\end{corollary}

\begin{remark} Note that the {\em deterministic} BC is a more general model than the {\em linear deterministic} BC model discussed in Section \ref{sec:deterministic}. As a direct consequence, Theorem \ref{thm: IndS linear deterministic} can be regarded as a special case of Corollary \ref{cor: deterministic IndS}.
\end{remark}

The region given in Theorem~\ref{pro: superposition} fails to achieve any positive rates if the condition $I(V;Y_i|U)\geq I(V;Z|U)$ is not satisfied for either $i=1$ or $i=2$. For instance, when $I(V;Y_1|U)>I(V;Z|U)>I(V;Y_2|U)$ for a given input probability distribution, the requirement of decoding randomness (in $V^n$ codewords) at the second receiver becomes excessive. To resolve this problem, we develop a Marton's coding approach in the following section, where we further introduce two individual satellite codewords $(V_1^n, V_2^n)$, and require $V_i^n$ to be decoded only at receiver $i$. This allows us to get a larger rate region for the \emph{mixed} scenarios where the eavesdropper is stronger than one legitimate receiver but weaker than the other one.

\subsection{Marton's coding} \label{sec: Marton's coding}

Although superposition coding demonstrates its optimality for some broadcast channels wherein one receiver is stronger than the other, it is not optimal in general. In fact, for broadcast channels, Marton's coding can outperform superposition coding by not requiring either receiver to recover both messages (for broadcast channels without any secrecy constraints) \cite{ElGamal:2012}. In the following, we consider achieving the indivudual secrey of the BC-RSI model, by utilizing the one-time pad signal as the cloud center, further information on both messages being carried in the satellite codewords, and additional information on each messages being conveyed in individual satellite codewords. This coding scheme is built on the previous superposition coding scheme but with one more layer that employs Marton's coding. As a direct result, it generalizes the rate region established by superpostion coding. 
Moreover, we provide a special case under which this Marton's coding approach outperforms the aforementioned superposition approach (the region given in Theorem~\ref{pro: superposition}).
\begin{theorem}\label{thm: Oma Marton}
The individual secrecy rate region for the BC-RSI with an external eavesdropper is achievable for the set of the non-negative rate pairs $(R_1, R_2)$ such that
\begin{align}\label{eqn: Oma Marton IndS region}
\begin{split}
	R_1  \leq &  I(V_0, V_1; Y_1|U)-I(V_0,V_1;Z|U)+\min\{R_2, \ I(U;Y_1)+I(V_0;Z|U)\};\\
	R_2  \leq &  I(V_0, V_2; Y_2|U)-I(V_0,V_2;Z|U)+\min\{R_1, \ I(U;Y_2)+I(V_0;Z|U)\},
\end{split}
\end{align} 
over any $p(u,v_0,v_1,v_2,x)=p(u)p(v_0|u)p(v_1,v_2|v_0)p(x|v_1,v_2)$ 
subject to $I(V_1;V_2|V_0) \leq I(V_1;Z|V_0)+I(V_2;Z|V_0)-I(V_1,V_2;Z|V_0),$
$I(V_i;Y_i|V_0) \geq I(V_i;Z|V_0)$ and $I(V_0, V_i;Y_i|U) \geq I(V_0, V_i;Z|U)$ for $i=1,2.$
\end{theorem}
\begin{IEEEproof}
The proof is given in Appendix~\ref{sec:App_Oma Marton}.
\end{IEEEproof}

The coding approach we develop here is built on the superposition coding which is discussed in Section \ref{subsec: superposition}, but with one additional coding layer that employs Marton's coding. That is, we split $M_i$ into $M_i=(M_{ik},M_{isk},M_{iss},M_{ism}),$ for $i=1,2,$ where $M_{ik}, M_{isk}, M_{iss}$ are encoded into $U^n, V_0^n$ codewords in the same way as by the superposition coding; while information on $M_{1sm}, M_{2sm}$ are carried by individual satellite codewords $V_1^n, V_2^n,$ respectively, via Marton's coding. Note that the secrecy of $M_{1sm}, M_{2sm}$ is ensured by additional randomness with the spirit of secrecy coding approach \cite{src:Wyner1975, Csisz'ar:Broadcast78}. 

As reflected in the obtained region in \eqref{eqn: Oma Marton IndS region}, for legitimate receiver $i$, part of the message, i.e., $(M_{ik},M_{isk}),$ is secured via one-time pad; while the other part, i.e., $(M_{iss},M_{ism}),$ is secured via secrecy coding. More specifically, for receiver 1, on one hand, $(M_{1k},M_{1sk})$ is secured via one-time pad (with key rate $R_2$) in the underneath superposition coding structure (at most $I(U;Y_i)$ bits in the cloud center $U^n$ and at most $I(V_0;Z|U)$ bits as randomness in the satellite codeword $V_0^n$). Thus, in total at most $\min \{R_2, \ I(U;Y_1)+I(V_0;Z|U)\}$ bits can be secured via one-time pad. On the other hand, $M_{1ss},M_{1sm}$ are secured via secrecy coding in $V_0^n$ and $V_1^n,$ respectively, which in total contribute $I(V_0, V_1; Y_1|U)-I(V_0,V_1;Z|U)$ secret bits. 

Furthermore, we have the following observations:
\begin{itemize}
	\item Letting $U=\emptyset$ and $R_k=R_{sk}=0$ in the proof as given in Appendix~\ref{sec:App_Oma Marton} and applying Fourier-Motzkin procedure, an achievable region under the joint secrecy constraint (follows from the secrecy proof in Appendix~\ref{sec:App_Oma Marton}) can be obtained. And this region (i.e., \eqref{eqn: JoS Marton}) improves the one given in \eqref{eqn: JoS Superposition} which coincides with the one established in \cite{Wyrembelski:Secrecy11}. 		
	\item If we set $V_1=V_2=V_0$, it reduces to the superposition coding approach and achieves the rate region in \eqref{eqn: superp SR} as given in Theorem~\ref{pro: superposition}.
	\item For the case where the eavesdropper's channel is less noisy than one legitimate receiver, but more noisy than the other, (e.g.: $Z$ is less noisy than $Y_2$, if $I(U;Z)\geq I(U;Y_2)$ for all $p(u)$ such that $U\to X \to (Y_2, Z)$ \cite{ElGamal:2012}), Marton's coding approach gives an achievable rate region by setting $U=V_0=V_2$ in \eqref{eqn: Oma Marton IndS region} as provided below.
\end{itemize}

\begin{corollary}
For the BC-RSI with an external eavesdropper, if $Z$ is less noisy than $Y_2$, then an achievable individual secrecy rate region is given by the union of non-negative rate pairs $(R_1,R_2)$ satisfying 
\begin{align}\label{eqn: region Y1Z2 R_{sk}=0}
\begin{split}
	R_1& \leq I(V_1;Y_1|U)-I(V_1;Z|U)+R_2;\\
	R_2& \leq \min\{I(U;Y_2), R_1\}, 
\end{split}
\end{align} 
where the union is taken over $p(u)p(v_1|u)p(x|v_1,u).$
\end{corollary}

We recall that superposition coding is optimal in cases of either a {\em strong} or {\em weak} eavesdropper (compared to both legitimate receivers). However, in the {\em mixed} case, where the eavesdropper's channel is less noisy than one legitimate receiver, but more noisy than the other, superposition coding is no longer optimal. 

For instance, consider the case where $Z$ is strictly less noisy than $Y_2$, i.e., $I(V;Z)>I(V;Y_2)$ for any $p(v)$ s.t. $V\to X \to (Y_2,Z).$ In order to apply superposition coding, one has to set $V=U$ to satisfy the condition that $I(V;Y_2|U)\geq I(V;Z|U)$ given in \eqref{eqn: superp SR} in Theorem~\ref{pro: superposition}. Therefore, the region in \eqref{eqn: superp SR} reduces to the set of the non-negative rate pairs $(R_1, R_2)$ such that 
\begin{align}
	\begin{split}\label{eqn: SS region Y1Z2 R_{sk}=0}
		R_1 &\leq R_2;\\
		R_2 &\leq \min\{I(U;Y_2), R_1\}.
	\end{split}
\end{align}
Compare the obtained region in \eqref{eqn: SS region Y1Z2 R_{sk}=0} by superposition coding with the one in \eqref{eqn: region Y1Z2 R_{sk}=0} by Marton's coding. It is easy to see that the Marton's coding outperforms in this case by not requiring the decoding of the corresponding individual satellite codeword at the weak receiver.  

\subsection{Joint secrecy rate region for BC-RSI with an external eavesdropper}
As a by-product, achievable joint secrecy rate regions can be obtained by letting $U=\emptyset$ and $R_k=R_{sk}=0$ in the superposition coding approach and Marton's coding approach proposed in previous subsections, which validity follows from the secrecy proof in Appendix~\ref{sec:AppSuperposition} for superposition coding; and the secrecy proof in Appendix~\ref{sec:App_Oma Marton} for Marton's coding, respectively. Note that the achievable joint secrecy rate region by Marton's coding, i.e., \eqref{eqn: JoS Marton}, is derived with the addition of a time-sharing random variable $Q.$ 
\begin{corollary} \label{Col: JoS Superposition}
(Achievable joint secrecy rate region by superposition coding)
	For BC-RSI with an external eavesdropper, an achievable region under the joint secrecy constraint can be obtained by superposition coding as the set of the non-negative rate pairs $(R_1, R_2)$ such that 
		\begin{align}\label{eqn: JoS Superposition}
		\begin{split}
			R_1& \leq I(V;Y_1)-I(V;Z)\\
			R_2& \leq I(V;Y_2)-I(V;Z)
		\end{split}
		\end{align}
	where $V\to X\to (Y_1, Y_2, Z)$ forms a Markov chain such that $I(V;Y_i)\geq I(V;Z)$ holds for $i=1,2.$
\end{corollary}

\begin{corollary} \label{Col: JoS Marton}
(Achievable joint secrecy rate region by Marton's coding)
	For BC-RSI with an external eavesdropper, an achievable region under the joint secrecy constraint can be obtained by Marton's coding as the set of the non-negative rate pairs $(R_1, R_2)$ such that 
		\begin{align}\label{eqn: JoS Marton}
		\begin{split}
			R_1& \leq I(V_0, V_1;Y_1|Q)-I(V_0, V_1;Z|Q)\\
			R_2& \leq I(V_0, V_2;Y_2|Q)-I(V_0, V_2;Z|Q)\\
			R_1+R_2& \leq I(V_0, V_1;Y_1|Q)+I(V_0, V_2;Y_2|Q)-2I(V_0;Z|Q)-I(V_1;V_2|V_0, Q)
		\end{split}
		\end{align}
	over any  $p(q, v_0,v_1,v_2,x)=p(q)p(v_0|q)p(v_1,v_2|v_0)p(x|v_1,v_2)$ subject to $I(V_1, V_2;Z|V_0)\leq I(V_1;Z|V_0)+I(V_2;Z|V_0)-I(V_1;V_2|V_0)$ and $I(V_i;Z|V_0)\leq I(V_i;Y_i|V_0)$ for $i=1,2.$
\end{corollary}

\begin{remark}
	The region by superposition coding given in \eqref{eqn: JoS Superposition} coincides with the one established in \cite{Wyrembelski:Secrecy11}. Note that \eqref{eqn: JoS Superposition} is included in \eqref{eqn: JoS Marton}, i.e., the region by Marton's coding, as a special case of \eqref{eqn: JoS Marton} by taking $V_1=V_2=V_0.$ 
\end{remark}

\section{Gaussian BC-RSI} \label{sec:gaussian}

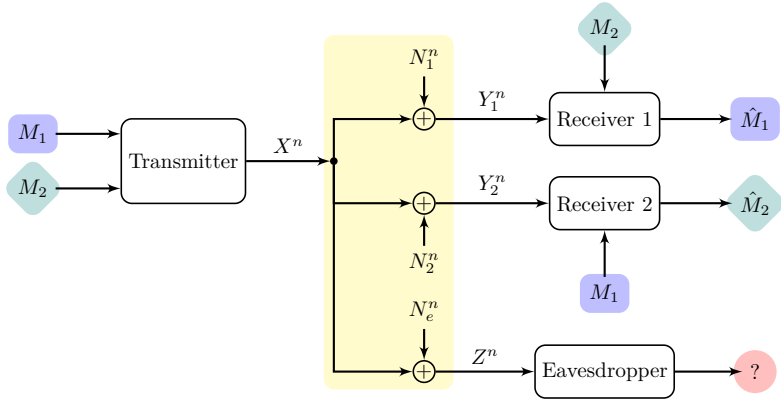
\begin{figure}
\centering
\begin{tikzpicture}[semithick, >=latex',scale=0.8, every node/.style={transform shape}]
 
  \begin{scope}[local bounding box=tx]
  	\node (enc) at (1.5,0) [transmitter_block] {Transmitter};
  \end{scope}
  
  \begin{scope}[local bounding box=Receiver 1]
  	\node (dec1) at (8.5,0.7) [receiver_block] {Receiver 1};
  \end{scope} 
  
  \begin{scope}[local bounding box=Receiver 2]
  	\node (dec2) at (8.5,-0.7) [receiver_block] {Receiver 2};
  \end{scope} 
   
  \begin{scope}[local bounding box=Eavesdropper]  
  	\node (eve) at (8.5,-3.5) [receiver_block] {Eavesdropper};
  \end{scope} 
  
  \begin{scope}[local bounding box=channel]
  	\node (dot1) at (4,0) [dot] {};
  	\node (add1) at (5.5, 0.7) [mul] {$+$};
  	\node (noise1) at (5.5, 1.7) {$N_1^n$};
  	\node (add2) at (5.5,-0.7) [mul] {$+$};
  	\node (noise2) at (5.5,-1.7) {$N_2^n$};
  	\node (add3) at (5.5,-3.5) [mul] {$+$};
  	\node (noise3) at (5.5,-2.5) {$N_e^n$};
  \end{scope}
  	\draw[->,thick] (enc) to node[above] {$X^n$} (dot1);
  	\draw[->,thick] (noise1) to (add1);
  	\draw[->,thick] (noise2) to (add2);
  	\draw[->,thick] (noise3) to (add3);
  	\draw[->,thick] (dot1) |- (add1);
  	\draw[->,thick] (dot1) |- (add2);
  	\draw[->,thick] (dot1) |- (add3);
 
 \begin{scope}[local bounding box=sideinfo1]
 	\node (sidedec1) at (8.5,2.2) {$M_2$};
 \end{scope}
 
 \begin{scope}[local bounding box=sideinfo2]
 	\node (sidedec2) at (8.5,-2.2) {$M_1$};
 \end{scope}
 
  \draw[->,thick] (add1) to node[above] {$Y_1^n$} (dec1);
  \draw[->,thick] (add2) to node[above] {$Y_2^n$} (dec2);
  
  \draw[->,thick] (sidedec1) to (dec1);
  \draw[->,thick] (sidedec2) to (dec2);
  
  \draw[->,thick] (add3) to node[above] {$Z^n$} (eve);
  
  \begin{scope}[local bounding box=message1]
  \node (mes1) at ([yshift=3ex, xshift=-2.5cm]enc) {$M_1$};
  \end{scope}
  
  \begin{scope}[local bounding box=message2]
  \node (mes2) at ([yshift=-3ex, xshift=-2.5cm]enc) {$M_2$};
  \end{scope}
  
  \path[->,thick] (mes1) edge ([yshift=3ex]enc.west);
  \path[->,thick] (mes2) edge ([yshift=-3ex]enc.west);
  
  \begin{scope}[local bounding box=decmessage1]
  \node (mess1out) at ([xshift=2.5cm]dec1) {$\hat{M}_1$};
  \end{scope}
  
  \begin{scope}[local bounding box=decmessage2]
  \node (mess2out) at ([xshift=2.5cm]dec2) {$\hat{M}_2$};
  \end{scope}
  
  \path[->,thick] (dec1.east) edge (mess1out);
  \path[->,thick] (dec2.east) edge (mess2out);

  \begin{scope}[local bounding box=evemessage]
  \node (eveOut) at ([xshift=2.5cm]eve) {?};
  
  \end{scope}
  \draw[->,thick] (eve) to (eveOut);

  \begin{pgfonlayer}{background}
  
 	\node [fill=yellow!25,fit=(channel),rounded corners] {};
  
 	\node [fill=blue!25,fit=(message1),rounded corners, scale=0.85] {};
  
 	\node [shape=diamond, fill=teal!25,fit=(message2),rounded corners, scale=0.6] {};
  
 	\node [fill=blue!25,fit=(decmessage1),rounded corners, scale=0.85] {};
  
 	\node [shape=diamond, fill=teal!25,fit=(decmessage2),rounded corners, scale=0.6] {};
  
 	\node [shape=diamond, fill=teal!25,fit=(sideinfo1),rounded corners, scale=0.6] {};
  
 	\node [fill=blue!25,fit=(sideinfo2),rounded corners, scale=0.85] {};
  
 	\node [shape=circle, fill=red!25,fit=(evemessage),rounded corners, scale=0.75] {};
  \end{pgfonlayer}
 \end{tikzpicture}

\caption{Gaussian BC-RSI with an external eavesdropper.}
\label{fig:gaussian}
\end{figure}

In this section, we consider Gaussian broadcast channel with receiver side information (Gaussian BC-RSI) as shown in Fig. \ref{fig:gaussian}. It is known that one can apply the discretization procedure \cite{ElGamal:2012} to extend the coding schemes for finite alphabet channels to their Gaussian counterpart. Using this technique, we obtain an achievable individual secrecy rate region for the Gaussian BC-RSI. Furthermore, we derive an outer bound to the secrecy capacity region, and, show that, in the high SNR regime, one can approach the individual secrecy capacity region for the Gaussian BC-RSI by employing the superposition coding. This observation is consistent with the results suggested by the linear deterministic approach analyzed in Section \ref{sec:deterministic}.

Suppose ${X}$ is the channel input with a power constraint $P$ on it and the signals received by both receivers and the eavesdropper are 
	\begin{align*}
		{Y}_1&={X}+{N}_{1};\\
		{Y}_2&={X}+{N}_{2};\\
		{Z}&={X}+{N}_{e},
	\end{align*}
where ${N}_{1}\sim \mathcal{N}(0, \sigma_1^2)$, ${N}_{2}\sim \mathcal{N}(0, \sigma_2^2)$ and ${N}_{e}\sim \mathcal{N}(0, \sigma_e^2)$ are additive white Gaussian noise (AWGN) independent of ${X}$.
According to the noise level in the channels to both receives and the eavesdropper, the overall channel can be regarded to be {\it stochastically} degraded in different orders. For simplicity, we only consider their corresponding {\it physically} degraded instances. The reason is that the same analysis can be easily extended to the stochastically degraded cases. So the following scenarios are of our interest (without loss of generality we assume $\sigma_1<\sigma_2$):
 \begin{enumerate}
 	\item $\sigma_e^2\geq \sigma_2^2 \geq \sigma_1^2$, i.e., $X\to Y_1\to Y_2 \to Z$ forms a Markov chain,
 	\item $\sigma_2^2\geq \sigma_1^2 \geq \sigma_e^2$, i.e., $X\to Z \to Y_1 \to Y_2$ forms a Markov chain, and
 	\item $\sigma_2^2\geq \sigma_e^2 \geq \sigma_1^2$, i.e., $X\to Y_1 \to Z \to Y_2$ forms a Markov chain.
 \end{enumerate}

The individual secrecy capacity of the first two cases can be easily derived by extending the results for discrete memoryless channel model to the Gaussian scenario. For the third case, we show in the following that we can approach the individual secrecy capacity region as $P\gg  \sigma_e^2$ or $P\ll \sigma_1^2$.

\subsection{An outer bound}

\begin{proposition}\label{prop: Gaussian upper bound}  An outer bound of the individual secrecy capacity region for the Gaussian BC-RSI when $X\to Y_1 \to Z \to Y_2$ forms a Markov chain is given by the set of the rate pairs $(R_{1}, R_{2})$ satisfying
 \begin{align*}
 		R_{2}\leq & C\left(\frac{(1-\gamma\alpha)P}{\gamma\alpha P+\sigma_2^2}\right);\\ 
 		R_{2}\leq R_{1}\leq & C\left(\frac{\alpha P}{\sigma_1^2}\right)-C\left(\frac{\alpha P}{\sigma_e^2}\right)+R_2,
 	\end{align*}
for some $\alpha, \gamma\in [0,1],$ and $C(x)=\frac{1}{2}\log (1+x)$ is the Gaussian capacity function.
\end{proposition}
\begin{IEEEproof} We observe that 
\begin{align*}
	\frac{n}{2}\log 2\pi e \sigma_e^2  & =h(Z^n|X^n)=h(Z^n|M_1, M_2, X^n)\\
								& \leq h(Z^n|M_1,M_2)\leq h(Z^n|M_2)\leq h(Z^n)\\
								& \stackrel{(a)}{\leq} \frac{n}{2}\log 2\pi e (P+\sigma_e^2),
\end{align*}
where $(a)$ is due to the fact that for a random variable with a fixed variance, Gaussian distribution maximizes the entropy. This shows that there exist $\alpha, \gamma\in [0,1],$ such that
\begin{align}
	h(Z^n|M_2)&=\frac{n}{2}\log 2\pi e (\alpha P+\sigma_e^2); \label{eqn: alpha}\\
	h(Z^n|M_1, M_2)&= \frac{n}{2}\log 2\pi e (\gamma\alpha P+\sigma_e^2). \label{eqn: gamma}
\end{align}
In particular, we have
	\begin{equation}\label{eqn: h(Z^n|M_1)}
		h(Z^n|M_1)= h(Z^n)-I(M_1;Z^n)\stackrel{(b)}{\geq}h(Z^n)-n\epsilon \geq h(Z^n|M_2)-n\epsilon=\frac{n}{2}\log 2\pi e (\alpha P+\sigma_e^2)-n\epsilon,
	\end{equation}	
where $(b)$ is due to the individual secrecy constraint.

Similarly, we have 
\begin{align*}
	\frac{n}{2}\log 2\pi e \sigma_2^2  
								& =h(Y_2^n|X^n)=h(Y_2^n|M_1, M_2, X^n)\\
								& \leq h(Y_2^n|M_1, M_2) \leq h(Y_2^n|M_1)\leq H(Y_2^n)\\
								& \stackrel{(a)}{\leq} \frac{n}{2}\log 2\pi e (P+\sigma_2^2).
\end{align*}
There must exist a $\beta$ such that
\begin{equation}\label{eqn: beta}
	h(Y_2^n|M_1, M_2)= \frac{n}{2}\log 2\pi e (\beta P+\sigma_2^2).
\end{equation}
Therefore, 
\begin{align}
nR_2&=	H(M_2)=H(M_2|M_1)\stackrel{(c)}{=}I(M_2;Y_2^n|M_1)+nO(\epsilon)\nonumber\\
	&= 	h(Y_2^n|M_1)-h(Y_2^n|M_1, M_2)+nO(\epsilon)\nonumber\\
	&\stackrel{(d)}{\leq} \frac{n}{2} \log \frac{P+\sigma_2^2}{\beta P+\sigma_2^2}+nO(\epsilon),
\label{eqn: R_2 upper bound beta}
\end{align}
where $(c)$ is due to the Fano's inequality and $(d)$ is due to \eqref{eqn: beta}.

Recall the Markov chain $(M_1, M_2)\to X^n\to Y_1^n\to Z^n \to Y_2^n.$ Applying the entropy power inequality (EPI) \cite{ElGamal:2012}, we obtain
\begin{equation*}
	h(Y_2^n|M_1, M_2)\geq \frac{n}{2} \log \left[ 2^{\frac{2}{n} h(Z^n|M_1, M_2)} + 2\pi e (\sigma_2^2-\sigma_e^2)\right].
\end{equation*}
Using \eqref{eqn: beta} here, we obtain
\begin{equation*}
	h(Z^n|M_1, M_2)\leq \frac{n}{2} \log  2\pi e (\beta P+\sigma_e^2).
\end{equation*}
Comparing to \eqref{eqn: gamma} which gives that $h(Z^n|M_1, M_2)= \frac{n}{2}\log 2\pi e (\gamma\alpha P+\sigma_e^2),$  we have 
\begin{equation} \label{eqn: gamma alpha <= beta}
\gamma\alpha\leq \beta.
\end{equation}
Recall \eqref{eqn: R_2 upper bound beta}, we have
\begin{equation*}
	nR_2\leq \frac{n}{2} \log \frac{P+\sigma_2^2}{\beta P+\sigma_2^2}+nO(\epsilon) \leq \frac{n}{2} \log \frac{P+\sigma_2^2}{\gamma\alpha P+\sigma_2^2}+nO(\epsilon)=nC\left(\frac{(1-\gamma\alpha)P}{\gamma\alpha P+\sigma_2^2}\right)+nO(\epsilon).
\end{equation*}
Letting $\epsilon\to 0,$ we obtain 
\begin{equation}\label{eqn: Upper R_2}
	R_2\leq C\left(\frac{(1-\gamma\alpha)P}{\gamma\alpha P+\sigma_2^2}\right).
\end{equation}

Now we proceed to bound $R_1.$ First we show $R_1\geq R_2$ as follows.
\allowdisplaybreaks
\begin{align*}
nR_1&=	H(M_1)=H(M_1|M_2)\\
	&\geq I(M_1;Y_1^n|M_2)\\
	&= I(M_1;Y_1^n, Z^n|M_2)-I(M_1; Z^n|M_2, Y_1^n)\\
	&\stackrel{(e)}{=} I(M_1; Z^n|M_2)+I(M_1; Y_1^n|M_2, Z^n)\\
	&= h(Z^n|M_2)-h(Z^n|M_1, M_2)+I(M_1; Y_1^n|M_2, Z^n)\\
	&\geq h(Z^n|M_2)-h(Z^n|M_1, M_2)\\
	&\stackrel{(f)}{\geq} h(Z^n)-h(Z^n|M_1, M_2)-nO(\epsilon)\\
	&\geq h(Z^n|M_1)-h(Z^n|M_1, M_2)-nO(\epsilon)\\
	&= I(M_2;Z^n|M_1)-nO(\epsilon)\\
	&\stackrel{(g)}{\geq}I(M_2;Y_2^n|M_1)-nO(\epsilon)\\
	&= H(M_2|M_1)-H(M_2|M_1,Y_2^n)-nO(\epsilon)\\
	&\stackrel{(h)}{\geq} nR_2-nO(\epsilon), 
\end{align*}
where 
	$(e)$ follows by the fact that $I(M_1; Z^n|M_2, Y_1^n)=0,$ which is implied by $I(M_1, M_2; Z^n|Y_1^n)=0$ due to the channel degradedness, i.e., the Markov chain $(M_1, M_2)\to X^n\to Y_1^n\to Z^n\to Y_2^n;$ 
	$(f)$ is due to the individual secrecy constraint; and 
	$(g)$ is due to the channel degradedness, i.e., $(M_1, M_2)\to X^n\to Y_1^n\to Z^n\to Y_2^n;$ 
	$(h)$ is due to the Fano's inequality. 
	
Finally, letting $\epsilon\to 0,$ we obtain 
	\begin{equation}\label{eqn: Upper R_1 low}
		R_1\geq R_2.
	\end{equation}

On the other hand, we have 
\begin{align}
nR_1&=	H(M_1)=H(M_1|M_2)\nonumber\\
	&\stackrel{(i)}{\leq}I(M_1;Y_1^n|M_2)+nO(\epsilon)\nonumber\\
	&= I(M_1;Y_1^n, Z^n|M_2)-I(M_1; Z^n|M_2, Y_1^n)+nO(\epsilon)\nonumber\\
	&\stackrel{(j)}{=} I(M_1; Z^n|M_2)+I(M_1; Y_1^n|M_2, Z^n)+nO(\epsilon)\nonumber\\
	&= h(Z^n|M_2)-h(Z^n|M_1, M_2)+I(M_1; Y_1^n|M_2, Z^n)+nO(\epsilon),\label{eqn: R_1 step 1}
\end{align}
where $(i)$ is due to the Fano's inequality and $(j)$ is due to the channel degradedness. Note that 
\begin{align}
	I(M_1; Y_1^n|M_2, Z^n)	&=h(Y_1^n|M_2, Z^n)-h(Y_1^n|M_1, M_2, Z^n)\nonumber\\
							&\leq h(Y_1^n|M_2, Z^n)-h(Y_1^n|M_1, M_2, X^n, Z^n)\nonumber\\
							&= h(Y_1^n|M_2, Z^n)-h(Y_1^n|X^n, Z^n)\nonumber\\
							&= h(Y_1^n, Z^n|M_2)-h(Z^n|M_2)-h(Y_1^n|X^n, Z^n)\nonumber\\
							&\stackrel{(k)}{=} h(Y_1^n|M_2)+h(Z^n|Y_1^n)-h(Z^n|M_2)-h(Y_1^n, Z^n|X^n)+h(Z^n|X^n)\nonumber\\
							&\stackrel{(k)}{=} h(Y_1^n|M_2)+h(Z^n|Y_1^n)-h(Z^n|M_2)-h(Y_1^n|X^n)-h(Z^n|Y_1^n)+h(Z^n|X^n)\nonumber\\
							&= h(Y_1^n|M_2)-h(Z^n|M_2)-h(Y_1^n|X^n)+h(Z^n|X^n), \label{eqn: R_1 step 2}
\end{align}
where $(k)$ follows by the fact that $h(Z^n|M_2, Y_1^n)=h(Z^n|Y_1^n)$ and $h(Z^n|X^n, Y_1^n)=h(Z^n|Y_1^n)$ due to the Markov chain $(M_1, M_2)\to X^n\to Y_1^n\to Z^n.$

Recall the Markov chain $(M_1, M_2)\to X^n\to Y_1^n\to Z^n \to Y_2^n.$ We apply the EPI and obtain
\begin{equation*}
	h(Z^n|M_2)\geq \frac{n}{2} \log \left[ 2^{\frac{n}{2} h(Y_1^n|M_2)} + 2\pi e (\sigma_e^2-\sigma_1^2)\right].
\end{equation*}
In addition to \eqref{eqn: alpha} which gives that $h(Z^n|M_2)=\frac{n}{2}\log 2\pi e (\alpha P+\sigma_e^2),$ we have
\begin{equation}
	h(Y_1^n|M_2)\leq \frac{n}{2} \log  2\pi e (\alpha P+\sigma_1^2). \label{eqn: h(Y_1^n|M_2)}
\end{equation}

Combining \eqref{eqn: R_1 step 1} and \eqref{eqn: R_1 step 2}, we have
\begin{align*}
	nR_1	&\leq h(Z^n|M_2)-h(Z^n|M_1, M_2)+I(M_1; Y_1^n|M_2, Z^n)\\
			&\leq h(Y_1^n|M_2)-h(Z^n|M_1, M_2)-h(Y_1^n|X^n)+h(Z^n|X^n)\\
			& = h(Z^n|M_1)-h(Z^n|M_1, M_2)+h(Y_1^n|M_2)-h(Z^n|M_1)-h(Y_1^n|X^n)+h(Z^n|X^n)\\
			&= I(M_2;Z^n|M_1)+h(Y_1^n|M_2)-h(Z^n|M_1)-h(Y_1^n|X^n)+h(Z^n|X^n)\\
			&\stackrel{(l)}{\leq} nR_2+h(Y_1^n|M_2)-h(Z^n|M_1)-h(Y_1^n|X^n)+h(Z^n|X^n)\\
			&= nR_2+h(Y_1^n|M_2)-h(Z^n|M_1)-h(N_1^n)+h(N_e^n)\\
			&\stackrel{(m)}{\leq} nR_2+\frac{n}{2} \log \frac{(\alpha P+\sigma_1^2)\sigma_e^2}{(\alpha P+\sigma_e^2)\sigma_1^2}+nO(\epsilon)\\
			&=nR_2+nC\left(\frac{\alpha P}{\sigma_1^2}\right)-nC\left(\frac{\alpha P}{\sigma_e^2}\right)+nO(\epsilon),
\end{align*}
where $(l)$ is due to the fact that $I(M_2;Z^n|M_1)\leq H(M_2)=nR_2;$ and $(m)$ is due to \eqref{eqn: h(Z^n|M_1)} and \eqref{eqn: h(Y_1^n|M_2)}. 

Finally, letting $\epsilon\to 0$, we have
\begin{equation}\label{eqn: Upper R_1 up}
	R_1\leq C\left(\frac{\alpha P}{\sigma_1^2}\right)-\left(\frac{\alpha P}{\sigma_e^2}\right)+R_2.
\end{equation}

Combining \eqref{eqn: Upper R_2}, \eqref{eqn: Upper R_1 low}, and \eqref{eqn: Upper R_1 up} establishes the outer bound.
\end{IEEEproof}

\begin{remark}
	Interestingly, $\gamma=1$ corresponds to the joint secrecy constraint, since $\gamma=1$ implies that 
			$h(Z^n|M_1, M_2)=h(Z^n)$
	according to \eqref{eqn: gamma}. However, in case of $(M_1, M_2)\to Y_1^n\to Z^n\to Y_2^n,$  we have
		  	$$nR_2=H(M_2)=I(M_2;Y_2^n|M_1)\leq I(M_1, M_2; Y_2^n)\leq I(M_1, M_2; Y_2^n, Z^n)=I(M_1, M_2; Z^n)=0$$
	under joint secrecy constraint. That is, only positive $R_1$ is possible. And,  $R_1\leq C\left({P}/{\sigma_1^2}\right)-C\left({P}/{\sigma_e^2}\right)$ is obtained by taking $\alpha=1$ via Wyner's secrecy coding. 
\end{remark}

\subsection{An inner bound}
\begin{proposition} \label{prop: Gaussian lower bound} An inner bound of the individual secrecy capacity region for the Gaussian BC-RSI when $X\to Y_1 \to Z \to Y_2$ forms a Markov chain is given by the set of the rate pairs $(R_{1}, R_{2})$ satisfying
 \begin{align*}
 		R_{2}\leq & C\left(\frac{(1-\gamma\alpha)P}{\gamma\alpha P+\sigma_2^2}\right);\\ 
 		R_{2}\leq R_{1}\leq & C\left(\frac{\gamma\alpha P}{\sigma_1^2}\right)-C\left(\frac{\gamma\alpha P}{\sigma_e^2}\right)+R_2,
 	\end{align*}
 where $\alpha, \gamma\in[0,1].$
\end{proposition}
\begin{IEEEproof}
For a fixed pair $\alpha, \gamma\in[0,1],$ one can derive an inner bound of $(R_{1}, R_{2})$ by applying superposition coding as described in the following. 

{\em Codebook generation}: Randomly and independently generate $2^{nR_2}$ sequences $u^n(k),$ $k\in [1: 2^{nR_2}],$ each i.i.d. $\mathcal{N}(0, (1-\gamma\alpha)P);$ and $2^{n(R_1-R_2+R_r)}$ sequences $v^n(s, r),$ $(s,r)\in [1:2^{n(R_1-R_2)}]\times [1: 2^{nR_r}],$ each i.i.d. $\mathcal{N}(0, \gamma\alpha P).$ 

{\em Encoding}: To send the message pair $(m_1, m_2)$ with $m_1=(m_{1k}, m_{1s}),$ where $m_{1k}$ is of the same length as $m_2,$ the encoder encapsulates $m_{1k}$ and $m_2$ in $m_k$ with $m_k\triangleq m_{1k}\oplus m_2,$ randomly chooses $r\in [1: 2^{nR_r}],$ and transmits $x^n(m_1, m_2)=u^n(m_{k})+v^n(m_{1s}, r).$ 

{\em Decoding}: Receiver 2 decodes $m_k$ from $y_2^n=u^n(m_{k})+(v^n(m_{1s}, r)+n_2^n)$ while treating $v^n(m_{1s}, r)$ as noise, and further recovers $m_2$ with his knowledge of $m_{1}$. The probability of decoding error tends to zero as $n\to \infty$ if $R_2\leq C\left(\frac{(1-\gamma\alpha)P}{\gamma\alpha P+\sigma_2^2}\right).$

Receiver 1 uses successive cancellation. It first decodes $m_k$ from $y_1^n=u^n(m_{k})+(v^n(m_{1s}, r)+n_1^n)$ while treating $v^n(m_{1s}, r)$ as noise, and recovers part of $m_1,$ i.e., $m_{1k},$ with the knowledge of $m_2.$ The probability of this decoding error tends to zero as $n\to \infty$ if $R_2\leq C\left(\frac{(1-\gamma\alpha)P}{\gamma\alpha P+\sigma_2^2}\right),$ since it implies that $R_2\leq C\left(\frac{(1-\gamma\alpha)P}{\gamma\alpha P+\sigma_1^2}\right)$ due to the fact that $\sigma_1^2\leq \sigma_2^2.$ (This implies that $R_2\leq R_1.$)  Then, it subtracts off $u^n(m_{k})$ and decodes $v^n(m_{1s}, r)+n_1^n$ to recover $(m_{1s}, r)$ and thus $m_{1s},$ i.e., the rest of $m_1.$ The probability of this decoding error tends to zero as $n\to \infty$ if $R_1-R_2+R_r\leq C\left(\frac{\gamma\alpha P}{\sigma_1^2}\right).$

{\em Secrecy}: The eavesdropper could decode $m_k$ from $z^n=u^n(m_{k})+(v^n(m_{1s}, r)+n_e^n).$ However, $m_k$ does not disclose any information about $m_{1s}$ and $m_2,$ individually. Subtracting off $u^n(m_{k})$ from $z^n,$ the eavesdropper gets a better observation $v^n(m_{1s}, r)+n_e^n,$ which actually does not help to recover $m_{1s}$ if $R_r\approx C\left(\frac{\gamma\alpha P}{\sigma_e^2}\right)$. In other words, the secrecy of $m_{1s}$ is guaranteed by the embedded secrecy coding in the choice of $v^n.$ The individual secrecy for $m_1$ then follows from an analysis similar to the previous sections.

As a conclusion, $(R_1, R_2)$ is achievable under the individual secrecy constraints, once $R_1, R_2, R_r$ satisfy
	\begin{align*}
		R_2 & \leq C\left(\frac{(1-\gamma\alpha)P}{\gamma\alpha P+\sigma_2^2}\right);\\
		R_1-R_2+R_r& \leq C\left(\frac{\gamma\alpha P}{\sigma_1^2}\right);\\
		R_r& \approx C\left(\frac{\gamma\alpha P}{\sigma_e^2}\right).
	\end{align*}
Eliminating $R_{r}$, we get the desired region of $(R_1, R_2),$ which concludes our proof of achievability.
\end{IEEEproof}

\subsection{Individual secrecy capacity region}

\begin{proposition}
When $\sigma_2^2\geq \sigma_e^2\geq \sigma_1^2,$ and $P\gg \sigma_2^2$ or $P\ll \sigma_1^2,$ the individual secrecy capacity region for the Gaussian BC-RSI is given as the set of $(R_1, R_2)$ satisfying 
			\begin{align*}				 
				R_{2}\leq & C\left(\frac{(1-\gamma\alpha)P}{\gamma\alpha P+\sigma_2^2}\right);\\ 
				R_{1}\leq & C\left(\frac{\gamma\alpha P}{\sigma_1^2}\right)-C\left(\frac{\gamma\alpha P}{\sigma_e^2}\right)+R_2,
			\end{align*}
	where $\gamma, \alpha\in[0,1].$
\end{proposition}
\begin{IEEEproof}
	Consider the gap between the inner and outer bounds derived in previous subsections. If we take the same choice of $\alpha, \gamma$ in both bounds, the gap occurs only in $R_1$, which is given by
		\begin{align*}
			C\left(\frac{\alpha P}{\sigma_1^2}\right)-C\left(\frac{\alpha P}{\sigma_e^2}\right)-C\left(\frac{\gamma\alpha P}{\sigma_1^2}\right)+C\left(\frac{\gamma\alpha P}{\sigma_e^2}\right)=\frac{1}{2}\log \frac{(\alpha P+\sigma_1^2)(\gamma\alpha P+\sigma_e^2)}{(\alpha P+\sigma_e^2)(\gamma\alpha P+\sigma_1^2)} \to 0, 
		\end{align*}
	as $P\gg \sigma_e^2$ or $P\ll \sigma_1^2$, regardless of the values of $\alpha, \gamma$.
\end{IEEEproof}

As a conclusion, we characterize the individual secrecy capacity region for the Gaussian BC-RSI as follows.

\begin{proposition}
	The individual secrecy capacity region for the Gaussian BC-RSI is given by the following set of $(R_1, R_2)$:
	\begin{itemize}
		\item If $\sigma_e^2\geq \sigma_2^2\geq \sigma_1^2:$
				\begin{align*}
					R_1&\leq \min\left\{C\left(\frac{P}{\sigma_1^2}\right)-C\left(\frac{P}{\sigma_e^2}\right)+R_2, C\left(\frac{P}{\sigma_1^2}\right)\right\};\\
					R_2&\leq \min\left\{C\left(\frac{P}{\sigma_2^2}\right)-C\left(\frac{P}{\sigma_e^2}\right)+R_1, C\left(\frac{P}{\sigma_2^2}\right)\right\},
				\end{align*}
		\item If $\sigma_2^2\geq \sigma_1^2\geq \sigma_e^2:$
						\begin{equation*}
							R_1=R_2\leq C\left(\frac{P}{\sigma_2^2}\right),
						\end{equation*}
		\item If $\sigma_2^2\geq \sigma_e^2\geq \sigma_1^2,$ and, $P\gg \sigma_2^2$ or $P\ll \sigma_1^2:$
			 \begin{align*}
			 		R_{1}\leq & C\left(\frac{\gamma P}{\sigma_1^2}\right)-C\left(\frac{\gamma P}{\sigma_e^2}\right)+R_2;\\
			 		R_{2}\leq & C\left(\frac{(1-\gamma)P}{\gamma P+\sigma_2^2}\right),\quad \mbox{where}\ \gamma \in[0,1].
			 	\end{align*}
	\end{itemize}
\end{proposition}


\section{Conclusion}
In this paper, we studied the problem of secure communication over BC-RSI under the individual secrecy constraints. 
We first characterized the individual secrecy capacity region for the linear deterministic channel model. Then, utilizing secret key, secrecy coding, superposition coding, and Marton's coding approaches, we derived achievable rate regions for the discrete memoryless model. Together with converse arguments, these techniques allow us to characterize the individual secrecy capacity region for some specific scenarios which include 1) the case of a {\it strong} eavesdropper (as a line on $(R_1, R_2)$ plane); 2) the case of a {\it weak} eavesdropper (as a rectangle with missing corners);
and 3) the case that the eavesdropper's channel is deterministic.
Our results exhibit the coupling between the communication rates. In particular, we observe that one can not arbitrarily decrease one user's rate without sacrificing the rate of the other. Moreover, we studied the corresponding Gaussian scenario, where, in addition to the capacity regions for strong and weak eavesdropper cases, we established the individual secrecy capacity region for the low and high SNR regimes when the eavesdropper channel is weaker than one of the legitimate receivers but stronger than the other.

We here point out some avenues for further research. First, the characterization of the individual secrecy capacity region for the general case remains as an open problem. In particular, the characterization of the capacity region for the \emph{mixed} case (where the eavesdropper channel is less noisy than one legitmate receiver but more noisy than the other) has resisted our best efforts thus far. (For the Gaussian case, we were able to establish low and high SNR individual secrecy capacity results in this scenario.) 
Remarkably, this mixed case is distinctive for the study on secure communication via broadcast channels (with RSI or without RSI) since in this case, positive rate pairs are attainable under the {\em individual secrecy} constraint but impossible under the {\em joint secrecy} constraint. We believe that our results will initiate the study of \emph{individual secrecy} for other multi-terminal models. During the preparation of this manuscript, we have noticed that the parallel work \cite{Mansour:Capacity15} has considered the extension of BC-RSI model to include common messages. Studying other channel models under the lens of individual secrecy and comparing this notion to other secrecy constraints will be of interest.


\appendices


\section{Upper bound on the individual secrecy rate}
\label{sec:UpperBoundAppendix}

An upper bound on the individual secrecy rate follows from the results for wiretap channel with shared key \cite{Kang:Wiretap10} as provided below.

\begin{lemma}\label{lemma: upper bound}
For any $R_2$ in the achievable region, $R_1$ is upper bounded as 
\begin{equation*}
R_1\leq \max\limits_{U\to V\to X\to (Y_1,Z)} \min\{I(V;Y_1|U)-I(V;Z|U)+R_2,I(V;Y_1)\}.
\end{equation*} 
If the channel to the legitimate receiver 1 is degraded
with respect to the channel to the eavesdropper, then for any $R_2$ in the achievable region, $R_1$ is upper bounded by 
\begin{equation*} 
R_1\leq \max\limits_{X\to Y_1\to Z}
\min\{I(X;Y_1)-I(X;Z)+R_2,I(X;Y_1)\}.
\end{equation*} 
Similar results hold for interchanging $1$ and $2$ above.
\end{lemma}
\begin{IEEEproof}
The proof follows by the result given for the wiretap channel with shared key \cite[Theorem 1]{Kang:Wiretap10}. As the rate for $M_2$ is $R_2$, then the secrecy rate for receiver $1$ can be upper bounded by the wiretap channel with shared key of rate $R_2$.
\end{IEEEproof}


\section{Achievability Proof for Theorem~\ref{thm: IndS linear deterministic}}
\label{sec:AppDeterministic}

For each scenario, a specific coding scheme is provided, where for a given $m_1=[m_1(1), \cdots, m_1(R_1)]$ and $m_2=[m_2(1), \cdots, m_2(R_2)]$, we construct the codeword $X=[x(1), x(2), \cdots, x(n_1)]^T$. 
\subsection{$q=n_1\geq n_2\geq n_e$}
\subsubsection{$R_1<R_2$ and $R_1<n_e$}
\begin{figure}[H]
					\centering
					\begin{tabular}{rcl}
							$m_1:$ &    & 
									$\begin{tikzpicture}
										\node[minimum height=1.6em, anchor=base, fill=blue!25] {$m_{1}(1), \cdots, m_1(R_1)$}; 
									\end{tikzpicture}$\\
							$m_2:$ & 	& 
									$\begin{tikzpicture}
										\node[minimum height=1.6em, anchor=base, fill=teal!25] {$m_{2}(1), \cdots, m_2(R_1), \cdots, m_2(R_2)$}; 
									\end{tikzpicture}$ \\
							$X^T:$ & 	&
									$
									\underbrace{
									\underbrace{
									\underbrace{
									\overbrace{
									\begin{tikzpicture}
										\node[minimum height=1.6em, anchor=base, fill=red!25] {$\quad m_{1}(k)\oplus m_{2}(k)\quad $};
									\end{tikzpicture}
									}^{R_{1}}
									\begin{tikzpicture}
										\node[minimum height=1.6em, anchor=base, fill=yellow!25] {$r(k)$};
									\end{tikzpicture}
									}_{n_e}
									\overbrace{
									\begin{tikzpicture}
										\node[minimum height=1.6em, anchor=base, fill=teal!25] {$m_{2}(k-n_e+R_1)$};
									\end{tikzpicture}
									}^{R_{2}-R_{1}}
									}_{\leq n_2}
									\begin{tikzpicture}
										\node[minimum height=1.6em, anchor=base, fill=yellow!25] {$r(k)$};
									\end{tikzpicture}
									}_{n_1}$		
							\end{tabular}
							\caption{Codeword $X$ for $n_1\geq n_2\geq n_e$ and $R_1<R_2, R_1<n_e.$}
							\label{fig: X^n 1<a}											
						\end{figure}
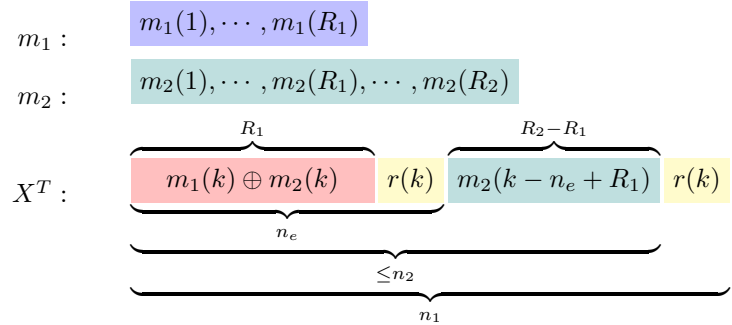

We have 
				\begin{equation}\label{eqn: Cs 111}
					 		R_1	< n_e; \quad R_1<	R_2	\leq n_2-n_e +R_1.
				\end{equation}
We set
				\begin{equation*}
					x(k)=\left\{
							\begin{array}{lll}
								m_1(k)\oplus m_2(k) & &  1\leq k\leq R_1\\
								r(k) &	& R_1< k\leq n_e  \\
								m_2(k-n_e+R_1) & & n_e< k\leq n_e+R_2-R_1\\
								r(k) &	& n_e+R_2-R_1< k\leq n_1 \\
							\end{array}
						\right. 
				\end{equation*}
				where $r(k)$ is randomly chosen from $\{0, 1\}.$ The construction of $X$ is illustrated in Fig. \ref{fig: X^n 1<a}. In this scenario, receiver 2 could recover $m_2$ completely only if $R_2-R_1+n_e\leq n_2$. Combining this with the aforementioned conditions, $R_1<R_2$ and $R_1<n_e,$ we obtain the desired region of $(R_1, R_2)$ as specified in \eqref{eqn: Cs 111}.		
	\subsubsection{$R_1<R_2$ and $n_e\leq R_1$}
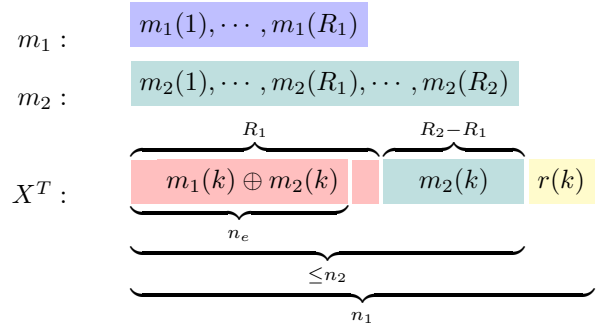
\begin{figure}[H]
									\centering
									\begin{tabular}{rcl}
										$m_1:$ &	& 
												$\begin{tikzpicture}
													\node[minimum height=1.6em, anchor=base, fill=blue!25] {$m_{1}(1), \cdots, m_1(R_1)$}; 
												\end{tikzpicture}$ \\
										$m_2:$ &	& 
												$\begin{tikzpicture}
													\node[minimum height=1.6em, anchor=base, fill=teal!25] {$m_{2}(1), \cdots, m_2(R_1), \cdots, m_2(R_2)$}; 
												\end{tikzpicture}$ \\
										$X^T:$ & 	&
												$
												\underbrace{
												\underbrace{
												\overbrace{
												\underbrace{
												\begin{tikzpicture}
													\node[minimum height=1.65em, anchor=base, fill=red!25] {$\ $};
												\end{tikzpicture}
												\begin{tikzpicture}
													\node[minimum height=1.6em, anchor=base, fill=red!25] {$m_{1}(k)\oplus m_{2}(k)$};
												\end{tikzpicture}
												}_{n_e}
												\begin{tikzpicture}
													\node[minimum height=1.65em, anchor=base, fill=red!25] {$\ $};
												\end{tikzpicture}
												}^{R_1}
												\overbrace{
												\begin{tikzpicture}
													\node[minimum height=1.6em, anchor=base, fill=teal!25] {$\quad m_{2}(k)\quad$};
												\end{tikzpicture}
												}^{R_{2}-R_{1}}
												}_{\leq n_2}
												\begin{tikzpicture}
													\node[minimum height=1.6em, anchor=base, fill=yellow!25] {$r(k)$};
												\end{tikzpicture}
												}_{n_1}
												$
										\end{tabular}
										\caption{Codeword $X$ for $n_1\geq n_2\geq n_e$ and $n_e\leq R_1<R_2.$}
										\label{fig: X^n 1<b}													
									\end{figure}

We have
					\begin{equation}\label{eqn: Cs 112}
								 n_e\leq R_1 	< R_2 \leq n_2.
					\end{equation}
We set
							\begin{equation*}
								x(k)=\left\{
										\begin{array}{lll}
											m_1(k)\oplus m_2(k) & &  1\leq k\leq R_1\\
											m_2(k) & & R_1< k\leq R_2\\
											r(k) &	& R_2< k\leq n_1, 
										\end{array}
									\right. 
							\end{equation*} 
			where $r(k)$ is randomly chosen from $\{0, 1\}.$ The construction of $X$ is illustrated in Fig. \ref{fig: X^n 1<b}. In this scenario, receiver 2 could recover $m_2$ completely only if $R_2\leq n_2.$ Combining this with the aforementioned conditions, $n_e\leq R_1<R_2,$ we obtain the desired region of $(R_1, R_2)$ as specified in \eqref{eqn: Cs 112}.			
\subsubsection{$R_1\geq R_2$ and $R_2<n_e$} 
\begin{figure}[H]
					\centering
					\begin{tabular}{rcl}
							$m_1:$ &	& 
									$\begin{tikzpicture}
										\node[minimum height=1.6em, anchor=base, fill=blue!25] {$m_{1}(1), \cdots, m_1(R_2), \cdots, m_1(R_1)$}; 
									\end{tikzpicture}$\\
							$m_2:$ & 	& 
									$\begin{tikzpicture}
										\node[minimum height=1.6em, anchor=base, fill=teal!25] {$m_{2}(1), \cdots, m_2(R_2)$}; 
									\end{tikzpicture}$\\
							$X^T:$ & 	&
									$
									\underbrace{
									\underbrace{
									\overbrace{
									\begin{tikzpicture}
										\node[minimum height=1.6em, anchor=base, fill=red!25] {$\quad m_{1}(k)\oplus m_{2}(k)\quad $};
									\end{tikzpicture}
									}^{R_{2}}
									\begin{tikzpicture}
										\node[minimum height=1.6em, anchor=base, fill=yellow!25] {$r(k)$};
									\end{tikzpicture}
									}_{n_e\leq n_2}
									\overbrace{
									\begin{tikzpicture}
										\node[minimum height=1.6em, anchor=base, fill=blue!25] {$m_{1}(k-n_e+R_2)$};
									\end{tikzpicture}
									}^{R_{1}-R_{2}}
									\begin{tikzpicture}
										\node[minimum height=1.6em, anchor=base, fill=yellow!25] {$r(k)$};
									\end{tikzpicture}
									}_{n_1}
									$
								\end{tabular}
								\caption{Codeword $X$ for  $n_1\geq n_2\geq n_e$ and $R_1\geq R_2, R_2<n_e.$}
								\label{fig: X^n 1>a}												
						\end{figure}
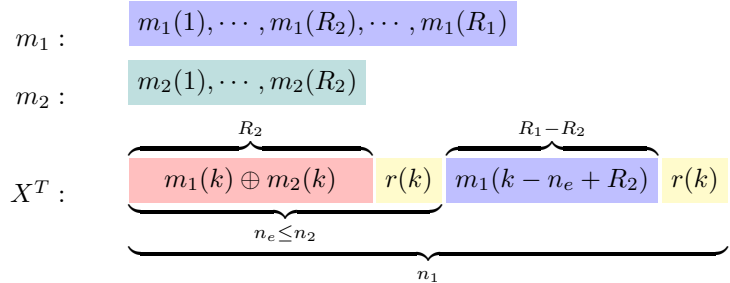

We have 
						\begin{equation}\label{eqn: Cs 121}
							R_2<n_e; \quad R_2\leq R_1\leq n_1-n_e+R_2.
						\end{equation}
We set
						\begin{equation*}
							x(k)=\left\{
									\begin{array}{lll}
										m_1(k)\oplus m_2(k) & &  1\leq k\leq R_2\\
										r(k) &	& R_2< k\leq n_e \\
										m_1(k-n_e+R_2) & & n_e+1\leq k\leq n_e+R_1-R_2\\
										r(k) &	& n_e+R_1-R_2< k\leq n_1 \\
									\end{array}
								\right. 
						\end{equation*}
						where $r(k)$ is randomly chosen from $\{0, 1\}.$ The construction of $X$ is illustrated in Fig. \ref{fig: X^n 1>a}. In this scenario, receiver 1 could recover $m_1$ completely only if $R_1-R_2+n_e\leq n_1$. Combining this with the aforementioned conditions, i.e.,  $R_1\geq R_2$ and $R_2<n_2$, we obtain the desired region of $(R_1, R_2)$ as specified in \eqref{eqn: Cs 121}.
\subsubsection{$R_1\geq R_2$ and $n_e\leq R_2$} 
\begin{figure}[H]
						\centering
							\begin{tabular}{rcl}
								$m_1:$ &	& 
										$\begin{tikzpicture}
											\node[minimum height=1.6em, anchor=base, fill=blue!25] {$m_{1}(1), \cdots, m_1(R_2), \cdots, m_1(R_1)$}; 
										\end{tikzpicture}$\\
								$m_2:$ &	& 
										$\begin{tikzpicture}
											\node[minimum height=1.6em, anchor=base, fill=teal!25] {$m_{2}(1), \cdots, m_2(R_2)$}; 
										\end{tikzpicture}$\\
								$X^T:$ &	&
											$
												\underbrace{
												\underbrace{
												\overbrace{
												\underbrace{
												\begin{tikzpicture}
													\node[minimum height=1.65em, anchor=base, fill=red!25] {$\ $};
												\end{tikzpicture}
												\begin{tikzpicture}
													\node[minimum height=1.6em, anchor=base, fill=red!25] {$m_{1}(k)\oplus m_{2}(k)$};
												\end{tikzpicture}
												}_{n_e}
												\begin{tikzpicture}
													\node[minimum height=1.65em, anchor=base, fill=red!25] {$\ $};
												\end{tikzpicture}
												}^{R_2}
												}_{\leq n_2}
												\overbrace{
												\begin{tikzpicture}
													\node[minimum height=1.6em, anchor=base, fill=blue!25] {$\quad m_{1}(k)\quad$};
												\end{tikzpicture}
												}^{R_{1}-R_{2}}
												\begin{tikzpicture}
													\node[minimum height=1.6em, anchor=base, fill=yellow!25] {$r(k)$};
												\end{tikzpicture}
												}_{n_1}
												$
											\end{tabular}
											\caption{Codeword $X$ for  $n_1\geq n_2\geq n_e$ and $n_e\leq R_2\leq R_1.$}
											\label{fig: X^n 1>b}													
									\end{figure}
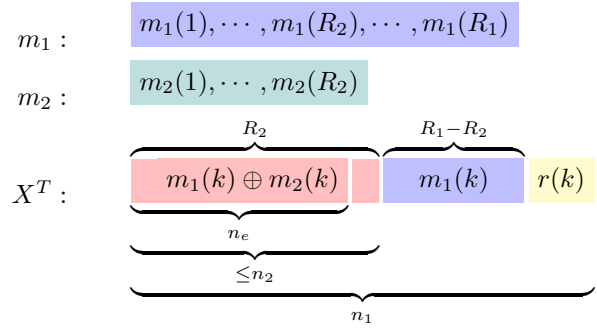

We have
						\begin{equation}\label{eqn: Cs 122}
							R_2\leq n_2; \quad n_e\leq R_2\leq R_1\leq n_1.
						\end{equation}
We set			
									\begin{equation*}
										x(k)=\left\{
												\begin{array}{lll}
													m_1(k)\oplus m_2(k) & &  1\leq k\leq R_2\\
													m_1(k) & & R_2< k\leq R_1\\
													r(k) &	& R_1< k\leq n_1,
												\end{array}
											\right. 
									\end{equation*} 
where $r(k)$ is randomly chosen from $\{0, 1\}.$ The construction of $X$ is illustrated in Fig. \ref{fig: X^n 1>b}. In this scenario, receiver 2 could recover $m_2$ completely only if $R_2\leq n_2;$ and receiver 1 could recover $m_1$ completely only if $R_1\leq n_1.$ Combining this with the aforementioned assumptions, $n_e\leq R_2\leq R_1$, we obtain the desired region of $(R_1, R_2)$ as specified in \eqref{eqn: Cs 122}. 	
\subsection{$q=n_1\geq n_e\geq n_2$} 
\begin{figure}[H]
					\centering
					\begin{tabular}{rcl}
							$m_1:$ &	& 
									$\begin{tikzpicture}
										\node[minimum height=1.6em, anchor=base, fill=blue!25] {$m_{1}(1), \cdots, m_1(R_2), \cdots, m_1(R_1)$}; 
									\end{tikzpicture}$\\
							$m_2:$ &	& 
									$\begin{tikzpicture}
										\node[minimum height=1.6em, anchor=base, fill=teal!25] {$m_{2}(1), \cdots, m_2(R_2)$}; 
									\end{tikzpicture}$\\
							$X^T:$ &	&
									$
									\underbrace{
									\underbrace{
									\underbrace{
									\overbrace{
									\begin{tikzpicture}
										\node[minimum height=1.6em, anchor=base, fill=red!25] {$\quad m_{1}(k)\oplus m_{2}(k)\quad$};
									\end{tikzpicture}
									}^{R_{2}}
									}_{\leq n_2}
									\begin{tikzpicture}
										\node[minimum height=1.6em, anchor=base, fill=yellow!25] {$r(k)$};
									\end{tikzpicture}
									}_{n_e\geq n_2}
									\overbrace{
									\begin{tikzpicture}
										\node[minimum height=1.6em, anchor=base, fill=blue!25] {$m_{1}(k-n_e+R_2)$};
									\end{tikzpicture}
									}^{R_{1}-R_{2}}
									\begin{tikzpicture}
										\node[minimum height=1.6em, anchor=base, fill=yellow!25] {$r(k)$};
									\end{tikzpicture}
									}_{n_1}
									$
								\end{tabular}
								\caption{Codeword $X$ for $n_1\geq n_e\geq n_2.$}	
								\label{fig: X^n 2}												
						\end{figure}
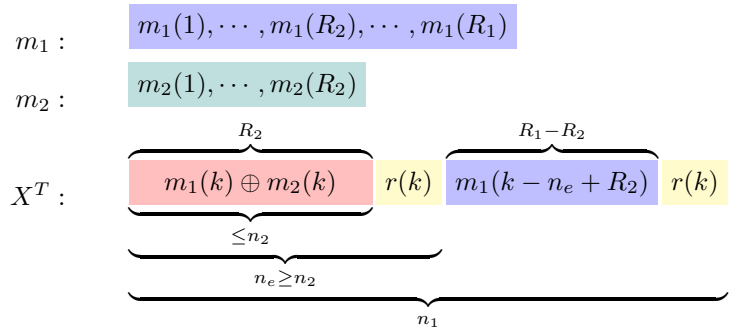

Since $R_2\leq n_2\leq n_e,$ we have 
		\begin{equation}\label{eqn: Cs 12}
				R_2\leq n_2; \quad R_2\leq R_1\leq n_1-n_e+R_2.
		\end{equation}
We set
						\begin{equation*}
							x(k)=\left\{
									\begin{array}{lll}
										m_1(k)\oplus m_2(k) & &  1\leq k\leq R_2\\
										r(k) &	& R_2< k\leq n_e\\
										m_1(k-n_e+R_2) & & n_e < k\leq n_e+R_1-R_2\\
										r(k) &	& n_e+R_1-R_2< k\leq n_1 \\
									\end{array}
								\right. 
						\end{equation*} 
				where $r(k)$ is randomly chosen from $\{0, 1\}.$ The construction of $X$ is illustrated in Fig. \ref{fig: X^n 2}. In this case, receiver 2 could recover $m_2$ completely only if $R_2\leq n_2;$ and receiver 1 could recover $m_1$ completely only if $R_1-R_2+n_e\leq n_1.$ Combining these with the fact $R_1-R_2\geq 0$, which is implied by the code construction, we obtain the desired region of $(R_1, R_2)$ as specified in \eqref{eqn: Cs 12}.
\subsection{$q=n_e\geq n_1\geq n_2$}
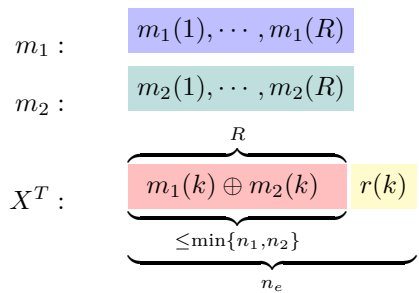
\begin{figure}[H]
					\centering
						\begin{tabular}{rcl}
							$m_1:$ &	& 
									$\begin{tikzpicture}
										\node[minimum height=1.6em, anchor=base, fill=blue!25] {$m_{1}(1), \cdots, m_1(R)$}; 
									\end{tikzpicture}$\\
							$m_2:$ &	& 
									$\begin{tikzpicture}
										\node[minimum height=1.6em, anchor=base, fill=teal!25] {$m_{2}(1), \cdots, m_2(R)$}; 
									\end{tikzpicture}$\\
							$X^T:$ &	&
									$
									\underbrace{
									\underbrace{
									\overbrace{
									\begin{tikzpicture}
										\node[minimum height=1.6em, anchor=base, fill=red!25] {$\  m_{1}(k)\oplus m_{2}(k)\ \ $};
									\end{tikzpicture}
									}^{R}
									}_{\leq \min\{n_1, n_2\}}
									\begin{tikzpicture}
										\node[minimum height=1.6em, anchor=base, fill=yellow!25] {$r(k)$};
									\end{tikzpicture}
									}_{n_e}
									$
								\end{tabular}
								\caption{Codeword $X$ for $n_e\geq n_1\geq n_2.$}	
								\label{fig: X^n 3}											
						\end{figure}

In this case, we have $R_1\leq R_2$ and $R_2\leq R_1$ both holds. This gives that 
	\begin{equation}\label{eqn: Cs 13}
		R_1=R_2=R\leq \min\{n_1, n_2\}.
	\end{equation}
We set
						\begin{equation*}
							x(k)=\left\{
									\begin{array}{lll}
										m_1(k)\oplus m_2(k) & &  1\leq k\leq R\\
										r(k) &	& R < k\leq n_e, 
									\end{array}
								\right. 
						\end{equation*} 
where $r(k)$ is randomly chosen from $\{0, 1\}.$ The construction of $X$ is illustrated in Fig. \ref{fig: X^n 3}. In this scenario, both receivers could recover $m_1, m_2,$ respectively, only if $R\leq \min\{n_1, n_2\}.$ Combining this with the fact $R_1=R_2=R$, which is implied by the code construction, we obtain the desired region of $(R_1, R_2)$ as specified in \eqref{eqn: Cs 13}.				
\begin{remark} Note that in our achievability schemes, the elements of the input vector $X$ are i.i.d. $\mbox{Bern}(\frac{1}{2})$ in all scenarios. That is, $\mbox{Bern}(\frac{1}{2})$ serves as an optimal input distribution to achieve the individual secrecy capacity. Nevertheless, this choice is not the only optimal one. As an alternative, instead of choosing as uniformly random, one can simply use zeros for the bits represented by $r(k)$ in our achievability schemes. 
\end{remark}

	 
\section{Proof of Theorem~\ref{pro: superposition}}
\label{sec:AppSuperposition}

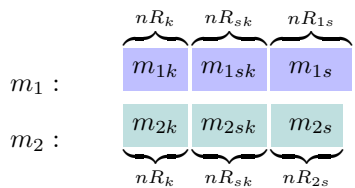
\begin{figure}[H]
\centering
\begin{tabular}{rcl}
				$m_1:$ &	& 
						$
						\overbrace{
						\begin{tikzpicture}
							\node[minimum height=1.6em, anchor=base, fill=blue!25] {$m_{1k}$}; 
						\end{tikzpicture}
							}^{nR_k}
						\overbrace{					
						\begin{tikzpicture}
							\node[minimum height=1.6em, anchor=base, fill=blue!25] {$m_{1sk}$};
						\end{tikzpicture} 
						}^{nR_{sk}}
						\overbrace{
						\begin{tikzpicture}
							\node[minimum height=1.6em, anchor=base, fill=blue!25] {\ $m_{1s}$\ \  }; 
						\end{tikzpicture}
						}^{nR_{1s}}
						$\\
				$m_2:$ &	&  
								$
									\underbrace{
										\begin{tikzpicture}
											\node[minimum height=1.6em, anchor=base, fill=teal!25] {$m_{2k}$}; 
										\end{tikzpicture}
											}_{nR_k}
										\underbrace{
										\begin{tikzpicture}
											\node[minimum height=1.6em, anchor=base, fill=teal!25] {$m_{2sk}$};
										\end{tikzpicture}
												}_{nR_{sk}}
										\underbrace{
											\begin{tikzpicture}
												\node[minimum height=1.6em, anchor=base, fill=teal!25] {\ $m_{2s}$}; 
											\end{tikzpicture}
										}_{nR_{2s}}
									$
			\end{tabular}
			\caption{Superposition coding: Rate splitting.}
			\label{eqn: superC RS}	
			\end{figure}

{\em Rate splitting:} As illustrated in Fig. \ref{eqn: superC RS}, we split $M_1=(M_{1k}, M_{1sk}, M_{1s})$ and $M_2=(M_{2k}, M_{2sk}, M_{2s}),$ with both $M_{1k}$ and $M_{2k}$ of entropy $nR_k$, both $M_{1sk}$ and $M_{2sk}$ of entropy $nR_{sk}$, $M_{1s}$ of entropy $nR_{1s}$ and $M_{2s}$ of entropy $nR_{2s}.$ Thus, we have $R_1=R_k+R_{sk}+R_{1s}$ and $R_2=R_k+R_{sk}+R_{2s}.$  
{\em Codebook generation:} Fix $p(u), p(v|u).$ First, randomly generate $2^{nR_k}$ i.i.d. sequences $u^{n}(k),$ $k\in[1:2^{nR_k}],$ according to $p(u).$
Secondly, for each $u^n(k),$ according to $p(v|u),$ randomly generate i.i.d. sequences $v^n(k, m_{sk}, m_{1s}, m_{2s}, r)$ with $(m_{sk}, m_{1s}, m_{2s}, r)\in [1: 2^{nR_{sk}}]\times [1: 2^{nR_{1s}}]\times [1: 2^{nR_{2s}}]\times [1: 2^{nR_r}].$ 
	
\begin{figure}[H]
	\centering
	\begin{tabular}{rcl}
			$u^n(k):$ &	& 
					$\overbrace{
						\begin{tikzpicture}
							\node[minimum height=1.6em, anchor=base, fill=red!25] {$m_{1k}\oplus m_{2k}$}; 
						\end{tikzpicture}
							}^{nR_k} $\\
			$v^n(k,m_{sk}, m_{1s}, m_{2s}, r):$ &	& 
											$
											\underbrace{
											\begin{tikzpicture}
												\node[minimum height=1.6em, anchor=base, fill=blue!25] {$m_{1s}$}; 
											\end{tikzpicture}
											}_{nR_{1s}}
											\overbrace{
											\begin{tikzpicture}
												\node[minimum height=1.6em, anchor=base, fill=red!25] {$m_{1sk}\oplus m_{2sk}$}; 
											\end{tikzpicture}
											}^{nR_{sk}}
			 								\overbrace{
											\begin{tikzpicture}
												\node[minimum height=1.6em, anchor=base, fill=red!25] {$\ r\ $}; 
											\end{tikzpicture}
											}^{nR_r}
											\underbrace{
											\begin{tikzpicture}
												\node[minimum height=1.6em, anchor=base, fill=teal!25] {$m_{2s}$}; 
											\end{tikzpicture}
										}_{nR_{2s}}
										$
			\end{tabular}
			\caption{Superposition coding: Encoding.}
			\label{fig: superC encoding}
		\end{figure}
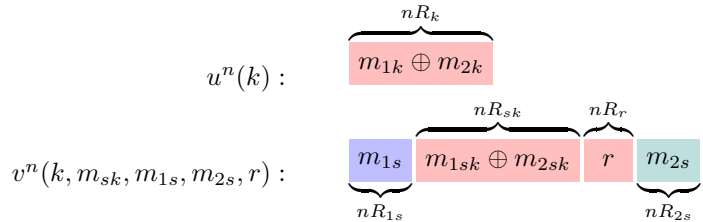

{\em Encoding:} To send messages $(m_1, m_2),$ choose $u^n(k),$ where $k=m_k\triangleq m_{1k}\oplus m_{2k}.$ Given $u^n(k),$ randomly choose $r\in [1: 2^{nR_r}]$ and find $v^n(k, m_{sk}, m_{1s}, m_{2s}, r),$ where $m_{sk}\triangleq m_{1sk}\oplus m_{2sk}.$ The choice of $u^n, v^n$ for given $(m_1, m_2)$ is illustrated in Fig. \ref{fig: superC encoding}. Generate $x^n$ according to $p(x|v),$ and transmit it to the channel.
{\em Decoding:} 
Receiver 1, upon receiving $y_1^n$ and with knowledge of $m_2,$ decodes $\hat{m}_1=(m_{2k}\oplus \hat{k}, m_{2sk}\oplus \hat{m}_{sk}, \hat{m}_{1s})$ if $(\hat{k}, \hat{m}_{sk}, \hat{m}_{1s}, m_{2s})$ is the unique quadruple such that $(u^n(\hat{k}), v^n(\hat{k}, \hat{m}_{sk}, \hat{m}_{1s}, m_{2s}, \hat{r}), y_1^n)$ is jointly typical. 

Receiver 2, upon receiving $y_2^n$ and with knowledge of $m_1,$  decodes $\tilde{m}_2=(m_{1k} \oplus \tilde{k}, m_{1sk}\oplus \tilde{m}_{sk}, \tilde{m}_{2s}),$ if $(\tilde{k}, \tilde{m}_{sk}, m_{1s}, \tilde{m}_{2s})$ is the unique quadruple such that $(u^n(\tilde{k}), v^n(\tilde{k}, \tilde{m}_{sk}, m_{1s}, \tilde{m}_{2s}, \tilde{r}), y_2^n)$ is jointly typical.

{\em Analysis of the error probability of decoding:} 
Assume that $(M_1,M_2)=(m_1,m_2)$ with $m_1=(m_{1k}, m_{1sk}, m_{1s}),$ $m_2=(m_{2k}, m_{2sk}, m_{2s})$ is sent. Or, more specifically, $k,  m_{sk},$  $m_{1s}$ and $m_{2s}$ are sent, where $k\triangleq m_{1k}\oplus m_{2k}$ and $m_{sk}=m_{1sk}\oplus m_{2sk}.$    
 
At receiver 1, i.e., for $P_{e,1}$, a decoding error happens if receiver 1's estimate is $u^n(\hat{k}),$ $v^n(\hat{k}, \hat{m}_{sk}, \hat{m}_{1s}, m_{2s}, \hat{r})$ with $(\hat{k}, \hat{m}_{sk}, \hat{m}_{1s})\neq (k, m_{sk}, m_{1s}).$ In more details, the error event can be partitioned into the followings:
	\begin{enumerate}
		\item Error event corresponds to $\hat{k} \neq k.$ Note that this event occurs with arbitrarily small probability if 
			\begin{equation}\label{eqn: SS cond pe11}
					R_1+R_r\leq I(U, V;Y_1)=I(V;Y_1).
			\end{equation}
		\item Error event corresponds to $\hat{k} = k,$ but $(\hat{m}_{1s}, \hat{m}_{sk}) \neq (m_{1s}, m_{sk})$ Note that this event occurs with arbitrarily small probability if 
			\begin{equation}\label{eqn: SS cond pe12}
					R_{sk}+R_{1s}+R_r\leq I(V;Y_1|U).
			\end{equation}	
	\end{enumerate}

Similar analysis can be done at the receiver 2, from which the decoding error probability  $P_{e,2}$ can be made arbitrarily small if
	\begin{align}
			R_2+R_r& \leq I(V;Y_2) \label{eqn: SS cond pe21}\\
			R_{sk}+R_{2s}+R_r&\leq I(V;Y_2|U) \label{eqn: SS cond pe22}
	\end{align} 

{\em Analysis of individual secrecy:} 
Due to the symmetric roles of receiver 1 and receiver 2, we only need to prove the secrecy of one message (e.g., $M_1$). The proof for the other case (e.g., the secrecy of $M_2$) follows similarly. For the secrecy of $M_1,$ we have 
	\begin{align*}
		 I(M_1; Z^n)	
			=	& I(M_{1k}, M_{1sk}, M_{1s}; Z^n)\\ 
			=	& I(M_{1k}; Z^n)+I(M_{1sk}, M_{1s};Z^n|M_{1k})\\
			\stackrel{(a)}{=}
				& I(M_{1sk}, M_{1s}; Z^n|M_{1k})\\
			=	& I(M_{1sk}; Z^n|M_{1k})+I(M_{1s};Z^n|M_{1k}, M_{1sk})\\
			\stackrel{(b)}{=}& I(M_{1s};Z^n|M_{1k}, M_{1sk})\\
			=	& H(M_{1s})-H(M_{1s}|M_{1k}, M_{1sk}, Z^n)\\
			\stackrel{(c)}{\leq} & nR_{1s}-H(M_{1s}|M_k, Z^n),
	\end{align*}
where 
	$(a)$ is due to the fact that $I(M_{1k}; Z^n)=0$  by $I(M_{1k}; Z^n)\leq I(M_{1k}; Z^n, M_k)= I(M_{1k}; M_k)=0,$ which follows by the Markov chain $M_{1k}\rightarrow M_k\rightarrow Z^n;$ 
	$(b)$ follows the fact that $I(M_{1sk}; Z^n|M_{1k})=0$ as $H(M_{1sk}|Z^n, M_{1k})	\geq H(M_{1sk}|Z^n, M_{1k}, M_{sk}) = H(M_{1sk}|M_{sk}) = H(M_{1sk})= H(M_{1sk}|M_{1k})$;
	$(c)$ is due to the fact that $H(M_{1s}|M_{1k}, M_{1sk}, Z^n)\geq H(M_{1s}|M_{1k}, M_{k}, M_{1sk}, Z^n)=H(M_{1s}|M_{k}, Z^n),$ where the last equality follows as $M_{1k}, M_{1sk}$ are independent of $M_{1s}$ given $M_k,Z^n,$ which is due to the Markov chain $M_{1s}\to (Z^n,M_{k})\to (M_{1k},M_{1sk})$.  

To complete the proof that $I(M_1;Z^n)\leq n\delta'(\epsilon)$, we show in the following that $H(M_{1s}, M_{2s}|M_k, Z^n)\geq n(R_{1s}+R_{2s})-n\delta'(\epsilon),$ which implies that $H(M_{1s}|M_k, Z^n)\geq nR_{1s}-n\delta'(\epsilon).$ 
\begin{align*}
 H(M_{1s}, M_{2s}|M_k, Z^n)
	\stackrel{(d)}{=}&	H(M_{1s}, M_{2s}|U^n, Z^n)\\
	=& H(M_{1s}, M_{2s}, Z^n|U^n) - H(Z^n|U^n) \\
	=& H(M_{1s}, M_{2s}, Z^n, V^n|U^n)- H(V^n|U^n, M_{1s}, M_{2s}, Z^n) - H(Z^n|U^n) \\
	=& H(V^n|U^n) + H(Z^n|U^n,V^n)  - H(V^n|U^n, M_{1s}, M_{2s}, Z^n)- H(Z^n|U^n)\\
	\stackrel{(e)}{\geq}& n(R_{sk}+R_{1s}+R_{2s}+R_{r})+nH(Z|U,V) -nH(Z|U)- n \epsilon\\
	\stackrel{(f)}{=} & n(R_{1s}+R_{2s})-n\delta'(\epsilon)
	\end{align*}
where 
	$(d)$ is due to the fact that $U^n$ is uniquely determined by $M_k;$ 
	$(e)$ follows by $H(V^n|U^n)=n(R_{sk}+R_{1s}+R_{2s}+R_{r})$ by the codebook construction and the choice of $V^n$ is randomly chosen based on $M_K, M_{sk}, M_{1s}, M_{2s}$ which are presumed to be uniformly distributed; Moreover, since the channel is discrete memoryless, we have $H(Z^n|U^n,V^n)=\sum_{i=1}^n H(Z_i|U_i,V_i)=nH(Z|U,V)$; and, $H(V^n|U^n, M_{1s}, M_{2s}, Z^n)\leq n\epsilon$ due to Fano's inequality by taking 
		\begin{equation}\label{eqn: SuperC Eve cond1}
			R_{sk}+R_r\leq I(V;Z|U)-\epsilon',
		\end{equation}
	since the eavesdropper can decode $V^n$ reliably by using typical set decoding given $(U^n, M_{1s}, M_{2s}, Z^n)$; and $H(Z^n|U^n)=\sum_{i=1}^n H(Z_i|Z^{i-1},U^n) \leq \sum_{i=1}^n H(Z_i|U_i)=nH(Z|U);$ 
	$(f)$ holds by taking 
		\begin{equation}\label{eqn: SuperC Eve cond2}
				R_{sk}+R_r\geq I(V;Z|U)-2\epsilon'
		\end{equation} 
	and $\delta'(\epsilon)=\epsilon+2\epsilon'.$

{\em Achievable rate region:} Combining the followings:
\begin{itemize}
	\item the non-negativity for rates, i.e.,
		\begin{equation*}
			R_{k}, R_{sk}, R_{1s}, R_{2s}, R_r \geq 0;
		\end{equation*}
	\item the rate relations imposed by rate splitting, i.e.,
		\begin{align*}
			R_{1}&=R_k+R_{sk}+R_{1s},\\
			R_{2}&=R_k+R_{sk}+R_{2s};
		\end{align*}
	\item the constraints for a reliable communication to both legitimate receivers, i.e., \eqref{eqn: SS cond pe11}-\eqref{eqn: SS cond pe22}:
			\begin{align*}
				R_1+R_{r} &\leq  I(V;Y_1),\\
				R_2+R_{r} &\leq  I(V;Y_2),\\
				R_{sk}+R_{1s}+R_{r} &\leq  I(V;Y_1|U),\\
				R_{sk}+R_{2s}+R_{r} &\leq  I(V;Y_2|U);
			\end{align*}
	\item  the constraints for individual secrecy of the messages at the eavesdropper, i.e., \eqref{eqn: SuperC Eve cond1}-\eqref{eqn: SuperC Eve cond2}:
	 	\begin{align*}
		 	R_{sk}+R_r &\approx  I(V;Z|U).
	 	\end{align*}
\end{itemize}
Eliminating $R_{r}, R_{k}, R_{sk}, R_{1s}, R_{2s}$ by applying Fourier-Motzkin procedure \cite{ElGamal:2012}, we obtain a region as the union of the set of non-negative $(R_1, R_2)$ pairs satisfying
	\begin{align}\label{eqn: Super SiM IndSR}
		\begin{split}
			R_1	&\leq \min\{I(V;Y_1), \ \ I(V;Y_1|U)-I(V;Z|U)+R_2\};\\
			R_2	&\leq \min\{I(V;Y_2), \ \ I(V;Y_2|U)-I(V;Z|U)+R_1\}.
		\end{split}
	\end{align} 
	where the union is taken over all $p(u)p(v|u)p(x|v)$ subject to $I(V;Y_i|U)\geq I(V;Z|U)$ for $i=1,2.$

\section{Discussion on Rate splitting in Superposition Coding}
\label{sec:App_Superposition_Rate Splitting}

\begin{figure}[H]
\centering
\begin{tabular}{rcl}
				$m_1:$ &	& 
						$
						\overbrace{
						\begin{tikzpicture}
							\node[minimum height=1.6em, anchor=base, fill=blue!25] {$m_{1k}$}; 
						\end{tikzpicture}
							}^{nR_k}
						\overbrace{					
						\begin{tikzpicture}
							\node[minimum height=1.6em, anchor=base, fill=blue!25] {$m_{1sk}$};
						\end{tikzpicture} 
						}^{nR_{sk}}
						\overbrace{
						\begin{tikzpicture}
							\node[minimum height=1.6em, anchor=base, fill=blue!25] {\ $m_{1s}$\ \  }; 
						\end{tikzpicture}
						}^{nR_{1s}}
						\overbrace{
						\begin{tikzpicture}
							\node[minimum height=1.6em, anchor=base, fill=blue!25] {\ $m_{1m}$\ \  }; 
						\end{tikzpicture}
						}^{nR_{1m}}
						$\\
				$m_2:$ &	&  
								$
									\underbrace{
										\begin{tikzpicture}
											\node[minimum height=1.6em, anchor=base, fill=teal!25] {$m_{2k}$}; 
										\end{tikzpicture}
											}_{nR_k}
										\underbrace{
										\begin{tikzpicture}
											\node[minimum height=1.6em, anchor=base, fill=teal!25] {$m_{2sk}$};
										\end{tikzpicture}
												}_{nR_{sk}}
										\underbrace{
											\begin{tikzpicture}
												\node[minimum height=1.6em, anchor=base, fill=teal!25] {\ $m_{2s}$}; 
											\end{tikzpicture}
										}_{nR_{2s}}
											\underbrace{
											\begin{tikzpicture}
												\node[minimum height=1.6em, anchor=base, fill=teal!25] {\ $m_{2m}$\ \  }; 
											\end{tikzpicture}
											}_{nR_{2m}}
									$
			\end{tabular}
			\caption{Superposition coding: Rate splitting.}
			\label{fig: Oma SS RS}	
			\end{figure}
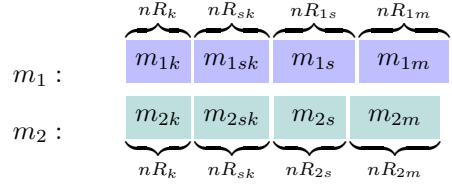

{\em Rate splitting:} As illustrated in Fig. \ref{fig: Oma SS RS}, we represent $M_1, M_2$ by $M_1=(M_{1k}, M_{1sk}, M_{1s}, M_{1m})$ and $M_2=(M_{2k}, M_{2sk}, M_{2s}, M_{2m})$ with $M_{1k}, M_{2k}$ of entropy $nR_k;$ $M_{1sk}, M_{2sk}$ of entropy $nR_{sk};$ while $M_{1s}, M_{1m}$ of entropy $nR_{1s}, nR_{1m};$ and $M_{2s}, M_{2m}$ of entropy $nR_{2s}, nR_{2m},$ respectively. For simplicity, we denote $M_k=M_{1k}\oplus M_{2k},$ $M_{sk}=M_{1sk}\oplus M_{2sk},$ $M_{ss}=(M_{1s}, M_{2s})$ and $M_{sm}=(M_{1m}, M_{2m}).$
{\em Codebook generation:} Fix $p(u), p(v|u)$ and $p(t|v).$ 

First, randomly generate $2^{nR_k}$ i.i.d. sequences $u^{n}(k),$ $k\in[1:2^{nR_k}],$ according to $p(u).$

For each $u^n(k),$  according to $p(v|u),$ randomly generate $2^{n(R_{1s}+R_{2s}+R_{sk}+R_{r})}$ i.i.d. sequences $v^n(k, ss_1, ss_2, sk, r)$ with $(ss_1, ss_2, sk, r)\in [1: 2^{nR_{1s}}]\times[1: 2^{nR_{2s}}]\times [1: 2^{nR_{sk}}]\times [1: 2^{nR_{r}}].$

For each fixed $u^{n}(k)$ and $v^n(k, ss_1, ss_2, sk, r), $ randomly generate $2^{n(R_{1m}+R_{2m}+R_{r_1})}$ i.i.d. sequences $t^n$ with indices $(k, ss_1, ss_2, sk, r, sm_1, sm_2, r_1),$ where $(sm_1, sm_2, r_1)\in [1: 2^{nR_{1m}}]\times [1: 2^{nR_{2m}}]\times [1: 2^{nR_{r_1}}]$, according to $p(t|v)$. 

{\em Encoding:} To send messages $(m_1, m_2),$ choose $u^n(k),$ where $k=m_k\triangleq m_{1k}\oplus m_{2k}.$ 

Given $u^n(k),$ randomly choose $r\in [1: 2^{nR_r}]$ and find the corresponding $v^n(k, m_{1s}, m_{2s}, m_{sk}, r),$ where $m_{sk}\triangleq m_{1sk}\oplus m_{2sk}.$ 

Given $u^n(k)$ and $v^n(k, m_{1s}, m_{2s}, m_{sk}, r),$ randomly choose $r_1\in [1: 2^{nR_{r_1}}],$ and send the corresponding codeword $t^n$ with index $(k, m_{1s}, m_{2s}, m_{sk}, r, m_{1m}, m_{2m}, r_1).$

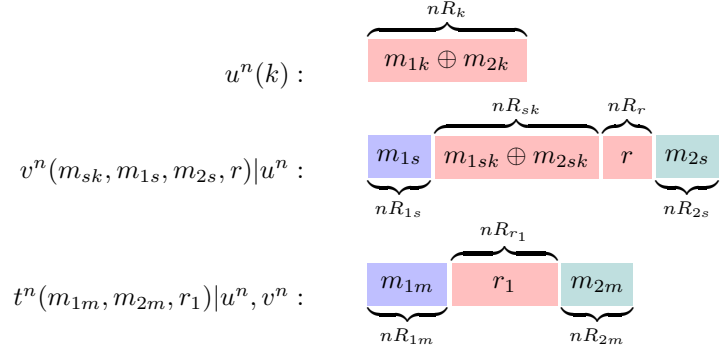
\begin{figure}[H]
\centering
	\begin{tabular}{rcl}
			$u^n(k):$ &	& 
						$
						\overbrace{
						\begin{tikzpicture}
							\node[minimum height=1.6em, minimum width=6em, anchor=base, fill=red!25] {$m_{1k}\oplus m_{2k}$}; 
						\end{tikzpicture}
							}^{nR_k}
						$\\
			$v^n( m_{sk}, m_{1s}, m_{2s},r)|u^n:$ &	& 
											$
											\underbrace{
											\begin{tikzpicture}
												\node[minimum height=1.6em, anchor=base, fill=blue!25] {$m_{1s}$}; 
											\end{tikzpicture}
											}_{nR_{1s}}
											\overbrace{
											\begin{tikzpicture}
												\node[minimum height=1.6em, anchor=base, fill=red!25] {$m_{1sk}\oplus m_{2sk}$}; 
											\end{tikzpicture}
											}^{nR_{sk}}
			 								\overbrace{
											\begin{tikzpicture}
												\node[minimum height=1.6em, anchor=base, fill=red!25] {$\ r\ $}; 
											\end{tikzpicture}
											}^{nR_r}
											\underbrace{
											\begin{tikzpicture}
												\node[minimum height=1.6em, anchor=base, fill=teal!25] {$m_{2s}$}; 
											\end{tikzpicture}
										}_{nR_{2s}}$ \\
			$t^n(m_{1m}, m_{2m}, r_1)|u^n, v^n:$ &	& 
											$
											\underbrace{
											\begin{tikzpicture}
												\node[minimum height=1.6em, minimum width=3em, anchor=base, fill=blue!25] {$m_{1m}$}; 
											\end{tikzpicture}
											}_{nR_{1m}}
											\overbrace{
											\begin{tikzpicture}
												\node[minimum height=1.6em, minimum width=4em, anchor=base, fill=red!25] {$r_1$}; 
											\end{tikzpicture}
											}^{nR_{r_1}}
											\underbrace{
											\begin{tikzpicture}
												\node[minimum height=1.6em, minimum width=2em, anchor=base, fill=teal!25] {$m_{2m}$}; 
											\end{tikzpicture}
										}_{nR_{2m}}
										$
			\end{tabular}
		\caption{Superposition coding: Encoding.}
		\label{fig: Oma Superposition encoding}
		\end{figure}

The choice of $u^n, v^n$ and $t^n$ for given $(m_1, m_2)$ is illustrated in Fig. \ref{fig: Oma Superposition encoding}. 

{\em Decoding:} Receiver 1, upon receiving $y_1^n$ and with the side information $m_2,$ decodes $\hat{m}_1=(m_{2k}\oplus \hat{k}, m_{2sk}\oplus \hat{m}_{sk}, \hat{m}_{1s}, \hat{m}_{1m}),$ if $(\hat{k}, \hat{m}_{sk}, \hat{m}_{1s}, m_{2s}, \hat{r}, \hat{m}_{1m},  m_{2m})$ is the unique tuple such that $(\hat{u}^n, \hat{v}^n, \hat{t}^n, y_1^n)$ is jointly typical, where $\hat{u}^n, \hat{v}^n, \hat{t}^n$ are with indices $(\hat{k},\hat{m}_{sk}, \hat{m}_{1s}, m_{2s}, \hat{r}, \hat{m}_{1m},  m_{2m}, \hat{r}_1)$. 

And, receiver 2, upon receiving $y_2^n$ and with the side information $m_1,$  decodes $\tilde{m}_2=(m_{1k}\oplus \tilde{k}, m_{1sk}\oplus \tilde{m}_{sk}, \tilde{m}_{2s}, \tilde{m}_{2m}),$ if $(\tilde{k}, \tilde{m}_{sk}, m_{1s}, \tilde{m}_{2s}, \tilde{r}, m_{1m},  \tilde{m}_{2m})$ is the unique tuple such that $(\tilde{u}^n, \tilde{v}^n, \tilde{t}^n, y_1^n)$ is jointly typical, where $\hat{u}^n, \hat{v}^n, \hat{t}^n$ are with indices $(\tilde{k}, \tilde{m}_{sk}, m_{1s}, \tilde{m}_{2s}, \tilde{r}, m_{1m}, \tilde{m}_{2m}, \tilde{r}_1)$. 

{\em Analysis of decoding error:}  Assume that $m_1=(m_{1k}, m_{1sk}, m_{1s}, m_{1m}),$ $m_2=(m_{2k}, m_{2sk}, m_{2s}, m_{2m})$ is sent, i.e., more specifically, $k,  m_{sk},  m_{1s}, m_{1m}$ and $m_{2s}, m_{2m}$ are sent, where $k\triangleq m_{1k}\oplus m_{2k}$ and $m_{sk}=m_{1sk}\oplus m_{2sk}.$  

For $P_{e,1}$, a decoding error happens if receiver 1's estimate is $(\hat{u}^n, \hat{v}^n, \hat{t}^n)$ with indices $(\hat{k}, \hat{m}_{1s}, m_{2s}, \hat{m}_{sk}, \hat{r}, \allowbreak \hat{m}_{1m}, m_{2m}, \hat{r}_1)$ such that $(\hat{k}, \hat{m}_{1s}, \hat{m}_{sk}, \hat{r}, \hat{m}_{1m})\neq (k, m_{1s}, m_{sk}, r, m_{1m}).$ In more details, the error event can be partitioned into the followings:
	\begin{enumerate}
		\item Error event corresponds to $\hat{k} \neq k.$ Note that this event occurs with arbitrarily small probability if 
			\begin{equation}\label{eqn: Oma SS cond pe11}
					R_1+R_r+R_{r_1}\leq I(U, V, T;Y_1)=I(T;Y_1).
			\end{equation}
		\item Error event corresponds to $\hat{k} = k,$ but $(\hat{m}_{1s}, \hat{m}_{sk}, \hat{r}) \neq (m_{1s}, m_{sk}, r)$ Note that this event occurs with arbitrarily small probability if 
			\begin{equation}\label{eqn: Oma SS cond pe12}
					R_1-R_k+R_r+R_{r_1}\leq I(V, T;Y_1|U)=I(T;Y_1|U).
			\end{equation}
		\item Error event corresponds to $(\hat{k}, \hat{m}_{1s}, \hat{m}_{sk}, \hat{r}) = (k, m_{1s}, m_{sk}, r)$ but $\hat{m}_{1m}\neq m_{1m}.$ Note that this event occurs with arbitrarily small probability if 
			\begin{equation}\label{eqn: Oma SS cond pe13}
					R_{1m}+R_{r_1}\leq I(T;Y_1|U, V)=I(T;Y_1|V).
			\end{equation}		
	\end{enumerate}

Similar analysis can be done at the receiver 2, from which the decoding error probability  $P_{e,2}$ can be made arbitrarily small if
	\begin{align}
			R_2+R_r+R_{r_1}& \leq I(T;Y_2) \label{eqn: Oma SS cond pe21}\\
			R_2-R_k+R_r+R_{r_1}&\leq I(T;Y_2|U) \label{eqn: Oma SS cond pe22}\\
			R_{2m}+R_{r_1}&\leq I(T;Y_2|V).\label{eqn: Oma SS cond pe23}
	\end{align} 

{\em Analysis of individual secrecy:} 
For the secrecy of $M_1$, we have  
	\begin{align*}
		 I(M_1; Z^n)	
			=& I(M_{1k}, M_{1sk}, M_{1s}, M_{1m}; Z^n)\\
			=& I(M_{1k}, M_{1sk}; Z^n)+I(M_{1s}, M_{1m};Z^n|M_{1k}, M_{1sk})\\
			\stackrel{(a)}{=}& I(M_{1s}, M_{1m};Z^n|M_{1k}, M_{1sk})\\
			=& H(M_{1s}, M_{1m})-H(M_{1s}, M_{1m}|M_{1k}, M_{1sk}, Z^n)\\
			\leq & H(M_{1s}, M_{1m})-H(M_{1s}, M_{1m}|M_k, M_{1k}, M_{1sk}, Z^n)\\
			\stackrel{(b)}{=} & nR_{1s}+ nR_{1m}-H(M_{1s}, M_{1m}|M_{k}, Z^n)
	\end{align*}
where 
	
	$(a)$ is due to the fact that $I(M_{1k}, M_{1sk}; Z^n) = 0$ since 
	\begin{align*}
		I(M_{1k}, M_{1sk}; Z^n) 
			&\leq I(M_{1k}, M_{1sk}; Z^n, M_{k}, M_{sk}, M_{1s}, M_{1m}, M_{2s}, M_{2m})\\
			&=I(M_{1k}, M_{1sk}; M_{k}, M_{sk}, M_{1s}, M_{1m}, M_{2s}, M_{2m})\\
			&=0,
	\end{align*}
	 where the first equality is by the Markov chain $(M_{1k}, M_{2k}, M_{1sk}, M_{2sk})\to (M_k,  M_{sk}, M_{ss}, M_{sm})\to Z^n$; 
	
	$(b)$ is due to the fact that $I(M_{1s}, M_{1m};  M_{1k}, M_{1sk}|Z^n, M_{k})=0,$ where the equality follows by: 
	\begin{enumerate}
		\item $I(M_{1s}, M_{1m};  M_{1k}, M_{1sk}|Z^n, M_{k})\geq 0;$ and
		\item $I(M_{1s}, M_{1m};  M_{1k}, M_{1sk}|Z^n, M_{k})\leq 0$ since 
		\begin{align*}
			H( M_{1k}, M_{1sk}|Z^n, M_{k}, M_{1s}, M_{1m})
				& \geq H( M_{1k}, M_{1sk}|Z^n, M_{k}, M_{sk}, M_{1s}, M_{1m})\\
				& =H(M_{1k}, M_{1sk}|M_k, M_{sk})\\
				& =H( M_{1k}, M_{1sk})\\
				& \geq H( M_{1k}, M_{1sk}|Z^n, M_{k}).
		\end{align*}
	\end{enumerate}
	
So far, we obtain 
	\begin{equation}\label{eqn: Oma SS secrecy step 1}
		I(M_1; Z^n)	\leq nR_{1s}+ nR_{1m}-H(M_{1s}, M_{1m}|M_{k}, Z^n).
	\end{equation} 
	
Similarly, for the secrecy of  $M_2,$ we have
	\begin{equation}\label{eqn: Oma SS secrecy step 1 for M2}
		I(M_2; Z^n)	\leq nR_{2s}+ nR_{2m}-H(M_{2s}, M_{2m}|M_{k}, Z^n).
	\end{equation} 	
	
In the following, we show that $H(M_{1s}, M_{1m}, M_{2s}, M_{2m}|M_k, Z^n)\geq n(R_{1s}+R_{1m}+R_{2s}+R_{2m})-n\delta'(\epsilon)$ holds if the rates satisfy \eqref{eqn: Oma SS IndS condiA} and \eqref{eqn: Oma SS IndS condiB}. This implies that $H(M_{1s}, M_{1m}|M_{k}, Z^n)\geq R_{1s}+R_{1m}-n\delta'(\epsilon)$ and $H(M_{2s}, M_{2m}|M_{k}, Z^n)\geq R_{2s}+R_{2m}-n\delta'(\epsilon).$ Further by (\ref{eqn: Oma SS secrecy step 1}) and \eqref{eqn: Oma SS secrecy step 1 for M2}, we obtain $I(M_1;Z^n)\leq n\delta'(\epsilon)$ and $I(M_2;Z^n)\leq n\delta'(\epsilon),$ thus completing the desired individual secrecy proof.

Note that 
\begin{align}
	& H(M_{1s}, M_{1m}, M_{2s}, M_{2m}|M_k, Z^n)\nonumber\\
		\stackrel{(a)}{=} & H(M_{ss}, M_{sm}|U^n, Z^n)\nonumber\\
		=& H(M_{ss}, M_{sm}, V^n|U^n, Z^n)-H(V^n|U^n, M_{ss}, M_{sm}, Z^n)\nonumber\\
		\geq & H(M_{ss}, M_{sm}, V^n|U^n,Z^n)-H(V^n|U^n, M_{ss}, Z^n)\nonumber\\
		\stackrel{(b)}{\geq} & H(V^n|U^n, Z^n)+H(M_{sm}|U^n, V^n, Z^n)-n(R_{sk}+R_{r}-I(V;Z|U))-n\epsilon_1, \label{eqn: Oma SS terms for inds}
\end{align}
where 
$(a)$ is due to the one-to-one correspondence between $M_k$ and $U^n;$ and the simplification by denoting $M_{ss}=(M_{1s}, M_{2s})$ and $M_{sm}=(M_{1m}, M_{2m});$
$(b)$ follows from \cite[Lemma 1]{src:Chia2012_BC} that $H(V^n|U^n, M_{ss}, Z^n)\leq n(R_{sk}+R_{r}-I(V;Z|U))+n\epsilon_1$ if
	\begin{equation}\label{eqn: Oma SS IndS condiA}
		R_{sk}+R_r\geq I(V;Z|U)+\epsilon.
	\end{equation}

For the first term in \eqref{eqn: Oma SS terms for inds}, i.e., $H(V^n|U^n, Z^n),$ we have
\begin{align}
	H(V^n|U^n, Z^n)
		& =H(V^n, Z^n |U^n)-H(Z^n|U^n)\nonumber\\
		& =H(V^n|U^n)+H(Z^n|U^n, V^n)-H(Z^n|U^n)\nonumber\\
		& =n(R_{1s}+R_{2s}+R_{sk}+R_r)+H(Z^n|U^n, V^n)-H(Z^n|U^n); \label{eqn: Oma SS term 1 for inds}
\end{align}
And, for the second term in \eqref{eqn: Oma SS terms for inds}, i.e., $H(M_{1m}, M_{2m}|U^n, V^n, Z^n),$ we have
\begin{align}
 	 & H(M_{1m}, M_{2m}|U^n, V^n, Z^n)=H(M_{sm}|U^n, V^n, Z^n)\nonumber\\
 	=	& H(M_{sm}, Z^n|U^n, V^n) - H(Z^n|U^n, V^n)\nonumber\\
	=	& H(M_{sm}, Z^n, T^n|U^n, V^n) - H(Z^n|U^n, V^n)-H(T^n|U^n, V^n, M_{sm}, Z^n) \nonumber\\
	\geq & H(T^n|U^n, V^n)+H(Z^n|U^n, V^n,T^n)- H(Z^n|U^n, V^n)-H(T^n|U^n, V^n, M_{sm}, Z^n)\nonumber\\
	=& n(R_{1m}+R_{2m}+R_{r_1})+ H(Z^n|U^n, V^n, T^n)- H(Z^n|U^n, V^n)-H(T^n|U^n, V^n, M_{sm}, Z^n)\nonumber\\
	\stackrel{(a)}{\geq}& n(R_{1m}+R_{2m}+R_{r_1})+ H(Z^n|U^n, V^n, T^n)- H(Z^n|U^n, V^n)-n(R_{r_1}-I(T;Z|U, V))-n\epsilon_2  \label{eqn: Oma SS term 2 for inds}
\end{align} 
where 
	$(a)$ follows from \cite[Lemma 1]{src:Chia2012_BC} that $H(T^n|U^n, V^n, M_{sm}, Z^n)\leq n(R_{r_1}-I(T;Z|U, V))+n\epsilon_2$ if
	\begin{equation}\label{eqn: Oma SS IndS condiB}
		R_{r_1}\geq I(T;Z|V)+\epsilon.
	\end{equation}

Combining \eqref{eqn: Oma SS term 1 for inds} and \eqref{eqn: Oma SS term 2 for inds} in \eqref{eqn: Oma SS terms for inds}, we have
\begin{align*}
	 & H(M_{1s}, M_{1m}, M_{2s}, M_{2m}|M_k, Z^n)\\
	\geq	& H(V^n|U^n, Z^n)+H(M_{1m}, M_{2m}|U^n, V^n, Z^n)-n(R_{sk}+R_{r}-I(V;Z|U))-n\epsilon_1\\
	\geq 	& n(R_{1s}+R_{2s}+R_{sk}+R_r)+H(Z^n|U^n, V^n)-H(Z^n|U^n)\\
	& +n(R_{1m}+R_{2m}+R_{r_1})+ H(Z^n|U^n, V^n, T^n)- H(Z^n|U^n, V^n)-n(R_{r_1}-I(T;Z|U, V))-n\epsilon_2\\
		& -n(R_{sk}+R_{r}-I(V;Z|U))-n\epsilon_1\\
	\stackrel{(a)}{\geq} & n(R_{1s}+R_{1m}+R_{2s}+R_{2m}) -nH(Z|U)+ nH(Z|U, V, T)+nI(V,T;Z|U)+ n\delta'(\epsilon)\\
	{=} & n(R_{1s}+R_{1m}+R_{2s}+R_{2m}) - n\delta'(\epsilon)	
\end{align*}
where 
	$(a)$ follows from $H(Z^n|U^n)\leq$  $\sum_{i=1}^{n} H(Z_i|U_i)$ $=nH(Z|U);$  $H(Z^n|U^n, V^n, T^n)=nH(Z|U, V, T)$; and $\delta'(\epsilon)=\epsilon_1+\epsilon_2.$
	
{\em Achievable rate region:} Combining the followings:
\begin{itemize}
	\item the non-negativity for rates, i.e.,
		\begin{equation*}
			R_{k}, R_{sk}, R_{1s}, R_{2s}, R_r \geq 0;
		\end{equation*}
	\item the rate relations imposed by rate splitting, i.e.,
		\begin{align*}
			R_{1}&=R_k+R_{sk}+R_{1s}+R_{1m},\\
			R_{2}&=R_k+R_{sk}+R_{2s}+R_{2m};
		\end{align*}
	\item the constraints for a reliable communication to both legitimate receivers, i.e., \eqref{eqn: Oma SS cond pe11}-\eqref{eqn: Oma SS cond pe23}:
	\allowdisplaybreaks
			\begin{align*}
			R_1+R_r+R_{r_1}& \leq I(T;Y_1) \\
			R_2+R_r+R_{r_1}& \leq I(T;Y_2) \\
			R_1-R_k+R_r+R_{r_1}&\leq I(T;Y_1|U) \\
			R_2-R_k+R_r+R_{r_1}&\leq I(T;Y_2|U) \\
			R_{1m}+R_{r_1}&\leq I(T;Y_1|V)\\
			R_{2m}+R_{r_1}&\leq I(T;Y_2|V).
			\end{align*}
	\item  the constraints for individual secrecy of the messages at the eavesdropper, i.e., \eqref{eqn: Oma SS IndS condiA} and \eqref{eqn: Oma SS IndS condiB}:
	 	\begin{align*}
		 		R_{sk}+R_r& \geq I(V;Z|U)\\
		 		R_{r_1}&\geq I(T;Z|V).
	 	\end{align*}
\end{itemize}

Applying Fourier-Motzkin procedure \cite{ElGamal:2012} to eliminate $R_{k}, R_{sk}, R_{1m}, R_{2m}, R_r, R_{r_1}$, we get an achievable region, which is the union of non-negative $(R_{1}, R_{2})$ pairs satisfying 
\begin{align}
		R_1	&\leq \min\{I(T;Y_1)-I(T;Z|V), \ \ I(T;Y_1|U)-I(T;Z|U)+R_2\};\\
		R_2	&\leq \min\{I(T;Y_2)-I(T;Z|V), \ \ I(T;Y_2|U)-I(T;Z|U)+R_1\},
\end{align} 
where the union is taken over probability distributions satisfying $U\to V\to T\to (Y_1, Y_2, Z)$ forming a Markov chain and for $i=1,2,$ both $I(T;Y_i|V) \geq I(T;Z|V)$ and $I(T;Y_i|U) \geq I(T;Z|U)$ hold.

Note that for fixed $p(t), p(t|u),$ the above region is outer bounded by the choice of $V=T,$ i.e., the outer bounding region is given by the union of non-negative $(R_{1}, R_{2})$ pairs satisfying 
\begin{align}
		R_1	&\leq \min\{I(V;Y_1), \ \ I(V;Y_1|U)-I(V;Z|U)+R_2\};\\
		R_2	&\leq \min\{I(V;Y_2), \ \ I(V;Y_2|U)-I(V;Z|U)+R_1\},
\end{align} 
where the union is taken over probability distributions satisfying $U\to V\to (Y_1, Y_2, Z)$ forming a Markov chain and $I(V;Y_i|U) \geq I(V;Z|U)$ holds for $i=1,2$. This reduces to the region provided in 
Theorem \ref{pro: superposition}.

\section{Proof of Converse for Theorem~\ref{thm: deterministic Z IndS}}
\label{app: Proof of Converse for deterministic BC-RSI with an eavesdropper}
Consider a BC-RSI with an external eavesdropper. In addition, the eavesdropper's channel is {\em deterministic} in the sense that $Z$ is a function of $X.$ For a reliable communication under individual secrecy constraint, we have
\begin{align*}
	nR_1= 	& H(M_1)=H(M_1|M_2)\\
		=	& I(M_1;Y_1^n|M_2)+H(M_1|M_2, Y_1^n)\\
		\stackrel{(a)}{\leq}& I(M_1;Y_1^n|M_2)+n\epsilon
\end{align*}
where $(a)$ is due to the reliability constraint, i.e., $H(M_1|M_2, Y_1^n)\leq n\epsilon_1$ by Fano's inequality.

On one hand, we have
\allowdisplaybreaks
\begin{align*}
	nR_1 \leq	& I(M_1;Y_1^n|M_2)+n\epsilon\\
		 = 	& \sum_{i=1}^{n}I(M_1;Y_{1i}|M_2, {Y_{1}}_{1}^{i-1})+n\epsilon\\
		 \leq	& \sum_{i=1}^{n}I(M_1, M_2, {Y_{1}}_{1}^{i-1};Y_{1i})+n\epsilon\\
		 \stackrel{(b)}{\leq}	& \sum_{i=1}^{n}I(X_i;Y_{1i})+n\epsilon\\
		 \stackrel{(c)}{\leq} & n I(X; Y_1)+n\epsilon
\end{align*}
where $(b)$ is due to the Markov chain $(M_1, M_2, {Y_{1}}_{1}^{i-1})\to X_i\to Y_{1i};$ $(c)$ is by introducing a time-sharing random variable $Q$ which is uniform over $1, 2\cdots, n$ and by taking $X=X_{Q}, Y_1=Y_{1, Q}.$

On the other hand, we have
	\begin{align*}
		nR_1 \leq	& I(M_1;Y_1^n|M_2)+n\epsilon\\
			 =		& \underbrace{I(M_1;Y_1^n|M_2)-I(M_1;Z^n|M_2)}_{nR_1^s}+\underbrace{I(M_1;Z^n|M_2)}_{nR_1^k}+n\epsilon.
	\end{align*}
The first term $R_1^s$ can be bounded as follows:
\begin{align*}
	nR_1^s 	=& I(M_1;Y_1^n|M_2)-I(M_1;Z^n|M_2)\\
			\leq & I(M_1; Y_1^n, Z^n|M_2)-I(M_1;Z^n|M_2)\\
			=&	I(M_1; Y_1^n|M_2, Z^n)\\
			= & \sum_{i=1}^n I(M_1; Y_{1i}|M_2, Z^n, Y_{1,1}^{i-1})\\
			\leq & \sum_{i=1}^n I(M_1, M_2, {Y_{1}}_{1}^{i-1}, Z_{1}^{i-1}, Z_{i+1}^{n}; Y_{1i}|Z_i)\\
			\stackrel{(d)}{\leq} &\sum_{i=1}^n I(X_i; Y_{1i}|Z_i)\\
			\stackrel{(e)}{\leq}& n I(X; Y_{1}|Z),
\end{align*}
where $(d)$ is due to the Markov chain $(M_1, M_2, {Y_{1}}_{1}^{i-1}, Z_{1}^{i-1}, Z_{i+1}^{n})\to X_i\to (Y_{1i}, Z_i);$
$(e)$ is by applying the time-sharing random variable $Q$ which is uniform over $1, 2\cdots, n$ and by taking $X=X_Q, Y_1=Y_{1, Q}, Z=Z_{Q}.$

And the second term $R_1^k$ can be bounded by 
\begin{align*}
	nR_1^k= & I(M_1;Z^n|M_2)\\
		\leq& I(M_1, M_2;Z^n)\\
		\stackrel{(d)}{\leq}& I(M_2;Z^n|M_1)+n\epsilon\\
		\leq & H(M_2|M_1)+n\epsilon\\
		=& nR_2+n\epsilon,
\end{align*}
where $(d)$ is due to the individual secrecy constraint, i.e., $I(M_1;Z^n)\leq n\epsilon.$ 

As a conclusion of above discussions, we have as $\epsilon\to 0$
	\begin{equation*} 
		R_1\leq \min\{I(X; Y_1), \ I(X; Y_1|Z)+R_2\}.
	\end{equation*}
A similar proof can be applied to $R_2$ and thus completes the proof of the converse.

\section{Proof of Theorem~\ref{thm: Oma Marton}}
\label{sec:App_Oma Marton}
\begin{figure}[H]
\centering
\begin{tabular}{rcl}
				$m_1:$ &	& 
						$
						\overbrace{
						\begin{tikzpicture}
							\node[minimum height=1.6em, anchor=base, fill=blue!25] {$m_{1k}$}; 
						\end{tikzpicture}
							}^{nR_k}
						\overbrace{					
						\begin{tikzpicture}
							\node[minimum height=1.6em, anchor=base, fill=blue!25] {$m_{1sk}$};
						\end{tikzpicture} 
						}^{nR_{sk}}
						\overbrace{
						\begin{tikzpicture}
							\node[minimum height=1.6em, anchor=base, fill=blue!25] {\ $m_{1ss}$\ \  }; 
						\end{tikzpicture}
						}^{nR_{1ss}}
						\overbrace{
						\begin{tikzpicture}
							\node[minimum height=1.6em, anchor=base, fill=blue!25] {\ $m_{1sm}$\ \  }; 
						\end{tikzpicture}
						}^{nR_{1sm}}
						$\\
				$m_2:$ &	&  
								$
									\underbrace{
										\begin{tikzpicture}
											\node[minimum height=1.6em, anchor=base, fill=teal!25] {$m_{2k}$}; 
										\end{tikzpicture}
											}_{nR_k}
										\underbrace{
										\begin{tikzpicture}
											\node[minimum height=1.6em, anchor=base, fill=teal!25] {$m_{2sk}$};
										\end{tikzpicture}
												}_{nR_{sk}}
										\underbrace{
											\begin{tikzpicture}
												\node[minimum height=1.6em, anchor=base, fill=teal!25] {\ $m_{2ss}$}; 
											\end{tikzpicture}
										}_{nR_{2ss}}
											\underbrace{
											\begin{tikzpicture}
												\node[minimum height=1.6em, anchor=base, fill=teal!25] {\ $m_{2sm}$\ \  }; 
											\end{tikzpicture}
											}_{nR_{2sm}}
									$
			\end{tabular}
			\caption{Martion's coding: Rate splitting.}
			\label{fig: Oma Marton RS}	
			\end{figure}
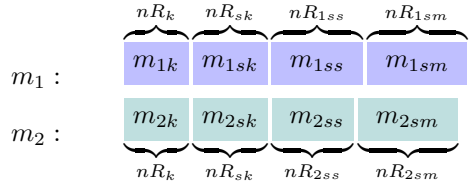

{\em Rate splitting:} As illustrated in Fig. \ref{fig: Oma Marton RS}, we represent $M_1, M_2$ by $M_1=(M_{1k}, M_{1sk}, M_{1ss}, M_{1sm})$ and $M_2=(M_{2k}, M_{2sk}, M_{2ss}, M_{2sm})$ with $M_{1k}, M_{2k}$ of entropy $nR_k;$ $M_{1sk}, M_{2sk}$ of entropy $nR_{sk};$ while $M_{1ss}, M_{1sm}$ of entropy $nR_{1ss}, nR_{1sm};$ and $M_{2ss}, M_{2sm}$ of entropy $nR_{2ss}, nR_{2sm},$ respectively. For simplicity, we denote $M_k=M_{1k}\oplus M_{2k},$ $M_{sk}=M_{1sk}\oplus M_{2sk},$ $M_{ss}=(M_{1ss}, M_{2ss})$ and $M_{sm}=(M_{1sm}, M_{2sm}).$
{\em Codebook generation:} Fix $p(u), p(v_0|u), p(v_1,v_2|v_0)$ and $p(x|v_1, v_2).$ 

First, randomly generate $2^{nR_k}$ i.i.d. sequences $u^{n}(k),$ $k\in[1:2^{nR_k}],$ according to $p(u).$

For each $u^n(k),$ randomly generate $2^{n(R_{1ss}+R_{2ss}+R_{sk}+R_{r})}$ i.i.d. sequences $v_0^n(k, m_{1ss}, m_{2ss}, m_{sk}, r)$ with $(m_{1ss}, m_{2ss}, m_{sk}, r)\in [1: 2^{nR_{1ss}}]\times[1: 2^{nR_{2ss}}]\times [1: 2^{nR_{sk}}]\times [1: 2^{nR_{r}}]$, according to $p(v_0|u)$;

For each fixed $v_0^n(k, m_{1ss}, m_{2ss}, m_{sk}, r),$ randomly generate $2^{n(R_{1sm}+R_{1r}+R_{1c})}$ i.i.d. sequences $v_1^n(k, m_{1ss}, \allowbreak m_{2ss}, m_{sk}, r, sm_1, r_1, c_1)$ with $(sm_1, r_1, c_1)\in [1: 2^{nR_{1sm}}]\times [1: 2^{nR_{1r}}]\times [1: 2^{nR_{1c}}]$, according to $p(v_1|v_0)$; and similarly generate $2^{n(R_{2sm}+R_{2r}+R_{2c})}$ i.i.d. sequences $v_2^n(k, m_{1ss}, m_{2ss}, m_{sk}, r, sm_2, r_2, c_2)$ with $(sm_2, r_2, c_2)\in [1: 2^{nR_{2sm}}]\times [1: 2^{nR_{2r}}]\times [1: 2^{nR_{2c}}],$ according to $p(v_2|v_0)$. 

{\em Encoding:} To send messages $(m_1, m_2),$ choose $u^n(k),$ where $k=m_k\triangleq m_{1k}\oplus m_{2k}.$ 

Given $u^n(k),$ randomly choose $r\in [1: 2^{nR_r}]$ and find $v_0^n(u^n, m_{1ss}, m_{2ss}, m_{sk}, r),$ where $m_{sk}\triangleq m_{1sk}\oplus m_{2sk}.$ 

Given $v_0^n(k, m_{1ss}, m_{2ss}, m_{sk}, r),$ randomly choose $(r_1, r_2)\in [1: 2^{nR_{1r}}] \times [1: 2^{nR_{2r}}],$ and pick $(c_1,c_2)$ such that $v_1^n(k, m_{1ss}, m_{2ss}, sk, r, sm_1, r_1, c_1)$ and $v_2^n(k, m_{1ss}, m_{2ss}, sk, r, sm_2, r_2, c_2)$ are jointly typical. (If there is more than one such jointly typical pair, choose one of them uniformly at random. 
This is possible with high probability, if 
\begin{equation}\label{eqn: Marton cond on cover index}
		R_{1c}+R_{2c}>I(V_1;V_2|V_0)
\end{equation}
(refer to \cite{src:ElGamal1981} for the proof). 

Finally, for the chosen jointly typical pair $(v_1^n,v_2^n)$, generate a codeword $x^n$ at random according to $p(x|v_1, v_2)$ and transmit it.

\begin{figure}[H]
\centering
	\begin{tabular}{rcl}
			$u^n(k):$ &	& 
						$
						\overbrace{
						\begin{tikzpicture}
							\node[minimum height=1.6em, minimum width=6em, anchor=base, fill=red!25] {$m_{1k}\oplus m_{2k}$}; 
						\end{tikzpicture}
							}^{nR_k}
						$\\
			$v_0^n(k, m_{1ss}, m_{2ss}, m_{sk}, r):$ &	& 
											$
											\underbrace{
											\begin{tikzpicture}
												\node[minimum height=1.6em, anchor=base, fill=blue!25] {$m_{1ss}$}; 
											\end{tikzpicture}
											}_{nR_{1ss}}
											\overbrace{
											\begin{tikzpicture}
												\node[minimum height=1.6em, anchor=base, fill=red!25] {$m_{1sk}\oplus m_{2sk}$}; 
											\end{tikzpicture}
											}^{nR_{sk}}
			 								\overbrace{
											\begin{tikzpicture}
												\node[minimum height=1.6em, anchor=base, fill=red!25] {$\ r\ $}; 
											\end{tikzpicture}
											}^{nR_r}
											\underbrace{
											\begin{tikzpicture}
												\node[minimum height=1.6em, anchor=base, fill=teal!25] {$m_{2ss}$}; 
											\end{tikzpicture}
										}_{nR_{2ss}}$ \\
			$v_1^n(k, m_{1ss}, m_{2ss}, sk, r, m_{1sm}, r_1, c_1):$ &	& 
														$
														\underbrace{
														\begin{tikzpicture}
															\node[minimum height=1.6em, minimum width=3em, anchor=base, fill=blue!25] {$m_{1sm}$}; 
														\end{tikzpicture}
														}_{nR_{1sm}}
														\overbrace{
														\begin{tikzpicture}
															\node[minimum height=1.6em, minimum width=2em, anchor=base, fill=red!25] {$r_{1}$}; 
														\end{tikzpicture}
														}^{nR_{1r}}
						 								\overbrace{
														\begin{tikzpicture}
															\node[minimum height=1.6em, minimum width=2em, anchor=base, fill=orange!25] {$c_{1}$}; 
														\end{tikzpicture}
														}^{nR_{1c}} 
													$\\
			$v_2^n(k, m_{1ss}, m_{2ss}, sk, r, m_{2sm}, r_2, c_2):$ &	& 
																	$
																	\underbrace{
																	\begin{tikzpicture}
																		\node[minimum height=1.6em, minimum width=3em, anchor=base, fill=teal!25] {$m_{2sm}$}; 
																	\end{tikzpicture}
																	}_{nR_{2sm}}
																	\overbrace{
																	\begin{tikzpicture}
																		\node[minimum height=1.6em, minimum width=2em, anchor=base, fill=red!25] {$r_{2}$}; 
																	\end{tikzpicture}
																	}^{nR_{2r}}
									 								\overbrace{
																	\begin{tikzpicture}
																		\node[minimum height=1.6em, minimum width=2em, anchor=base, fill=orange!25] {$c_{2}$}; 
																	\end{tikzpicture}
																	}^{nR_{2c}} 
																$
			\end{tabular}
		\caption{Marton's coding: Encoding.}
		\label{fig: Oma Marton encoding}
		\end{figure}
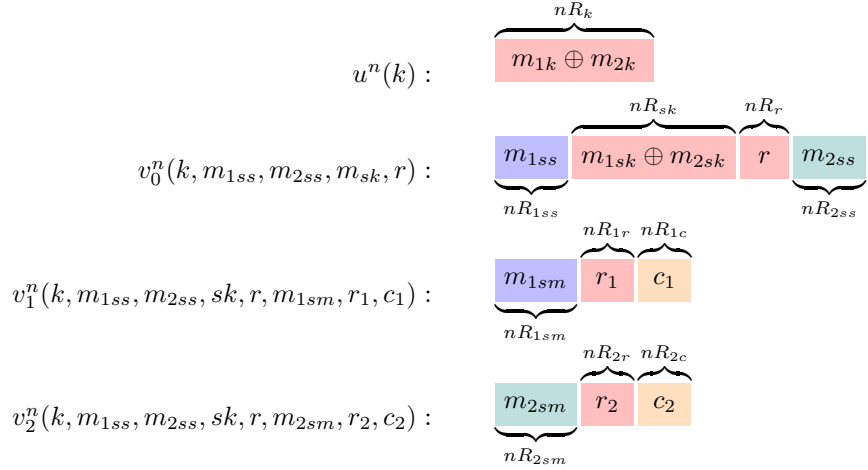

The choice of $u^n, v_0^n, v_1^n, v_2^n$ for given $(m_1, m_2)$ is illustrated in Fig. \ref{fig: Oma Marton encoding}. 

{\em Decoding:} Receiver 1, upon receiving $y_1^n,$ finds a unique $v_1^n(\hat{k}, \hat{m}_{1ss}, m_{2ss}, \hat{m}_{sk}, \hat{r}, \hat{m}_{1sm}, \hat{r}_1, \hat{c}_1)$ such that $(v_1^n, y_1^n)$ is jointly typical. 
And, receiver 2, upon receiving $y_2^n,$ finds a unique $v_2^n(\tilde{k}, m_{1ss}, \tilde{m}_{2ss}, \tilde{m}_{sk}, \tilde{r}, m_{1sm}, \allowbreak \tilde{r}_2, \tilde{c}_2)$ such that $(v_2^n, y_2^n)$ is jointly typical. 

{\em Analysis of decoding error:}  Assume that $m_1=(m_{1k}, m_{1sk}, m_{1ss}, m_{1sm}),$ $m_2=(m_{2k}, m_{2sk}, m_{2ss}, m_{2sm})$ is sent, i.e., more specifically, $k,  m_{sk},  m_{1ss}, m_{1sm}$ and $m_{2ss}, m_{2sm}$ are sent, where $k\triangleq m_{1k}\oplus m_{2k}$ and $m_{sk}=m_{1sk}\oplus m_{2sk}.$  For $P_{e,1}$, a decoding error happens if receiver 1's estimate is $u^n(\hat{k}),$ $v_0^n(u^n, \hat{m}_{1ss}, m_{2ss}, \hat{m}_{sk}, \hat{r}),$ $v_1^n(v_0^n, \hat{m}_{1sm}, \hat{r}_1, \hat{c}_1)$ with $(\hat{k}, \hat{m}_{1ss}, \hat{m}_{sk}, \hat{r}, \hat{m}_{1sm}, \hat{r}_1, \hat{c}_1)\neq (k, m_{1ss}, m_{sk}, r, m_{1sm}, r_1, c_1).$ In more details, the error event can be partitioned into the followings:
	\begin{enumerate}
		\item Error event corresponds to $\hat{k} \neq k.$ Note that this event occurs with arbitrarily small probability if 
			\begin{equation}\label{eqn: cond Marton pe11}
					R_1+R_r+R_{1r}+R_{1c}\leq I(U, V_0, V_1;Y_1)=I(V_0, V_1;Y_1).
			\end{equation}
		\item Error event corresponds to $\hat{k} = k,$ but $(\hat{m}_{1ss}, \hat{m}_{sk}, \hat{r}) \neq (m_{1ss}, m_{sk}, r)$ Note that this event occurs with arbitrarily small probability if 
			\begin{equation}\label{eqn: cond Marton pe12}
					R_1-R_k+R_r+R_{1r}+R_{1c}\leq I(V_0, V_1;Y_1|U).
			\end{equation}
		\item Error event corresponds to $(\hat{k}, \hat{m}_{1ss}, \hat{m}_{sk}, \hat{r}) = (k, m_{1ss}, m_{sk}, r)$ but $(\hat{m}_{1sm}, \hat{r}_1, \hat{c}_1)\neq (m_{1sm}, r_1, c_1).$ Note that this event occurs with arbitrarily small probability if 
			\begin{equation}\label{eqn: cond Marton pe13}
					R_{1sm}+R_{1r}+R_{1c}\leq I(V_1;Y_1|U, V_0)=I(V_1;Y_1|V_0).
			\end{equation}		
	\end{enumerate}

Similar analysis can be done at the receiver 2, from which the decoding error probability  $P_{e,2}$ can be made arbitrarily small if
	\begin{align}
			R_2+R_r+R_{2r}+R_{2c}& \leq I(V_0, V_2;Y_2) \label{eqn: cond Marton pe21}\\
			R_2-R_k+R_r+R_{2r}+R_{2c}&\leq I(V_0, V_2;Y_2|U) \label{eqn: cond Marton pe22}\\
			R_{2sm}+R_{2r}+R_{2c}&\leq I(V_2;Y_2|U, V_0)=I(V_2;Y_2|V_0).\label{eqn: cond Marton pe23}
	\end{align} 

{\em Analysis of individual secrecy:} 
For the secrecy of $M_1$, we have
\allowdisplaybreaks  
	\begin{align*}
		 I(M_1; Z^n)	
			=& I(M_{1k}, M_{1sk}, M_{1ss}, M_{1sm}; Z^n)\\
			=& I(M_{1k}, M_{1sk}; Z^n)+I(M_{1ss}, M_{1sm};Z^n|M_{1k}, M_{1sk})\\
			\stackrel{(a)}{=}& I(M_{1ss}, M_{1sm};Z^n|M_{1k}, M_{1sk})\\
			=& H(M_{1ss}, M_{1sm})-H(M_{1ss}, M_{1sm}|M_{1k}, M_{1sk}, Z^n)\\
			\leq & H(M_{1ss}, M_{1sm})-H(M_{1ss}, M_{1sm}|M_k, M_{1k}, M_{1sk}, Z^n)\\
			\stackrel{(b)}{=} & nR_{1ss}+ nR_{1sm}-H(M_{1ss}, M_{1sm}|M_{k}, Z^n)
	\end{align*}
where 
	
	$(a)$ is due to the fact that $I(M_{1k}, M_{1sk}; Z^n) = 0$ since 
	\begin{align*}
		I(M_{1k}, M_{1sk}; Z^n) 
			&\leq I(M_{1k}, M_{1sk}; Z^n, M_{k}, M_{sk}, M_{1ss}, M_{1sm}, M_{2ss}, M_{2sm})\\
			&=I(M_{1k}, M_{1sk}; M_{k}, M_{sk}, M_{1ss}, M_{1sm}, M_{2ss}, M_{2sm})\\
			&=0,
	\end{align*}
	 where the first equality is by the Markov chain $(M_{1k}, M_{2k}, M_{1sk}, M_{2sk})\to (M_k,  M_{sk}, M_{ss}, M_{sm})\to Z^n$; 
	
	$(b)$ is due to the fact that $I(M_{1ss}, M_{1sm};  M_{1k}, M_{1sk}|Z^n, M_{k})=0,$ where the equality follows by: 
	\begin{enumerate}
		\item $I(M_{1ss}, M_{1sm};  M_{1k}, M_{1sk}|Z^n, M_{k})\geq 0;$ and
		\item $I(M_{1ss}, M_{1sm};  M_{1k}, M_{1sk}|Z^n, M_{k})\leq 0$ since 
		\begin{align*}
			H( M_{1k}, M_{1sk}|Z^n, M_{k}, M_{1ss}, M_{1sm})
				& \geq H( M_{1k}, M_{1sk}|Z^n, M_{k}, M_{sk}, M_{1ss}, M_{1sm})\\
				& =H(M_{1k}, M_{1sk}|M_k, M_{sk})\\
				& =H( M_{1k}, M_{1sk})\\
				& \geq H( M_{1k}, M_{1sk}|Z^n, M_{k}).
		\end{align*}
	\end{enumerate}
	
So far, we obtain 
	\begin{equation}\label{eqn: marton secrecy step 1}
		I(M_1; Z^n)	\leq nR_{1ss}+ nR_{1sm}-H(M_{1ss}, M_{1sm}|M_{k}, Z^n).
	\end{equation} 
	
Similarly, for the secrecy of  $M_2,$ we have
	\begin{equation}\label{eqn: marton secrecy step 1 for M2}
		I(M_2; Z^n)	\leq nR_{2ss}+ nR_{2sm}-H(M_{2ss}, M_{2sm}|M_{k}, Z^n).
	\end{equation} 	
	
In the following, we show that $H(M_{1ss}, M_{1sm}, M_{2ss}, M_{2sm}|M_k, Z^n)\geq n(R_{1ss}+R_{1sm}+R_{2ss}+R_{1sm})-n\delta'(\epsilon)$ holds if the rates satisfy \eqref{eqn: IndS condiA}, \eqref{eqn: IndS condiB}, \eqref{eqn: IndS condiC} and \eqref{eqn: IndS condiD}. This implies that $H(M_{1ss}, M_{1sm}|M_{k}, Z^n)\geq R_{1ss}+R_{1sm}-n\delta'(\epsilon)$ and $H(M_{2ss}, M_{2sm}|M_{k}, Z^n)\geq R_{2ss}+R_{2sm}-n\delta'(\epsilon).$ Further by (\ref{eqn: marton secrecy step 1}) and \eqref{eqn: marton secrecy step 1 for M2}, we obtain $I(M_1;Z^n)\leq n\delta'(\epsilon)$ and $I(M_2;Z^n)\leq n\delta'(\epsilon),$ thus completing the desired individual secrecy proof.

Note that 
\begin{align}
	H(M_{1ss}, M_{1sm}, M_{2ss}, M_{2sm}|M_k, Z^n)
		\stackrel{(a)}{=}& H(M_{ss}, M_{sm}|U^n, Z^n)\nonumber\\
		=& H(M_{ss}, M_{sm}, V_0^n|U^n, Z^n)-H(V_0^n|U^n, M_{ss}, M_{sm}, Z^n)\nonumber\\
		\geq & H(M_{ss}, M_{sm}, V_0^n|U^n,Z^n)-H(V_0^n|U^n, M_{ss}, Z^n)\nonumber\\
		\geq & H(V_0^n|U^n, Z^n)+H(M_{sm}|U^n, V_0^n, Z^n)-H(V_0^n|U^n, M_{ss}, Z^n). \label{eqn: terms for inds}
\end{align}

We now bound the terms above.

For the first term in \eqref{eqn: terms for inds}, i.e., $H(V_0^n|U^n, Z^n),$ we have
\begin{align}
	H(V_0^n|U^n, Z^n)
		& =H(V_0^n, Z^n |U^n)-H(Z^n|U^n)\nonumber\\
		& =H(V_0^n|U^n)+H(Z^n|U^n, V_0^n)-H(Z^n|U^n)\nonumber\\
		& =n(R_{1ss}+R_{2ss}+R_{sk}+R_r)+H(Z^n|U^n, V_0^n)-H(Z^n|U^n); \label{eqn: term 1 for inds}
\end{align}

And, for the second term in \eqref{eqn: terms for inds}, i.e., $H(M_{1sm}, M_{2sm}|U^n, V_0^n, Z^n),$ we have
\begin{align}\label{eqn: term 2 for inds}
 	 H&(M_{1sm},M_{2sm}|U^n, V_0^n, Z^n)\nonumber\\
	 =& H(M_{1sm},M_{2sm},Z^n|U^n, V_0^n)-H(Z^n|U^n, V_0^n)\nonumber\\
	 =& H(M_{1sm},M_{2sm},Z^n,V_1^n,V_2^n|U^n, V_0^n)-H(Z^n|U^n, V_0^n)-H(V_1^n,V_2^n|U^n, V_0^n, M_{1sm},M_{2sm}, Z^n)\nonumber\\
	 =& H(Z^n,V_1^n,V_2^n|U^n, V_0^n)-H(Z^n|U^n, V_0^n)-H(V_1^n,V_2^n|U^n, V_0^n, M_{1sm},M_{2sm}, Z^n)\nonumber\\
	 =& H(Z^n|U^n,V_0^n,V_1^n,V_2^n)+H(V_1^n,V_2^n|U^n,V_0^n)-H(Z^n|U^n, V_0^n)-H(V_1^n,V_2^n|U^n, V_0^n, M_{1sm},M_{2sm}, Z^n)\nonumber\\
	\stackrel{(a)}{\geq} & H(Z^n|U^n,V_0^n,V_1^n,V_2^n)+H(V_1^n,V_2^n|U^n,V_0^n)-H(Z^n|U^n, V_0^n)\nonumber\\
	&\:{}-H(V_1^n|U^n, V_0^n, M_{1sm},M_{2sm}, Z^n)-H(V_2^n|U^n, V_0^n, M_{1sm},M_{2sm}, Z^n)\nonumber\\
	\stackrel{(b)}{\geq} & H(Z^n|U^n,V_0^n,V_1^n,V_2^n)+n(R_{1sm}+R_{1r}+R_{2sm}+R_{2r})-H(Z^n|U^n, V_0^n)\nonumber\\
	&\:{}-H(V_1^n|U^n, V_0^n, M_{1sm}, Z^n)-H(V_2^n|U^n, V_0^n, M_{2sm}, Z^n)\nonumber\\
	\stackrel{(c)}{\geq} & H(Z^n|U^n,V_0^n,V_1^n,V_2^n)+n(R_{1sm}+R_{1r}+R_{2sm}+R_{2r})-H(Z^n|U^n, V_0^n)\nonumber\\
	&\:{}-n(R_{1r}+R_{1c}-I(V_1;Z|V_0))-n\epsilon_1-n(R_{2r}+R_{2c}-I(V_2;Z|V_0))-n\epsilon_1
\end{align} 
where 
$(a)$ is due to the fact that $H(A,B|C)\leq H(A|C)+H(B|C)$,
$(b)$ follows as
\begin{align*}
 	H(V_1^n,V_2^n|U^n,V_0^n)& =H(M_{1sm},M_{2sm},M_{1r},M_{2r}|U^n,V_0^n)+H(V_1^n,V_2^n|U^n,V_0^n,M_{1sm},M_{2sm},M_{1r},M_{2r})\\
 	& \geq H(M_{1sm},M_{2sm},M_{1r},M_{2r}|U^n,V_0^n)=H(M_{1sm},M_{2sm},M_{1r},M_{2r})\\
 	& =n(R_{1sm}+R_{1r}+R_{2sm}+R_{2r})
\end{align*}
since the choice of $M_{1sm},M_{2sm},M_{1r},M_{2r}$ are independent of $U^n,V_0^n;$ 
$(c)$ is due to the followings:
First, we have
\begin{align}
H(V_1^n|U^n, V_0^n, M_{1sm}, Z^n)=H(V_1^n|V_0^n, M_{1sm}, Z^n) \leq n(R_{1r}+R_{1c}-I(V_1;Z|V_0))+n\epsilon_1
\end{align}
if, for an arbitrarily small $\epsilon>0$,
\begin{equation}\label{eqn: IndS condiA}
		R_{1r}+R_{1c}\geq I(V_1;Z|V_0)+\epsilon.
\end{equation}
This follows from \cite[Lemma 1]{src:Chia2012_BC}. And, similarly, we have
\begin{align}
H(V_2^n|U^n, V_0^n, M_{2sm}, Z^n) = H(V_2^n|V_0^n, M_{2sm}, Z^n) \leq n(R_{2r}+R_{2c}-I(V_2;Z|V_0))+n\epsilon_1
\end{align}
if, for an arbitrarily small $\epsilon>0$,
\begin{equation}\label{eqn: IndS condiB}
		R_{2r}+R_{2c}\geq I(V_2;Z|V_0)+\epsilon.
\end{equation}

Finally, for the last term in \eqref{eqn: terms for inds}, i.e., $H(V_0^n|U^n, M_{ss}, Z^n),$ we have
\begin{align}
H(V_0^n|U^n, M_{ss}, Z^n)\leq n(R_{sk}+R_{r}-I(V_0;Z|U))+n\epsilon_1, \label{eqn: term 3 for inds}
\end{align}
if, for an arbitrarily small $\epsilon>0$,
\begin{equation}\label{eqn: IndS condiC}
		R_{sk}+R_r\geq I(V_0;Z|U)+\epsilon.
\end{equation}
This follows from \cite[Lemma 1]{src:Chia2012_BC}.

Combining \eqref{eqn: term 1 for inds}, \eqref{eqn: term 2 for inds} and \eqref{eqn: term 3 for inds} in \eqref{eqn: terms for inds}, we have
\begin{align*}
	 H&(M_{1ss}, M_{1sm}, M_{2ss}, M_{2sm}|M_k, Z^n)\nonumber\\
	\geq	& H(V_0^n|U^n, Z^n)+H(M_{sm}|U^n, V_0^n, Z^n)-H(V_0^n|U^n, M_{ss}, Z^n)\\
	\geq & n(R_{1ss}+R_{2ss}+R_{sk}+R_r)+H(Z^n|U^n, V_0^n)-H(Z^n|U^n)\nonumber\\
	&\:{}+H(Z^n|U^n,V_0^n,V_1^n,V_2^n)+n(R_{1sm}+R_{1r}+R_{2sm}+R_{2r})-H(Z^n|U^n, V_0^n)\nonumber\\
	&\:{}-n(R_{1r}+R_{1c}-I(V_1;Z|V_0))-n\epsilon_1-n(R_{2r}+R_{2c}-I(V_2;Z|V_0))-n\epsilon_1\nonumber\\
	&\:{}-n(R_{sk}+R_{r}-I(V_0;Z|U))-n\epsilon_1\\
	\stackrel{(a)}{\geq} & n(R_{1ss}+R_{1sm}+R_{2ss}+R_{2sm}) \nonumber\\
	&\:{} +n(-R_{1c}-R_{2c}-H(Z|U)+H(Z|U,V_0,V_1,V_2)+I(V_1;Z|V_0)+I(V_2;Z|V_0)+I(V_0;Z|U)-3\epsilon_1)\\
	\stackrel{(b)}{\geq} & n(R_{1ss}+R_{1sm}+R_{2ss}+R_{2sm}) - n\delta(\epsilon)	
\end{align*}
where 
	$(a)$ follows from $H(Z^n|U^n)\leq$  $\sum_{i=1}^{n} H(Z_i|U_i)$ $=nH(Z|U)$ and the fact that $H(Z^n|U^n, V_0^n, V_1^n, V_2^n)=\allowbreak nH(Z|U, V_0, V_1, V_2)$; and 
	$(b)$ follows by the rate choice 
	\begin{equation}\label{eqn: IndS condiD}
		R_{1c}+R_{2c} \leq I(V_1;Z|V_0)+I(V_2;Z|V_0)-I(V_1,V_2;Z|V_0)
	\end{equation}
with $\delta(\epsilon)=3\epsilon_1$. 

{\em Achievable rate region:} Combining the non-negativity for rates, the conditions for reliable communication at both legitimate receivers, i.e., \eqref{eqn: Marton cond on cover index}, \eqref{eqn: cond Marton pe11}-\eqref{eqn: cond Marton pe23}, and individual secrecy at the eavesdropper, i.e., \eqref{eqn: IndS condiA}, \eqref{eqn: IndS condiB}, \eqref{eqn: IndS condiC} and \eqref{eqn: IndS condiD}, 
we obtain the followings: 
\begin{equation}\label{eqn: FM 0}
		R_{k}, R_{sk}, R_{1ss},R_{1sm},R_{2ss},R_{2sm}, R_r, R_{1c}, R_{2c}, R_{1r}, R_{2r} \geq 0 
\end{equation}
	\begin{eqnarray}
		R_1&=& R_{k}+R_{sk}+R_{1ss}+R_{1sm}				\label{eqn: FM 1}\\
		R_2&=& R_{k}+R_{sk}+R_{2ss}+R_{2sm} 			\label{eqn: FM 2}\\
		R_{1c}+R_{2c} &> & I(V_1;V_2|V_0) 			\label{eqn: FM 3}\\
		R_{1}+R_r+R_{1r}+R_{1c}&\leq& I(V_0, V_1;Y_1) \label{eqn: FM 4}\\
		R_1-R_k+R_r+R_{1r}+R_{1c}&\leq& I(V_0,V_1;Y_1|U) 	\label{eqn: FM 5} \\
		R_{1sm}+R_{1r}+R_{1c}&\leq& I(V_1;Y_1|V_0)	 	\label{eqn: FM 6}\\		
		R_{2}+R_r+R_{2r}+R_{2c}&\leq& I(V_0, V_2;Y_2)		\label{eqn: FM 7}\\
		R_2-R_k+R_r+R_{2r}+R_{2c}&\leq& I(V_0,V_2;Y_2|U) 	\label{eqn: FM 8}\\
		R_{2sm}+R_{2r}+R_{2c}&\leq& I(V_2;Y_2|V_0)		\label{eqn: FM 9}\\		
		R_{sk}+R_r & \geq & I(V_0;Z|U)					\label{eqn: FM 10}\\
		R_{1r}+R_{1c} &\geq & I(V_1;Z|V_0)				\label{eqn: FM 11}\\
		R_{2r}+R_{2c} &\geq & I(V_2;Z|V_0)				\label{eqn: FM 12}\\
		R_{1c}+R_{2c} &\leq& I(V_1;Z|V_0)+I(V_2;Z|V_0)-I(V_1,V_2;Z|V_0), \label{eqn: FM 13}
	\end{eqnarray}  
where the union is taken over probability distributions satisfying $$p(q, u,v_0,v_1,v_2,x)=p(q)p(u|q)p(v_0|u)p(v_1,v_2|v_0)p(x|v_1,v_2).$$

Eliminating $R_{1c}, R_{2c}, R_{1r}, R_{2r}, R_r, R_{1sm}, R_{2sm}, R_{1ss},R_{2ss}, R_{k}, R_{sk},$ by applying Fourier-Motzkin procedure \cite{ElGamal:2012}, we obtain the region of $(R_1, R_2)$ as given in \eqref{eqn: Oma Marton IndS region} in Theorem \ref{thm: Oma Marton}. Note that a sketch of this Fourier-Motzkin procedure is provided in Appendix \ref{App: FM}.



\section{Fourier-Motzkin Elimination for Theorem \ref{thm: Oma Marton}}\label{App: FM}
Here we briefly outline the Fourier-Motzkin procedure in the proof of Theorem \ref{thm: Oma Marton}. 
\begin{itemize}
	\item To eliminate $R_{1ss},$ we consider the non-negativity of the rate and the equality \eqref{eqn: FM 1}. We end up with
		\begin{align}
			R_{1sm}+R_{k}+R_{sk} 			& \leq R_1				\label{eqn: FM 14}
		\end{align}

	\item To eliminate $R_{2ss},$ we consider the non-negativity of the rate and the equality \eqref{eqn: FM 2}. We end up with
		\begin{align}
			R_{2sm}+R_{k}+R_{sk} 			&\leq  R_2				\label{eqn: FM 15}						
		\end{align}	
		
	\item To eliminate $R_{1sm},$ we consider the non-negativity of the rate and the inequalities 
	\eqref{eqn: FM 6} and \eqref{eqn: FM 14} which involve $R_{1sm}.$ We end up with
		\begin{align}
			R_{1r}+R_{1c} 			& \leq 	I(V_1;Y_1|V_0) 		\label{eqn: FM 24}\\
			R_{k}+R_{sk}			& \leq 	R_1						  \label{eqn: FM 25}
		\end{align}

	\item To eliminate $R_{2sm},$ we consider the non-negativity of the rate and the inequalities 
	\eqref{eqn: FM 9} and \eqref{eqn: FM 15} which involve $R_{2sm}.$ We end up with
		\begin{align}
			R_{2r}+R_{2c} 			& \leq 	I(V_2;Y_2|V_0) 		\label{eqn: FM 26}\\
			R_{k}+R_{sk}			& \leq 	R_2						  \label{eqn: FM 27}
		\end{align}

	\item To eliminate $R_{r},$ we consider the non-negativity of the rate and the inequalities \eqref{eqn: FM 4}, \eqref{eqn: FM 5}, \eqref{eqn: FM 7}, \eqref{eqn: FM 8}, \eqref{eqn: FM 10} which involve $R_{r}.$ We end up with 
		\begin{align}
			R_{1}+R_{1r}+R_{1c} 					& \leq I(V_0, V_1;Y_1)			\label{eqn: FM 16}\\				
			R_{1}-R_{k}+R_{1r}+R_{1c} 			& \leq I(V_0, V_1;Y_1|U)		\label{eqn: FM 17}\\		
			R_{2}+R_{2r}+R_{2c} 					& \leq I(V_0, V_2;Y_2)			\label{eqn: FM 18}\\				
			R_{2}-R_{k}+R_{2r}+R_{2c} 			& \leq I(V_0, V_2;Y_2|U)		\label{eqn: FM 19}\\		
			R_{1}-R_{sk}+R_{1r}+R_{1c} 					& \leq I(V_0, V_1;Y_1)-I(V_0;Z|U)			\label{eqn: FM 20}\\				
			R_{1}-R_{k}-R_{sk}+R_{1r}+R_{1c} 		& \leq I(V_0, V_1;Y_1|U)-I(V_0;Z|U)		\label{eqn: FM 21}\\		
			R_{2}-R_{sk}+R_{2r}+R_{2c} 					& \leq I(V_0, V_2;Y_2)-I(V_0;Z|U)			\label{eqn: FM 22}\\				
			R_{2}-R_{k}-R_{sk}+R_{2r}+R_{2c} 		& \leq I(V_0, V_2;Y_2|U)-I(V_0;Z|U)		\label{eqn: FM 23}
		\end{align}

	\item To eliminate $R_{k},$ we consider the non-negativity of the rate and the inequalities \eqref{eqn: FM 17}, \eqref{eqn: FM 19}, \eqref{eqn: FM 21}, \eqref{eqn: FM 23}, \eqref{eqn: FM 25}, \eqref{eqn: FM 27} which involve $R_{k}.$ We end up with
	\allowdisplaybreaks 
		\begin{align}
			R_{sk} 			&\leq 	R_1 			\label{eqn: FM 28}\\
			R_{sk} 			&\leq 	R_2 			\label{eqn: FM 29}\\			
			R_{sk}+R_{1r}+R_{1c} 					&\leq I(V_0, V_1; Y_1|U)		\label{eqn: FM 30}\\
			R_1-R_2+R_{sk}+R_{1r}+R_{1c} 	&\leq I(V_0, V_1; Y_1|U)		\label{eqn: FM 31}\\	
			R_2-R_1+R_{sk}+R_{2r}+R_{2c} 	&\leq I(V_0, V_2; Y_2|U)		\label{eqn: FM 32}\\
			R_{sk}+R_{2r}+R_{2c} 					&\leq I(V_0, V_2; Y_2|U)		\label{eqn: FM 33}\\
			R_{1r}+R_{1c} 					&\leq I(V_0, V_1; Y_1|U)-I(V_0;Z|U)		\label{eqn: FM 34}\\
			R_1-R_2+R_{1r}+R_{1c} 	&\leq I(V_0, V_1; Y_1|U)-I(V_0;Z|U)		\label{eqn: FM 35}\\	
			R_2-R_1+R_{2r}+R_{2c} 	&\leq I(V_0, V_2; Y_2|U)-I(V_0;Z|U)		\label{eqn: FM 36}\\
			R_{2r}+R_{2c} 					&\leq I(V_0, V_2; Y_2|U)-I(V_0;Z|U)		\label{eqn: FM 37}									
		\end{align}
	
	\item To eliminate $R_{sk},$ we consider the non-negativity of the rate and the inequalities \eqref{eqn: FM 20}, \eqref{eqn: FM 22},  \eqref{eqn: FM 28}, \eqref{eqn: FM 29}, \eqref{eqn: FM 30}, \eqref{eqn: FM 31}, \eqref{eqn: FM 32}, \eqref{eqn: FM 33} which involve $R_{sk}.$ We end up with  the following inequalities after cancelling the redundant ones. 
		\begin{align}
			R_{1} & \geq 0 \label{eqn: FM 38}\\
			R_{2} & \geq 0 \label{eqn: FM 39}
		\end{align}
	
	\item To eliminate $R_{1r},$ we consider the non-negativity of the rate and the inequalities \eqref{eqn: FM 11}, \eqref{eqn: FM 16},  \eqref{eqn: FM 24},  \eqref{eqn: FM 34},  \eqref{eqn: FM 35} which involve $R_{1r}$. We end up with the following inequalities after cancelling the redundant ones.
			\begin{align}
				R_{1}+R_{1c}	 & \leq I(V_0, V_1;Y_1) \label{eqn: FM 40}\\
				R_{1c}	 			 & \leq I(V_1;Y_1|V_0) +[I(V_0;Y_1|U)-I(V_0;Z|U)]^{-}\label{eqn: FM 41}\\
				R_{1}-R_{2}+R_{1c}	 			& \leq I(V_0, V_1; Y_1|U)-I(V_0;Z|U)	 \label{eqn: FM 43}\\
				R_{1}  				  & \leq I(V_0, V_1;Y_1)-I(V_1;Z|V_0) 	\label{eqn: FM 44}\\
				I(V_1;Z|V_0)	  & \leq I(V_1;Y_1|V_0) 						\label{eqn: FM C1}\\
				I(V_0,V_1;Z|U)	 & \leq I(V_0, V_1;Y_1|U) 					\label{eqn: FM C2}\\
				R_{1}-R_{2}		  & \leq I(V_0, V_1; Y_1|U)-I(V_0, V_1;Z|U)	 \label{eqn: FM 45}
			\end{align}
		where $[a]^{-}=\min\{0, a\}.$
	\item To eliminate $R_{2r},$ we consider the non-negativity of the rate and the inequalities \eqref{eqn: FM 12}, \eqref{eqn: FM 18},  \eqref{eqn: FM 26},  \eqref{eqn: FM 36},  \eqref{eqn: FM 37} which involve $R_{2r}$. We end up with the following inequalities after canceling the redundant ones.
			\begin{align}
				R_{2}+R_{2c}	 & \leq I(V_0, V_2;Y_2) \label{eqn: FM 46}\\
				R_{2c}	 			 & \leq I(V_2;Y_2|V_0)+|I(V_0;Y_2|U)-I(V_0;Z|U)|^{-} \label{eqn: FM 47}\\
				R_{2}-R_{1}+R_{2c}	 			& \leq I(V_0, V_2; Y_2|U)-I(V_0;Z|U)	 \label{eqn: FM 48}\\
				R_{2}  				  & \leq I(V_0, V_2;Y_2)-I(V_2;Z|V_0) 	\label{eqn: FM 50}\\
				I(V_2;Z|V_0)	  & \leq I(V_2;Y_2|V_0) 						\label{eqn: FM C3}\\
				I(V_0,V_2;Z|U)	 & \leq I(V_0, V_2;Y_2|U) 					\label{eqn: FM C4}\\
				R_{2}-R_{1}		  & \leq I(V_0, V_2; Y_2|U)-I(V_0, V_2;Z|U)	 \label{eqn: FM 51}
			\end{align}
		
	\item To eliminate $R_{1c},$ we consider the non-negativity of the rate and the inequalities \eqref{eqn: FM 3}, \eqref{eqn: FM 13}, \eqref{eqn: FM 40}, \eqref{eqn: FM 41} and \eqref{eqn: FM 43} which involve $R_{1c}$. We end up with the following inequalities after canceling the redundant ones.
		\begin{align}
			R_{2c}					& \leq 	I(V_1; Z|V_0)+I(V_2;Z|V_0)-I(V_1,V_2;Z|V_0) 		\label{eqn: FM 52}\\
			I(V_1;V_2|V_0) 		& \leq 	I(V_1; Z|V_0)+I(V_2;Z|V_0)-I(V_1,V_2;Z|V_0)	\label{eqn: FM C5}\\
			R_1-R_{2c}			& \leq 	I(V_0,V_1;Y_1)-I(V_1;V_2|V_0)								\label{eqn: FM 53}\\
			R_{2c}					& \geq I(V_1;V_2|V_0)-I(V_1;Y_1|V_0)-[I(V_0;Y_1|U)-I(V_0;Z|U)]^{-}			\label{eqn: FM 54}\\
			R_1-R_2-R_{2c}	& \leq I(V_0,V_1;Y_1|U)-I(V_0;Z|U)-I(V_1;V_2|V_0)	\label{eqn: FM 55}
		\end{align}	
	
	\item To eliminate $R_{2c},$ we consider the non-negativity of the rate and the inequalities \eqref{eqn: FM 46}, \eqref{eqn: FM 47}, \eqref{eqn: FM 48}, \eqref{eqn: FM 52},  \eqref{eqn: FM 53}, \eqref{eqn: FM 54},  \eqref{eqn: FM 55} which involve $R_{2c}$. We end up with the following inequalities after cancelling some redundant ones.
		\begin{align}
			R_{1}+R_{2}			& \leq I(V_0,V_1;Y_1)+I(V_0,V_2;Y_2)-I(V_1;V_2|V_0)			\label{eqn: FM 56}\\
			R_{1}					& \leq 	I(V_0, V_1;Y_1)-I(V_1;V_2|V_0)+I(V_2;Y_2|V_0)+[I(V_0;Y_2|U)-I(V_0;Z|U)]^{-} 		\label{eqn: FM 57}\\
			R_{2}					& \leq 	I(V_0,V_2;Y_2|U)-I(V_1;V_2|V_0)-I(V_0;Z|U)+I(V_0,V_1;Y_1) \label{eqn: FM 58}\\
			R_{1}					& \leq 	I(V_0, V_1;Y_1)-I(V_1;V_2|V_0)+I(V_1;Z|V_0)+I(V_2;Z|V_0)-I(V_1,V_2;Z|V_0)	 \label{eqn: FM 59}\\
			R_{2}					& \leq 	I(V_0,V_2;Y_2)-I(V_1;V_2|V_0)+I(V_1;Y_1|V_0)+[I(V_0;Y_1|U)-I(V_0;Z|U)]^{-} 		\label{eqn: FM 60}\\
			R_{2}-R_{1}			& \leq 	I(V_0,V_2;Y_2|U)-I(V_1;V_2|V_0)-I(V_0;Z|U)+I(V_1;Y_1|V_0)\nonumber\\
										&\quad +[I(V_0;Y_1|U)-I(V_0;Z|U)]^{-}  \label{eqn: FM 61}\\
			R_{1}					& \leq  I(V_0,V_1;Y_1|U)-I(V_1;V_2|V_0)-I(V_0;Z|U)+I(V_0,V_2;Y_2)				\label{eqn: FM 62}\\
			R_{1}-R_{2}			& \leq  I(V_0,V_1;Y_1|U)+I(V_2;Y_2|V_0)-I(V_1;V_2|V_0)-I(V_0;Z|U)\nonumber\\
										&\quad +[I(V_0;Y_2|U)-I(V_0;Z|U)]^{-}	 \label{eqn: FM 63}\\
			R_{1}-R_{2}			& \leq  I(V_0,V_1;Y_1|U)-I(V_1;V_2|V_0)-I(V_0;Z|U) \nonumber\\
										&\quad +I(V_1;Z|V_0)+I(V_2;Z|V_0)-I(V_1,V_2;Z|V_0)		\label{eqn: FM 64}
		\end{align}	
		Note that 
		\eqref{eqn: FM 56} is redundant due to \eqref{eqn: FM 44}, \eqref{eqn: FM 50} and \eqref{eqn: FM C5}; 				
		\eqref{eqn: FM 58} is redundant due to  \eqref{eqn: FM 51}, \eqref{eqn: FM 44} and \eqref{eqn: FM C5};  
		\eqref{eqn: FM 59} is redundant due to \eqref{eqn: FM 44} and \eqref{eqn: FM C5}; 
		\eqref{eqn: FM 61} is redundant due to \eqref{eqn: FM 51}, \eqref{eqn: FM C1} or \eqref{eqn: FM C2}, and \eqref{eqn: FM C5}; 		
		\eqref{eqn: FM 62} is redundant due to \eqref{eqn: FM 45}, \eqref{eqn: FM 50} and \eqref{eqn: FM C5}; 		
		\eqref{eqn: FM 63} is redundant due to \eqref{eqn: FM 45}, \eqref{eqn: FM C3} or \eqref{eqn: FM C4}, and \eqref{eqn: FM C5}; 		
		\eqref{eqn: FM 64} is redundant due to \eqref{eqn: FM 45} and \eqref{eqn: FM C5}.
		
\end{itemize}
		So far, we have for $R_1$  the inequalities \eqref{eqn: FM 38}, \eqref{eqn: FM 44}, \eqref{eqn: FM 45}, \eqref{eqn: FM 57} and for $R_2$ the inequalities \eqref{eqn: FM 39}, \eqref{eqn: FM 50}, \eqref{eqn: FM 51}, \eqref{eqn: FM 60}. An individual secrecy rate region is obtained as a set of the non-negative rate pairs $(R_1,R_2)$ such that
	\begin{align*}
		R_1  \leq & \min \{I(V_0, V_1; Y_1)-I_1,\ \  I(V_0, V_1; Y_1|U)-I(V_0,V_1;Z|U)+R_2\};\\
		R_2  \leq & \min \{I(V_0, V_2; Y_2)-I_2,\ \  I(V_0, V_2; Y_2|U)-I(V_0,V_2;Z|U)+R_1\},
	\end{align*} 
with 
	\begin{align*}
		I_1&=\max \{I(V_1;Z|V_0), \ I(V_1;V_2|V_0)-I(V_2;Y_2|V_0), \ I(V_1;V_2|V_0)+I(V_0;Z|U)-I(V_0, V_2;Y_2|U)\};\\
		I_2&=\max \{I(V_2;Z|V_0), \ I(V_1;V_2|V_0)-I(V_1;Y_1|V_0), \ I(V_1;V_2|V_0)+I(V_0;Z|U)-I(V_0, V_1;Y_1|U)\},
	\end{align*}
Note that $U\to V_0\to (V_1, V_2)\to (Y_1, Y_2, Z)$ forms a Markov chain such that \eqref{eqn: FM C1}, \eqref{eqn: FM C2}, \eqref{eqn: FM C3}, \eqref{eqn: FM C4}, \eqref{eqn: FM C5} hold. 
Further, we notice that $I_1=I(V_1;Z|V_0)$ since
\begin{align*}
	I(V_1;V_2|V_0)-I(V_2;Y_2|V_0) &\stackrel{(a)}{\leq}  I(V_1;V_2|V_0)-I(V_2;Z|V_0)\stackrel{(b)}{\leq} I(V_1;Z|V_0); \\
	I(V_1;V_2|V_0)+I(V_0;Z|U)-I(V_0, V_2;Y_2|U) &\stackrel{(c)}{\leq}  I(V_1;V_2|V_0)+I(V_0;Z|U)-I(V_0, V_2;Z|U)\stackrel{(b)}{\leq} I(V_1;Z|V_0),
\end{align*}
where 
	$(a)$ is due to \eqref{eqn: FM C3}; 
	$(b)$ is due to \eqref{eqn: FM C5}; and 
	$(c)$ is due to \eqref{eqn: FM C4}.
Similarly, we have $I_2=I(V_2;Z|V_0).$ Thus the region could be simplified into
	\begin{align*}
		R_1  \leq & \min \{I(V_0, V_1; Y_1)-I(V_1;Z|V_0),\ \  I(V_0, V_1; Y_1|U)-I(V_0,V_1;Z|U)+R_2\}\\
				= & I(V_0, V_1; Y_1|U)-I(V_0,V_1;Z|U)+\min\{R_2, \ I(U;Y_1)+I(V_0;Z|U)\};\\
		R_2  \leq & \min \{I(V_0, V_2; Y_2)-I(V_2;Z|V_0),\ \  I(V_0, V_2; Y_2|U)-I(V_0,V_2;Z|U)+R_1\}\\
				= &	I(V_0, V_2; Y_2|U)-I(V_0,V_2;Z|U)+\min\{R_1, \ I(U;Y_2)+I(V_0;Z|U)\}.
	\end{align*}  
This is the desired region in \eqref{eqn: Oma Marton IndS region} in Theorem \ref{thm: Oma Marton}.

%

\bibliographystyle{IEEEtran}
\bibliography{IEEEabrv,references}

\end{document}